\documentclass[12pt]{article}

\usepackage{graphicx}

\usepackage[letterpaper,margin=1in]{geometry}

\renewenvironment{abstract}
	{\quotation}
	{\endquotation}

\date{}

\makeatletter
\renewcommand{\fnum@figure}{\textbf{Figure \thefigure}}
\renewcommand{\fnum@table}{\textbf{Table \thetable}}
\makeatother

\usepackage{scicite}

\usepackage{url}
\usepackage{array}
\usepackage{algorithm}
\usepackage{algpseudocode} 
\usepackage{subcaption}
\usepackage{diagbox}
\usepackage{comment}
\usepackage{xcolor}
\usepackage{amsmath}
\usepackage{enumitem}
\usepackage{makecell}
\usepackage{booktabs}    
\usepackage{csvsimple}   
\usepackage{multirow}
\usepackage{amssymb}

\newcommand{\cmark}{\textcolor{black}{$\mathbf{\checkmark}$}} 
\newcommand{\xmark}{\textcolor{black}{$\mathbf{\times}$}}

\newenvironment{proof}[1][Proof]{\par\noindent\textbf{#1.} }{\hfill$\square$\par}

\def\scititle{
 \vspace{-0.5cm}
  Dimensionless learning based on information 
}
\title{\bfseries \boldmath \scititle}

\author{
	Yuan Yuan$^{1\ast}$ \&
	Adri\'an Lozano-Dur\'{a}n$^{1,2}$ \and
	\small$^{1}$Department of Aeronautics and Astronautics, Massachusetts Institute of Technology, Cambridge, MA \and 
	\small$^{2}$Graduate Aerospace Laboratories, California Institute of Technology, Pasadena, CA. \and 
	\small$^\ast$Corresponding author. Email: yuany999@mit.edu\and
}

\makeindex
\begin{document} 

\maketitle

\begin{abstract} 
\noindent Dimensional analysis is one of the most fundamental tools
for understanding physical systems. However, the construction of
dimensionless variables, as guided by the Buckingham-$\pi$ theorem, is
not uniquely determined. Here, we introduce IT-$\pi$, a model-free
method that combines dimensionless learning with the principles of
information theory. Grounded in the irreducible error theorem,
IT-$\pi$ identifies dimensionless variables with the highest
predictive power by measuring their shared information content.  The
approach is able to rank variables by predictability, identify distinct
physical regimes, uncover self-similar variables, determine the
characteristic scales of the problem, and extract its dimensionless
parameters. IT-$\pi$ also provides a bound of the minimum predictive
error achievable across all possible models, from simple linear
regression to advanced deep learning techniques, naturally enabling a
definition of model efficiency.  We benchmark IT-$\pi$ across
different cases and demonstrate that it offers superior performance
and capabilities compared to existing tools. The method is also
applied to conduct dimensionless learning for supersonic turbulence,
aerodynamic drag on both smooth and irregular surfaces,
magnetohydrodynamic power generation, and laser-metal interaction.
\end{abstract}


\vspace{-0.5cm}
\noindent
\subsection*{Introduction}

Physical laws and models must adhere to the principle of dimensional
homogeneity~\cite{buckingham1914physically, szirtes2007applied}, i.e.,
they must be independent of the units used to express their variables.
A similar idea was first introduced by Newton in his
\textit{Principia}, under the term ``dynamically similar
systems''~\cite{sterrett2017physically}, although Galileo had already
employed the notion of similar systems when discussing pendulum
motions. Over the following centuries, the concept of similar systems
was loosely applied in a variety of fields, including engineering
(Froude, Bertrand, Reech), theoretical physics (van der Waals, Onnes,
Lorentz, Maxwell, Boltzmann), and theoretical and experimental
hydrodynamics (Stokes, Helmholtz, Reynolds, Prandtl,
Rayleigh)~\cite{sterrett2017physically}.  The approach was formally
articulated in the early 20th century, laying the foundation for what
is known today as dimensional analysis~\cite{buckingham1914physically,
  palmer2008dimensional, lysik1990dimensional}.

The applications of dimensional analysis extend beyond the
construction of dimensionally consistent physical laws. It provides
the conditions under which two systems share identical behavior
(dynamic similarity), allowing predictions from laboratory experiments
to be extended to real-world
applications~\cite{sterrett2017physically,
  reynolds1883xxix}. Dimensional analysis also facilitates
dimensionality reduction, \textcolor{black}{i.e., simplifying physical
  problems to their most fundamental forms}, thus decreasing the
amount of data required for
analysis~\cite{sonin2004generalization,segel1972simplification}. Another
application is the discovery of self-similar variables, for which
systems exhibit invariant solutions under appropriate scaling
\cite{barenblatt1996scaling}. Finally, dimensional analysis can reveal
the physical regimes in which different phenomena dominate (e.g.,
incompressible versus compressible flow), enabling researchers to
identify the most influential physical mechanisms in a given
context~\cite{callaham2021learning}.

A key landmark of dimensional analysis is the Buckingham-$\pi$
theorem~\cite{buckingham1914physically}, which offers a systematic
framework for deriving dimensionless variables. However, the solution
is not unique, as there are infinitely many possible ways to construct
these variables.  To address this limitation, recent studies have
developed data-driven tools to identify unique dimensionless variables
that minimize the error for a given model structure, particularly in
the data-rich context enabled by modern simulations and experiments.
These methods combine dimensional analysis with machine learning
techniques to identify dimensionless variables using multivariate
linear regression\cite{mendez2005scaling}, polynomial
regression\cite{xie2022data, udrescu2020ai,jofre2020data}, ridge
regression\cite{bakarji2022dimensionally}, hypothesis
testing\cite{del2019lurking}, Gaussian process
regression\cite{constantine2017data}, neural
networks\cite{udrescu2020ai, bakarji2022dimensionally,
  xu2022artificial, watanabe2025data}, sparse identification of
nonlinear dynamics\cite{xie2022data,bakarji2022dimensionally},
clustering\cite{zhang2024clustering}, symbolic
regression\cite{udrescu2020ai,bempedelis2024extracting,
  schmidt2009distilling} and entropic principles~\cite{ma2024building,
  zhang2025}.  Table~\ref{tab:method_comparison} offers a
non-exhaustive comparative overview of data-driven methods for
discovering dimensionless variables, highlighting their respective
capabilities. These include applicability to ordinary and partial
differential equations (ODEs/PDEs), the ability to rank dimensionless
variables by predictability {(and thus relative
  importance)}, identification of distinct physical regimes, detection
of self-similar behavior, and extraction of characteristic scales
(e.g., length and time scales of the system).  Another key capability
is whether the method can determine a bound on the minimum possible
error across all models--which in turn enables the definition of a
model efficiency. The comparison in Table~\ref{tab:method_comparison}
indicates that, although many approaches incorporate several of these
properties, no single method currently supports all of these
capabilities simultaneously. One notable shortcoming of previous
{data-driven} methods is that they are not model-free;
i.e., the discovery of dimensionless variables relies on a predefined
model structure (e.g., linear regressions, neural networks,...).  This
can lead to potentially biased results, as the dimensionless variables
identified may not be optimal for other models.  {A
  detailed account of the challenges in dimensionless learning can be
  found in the Supplementary Materials.}
\begin{table}[h!]
\centering
\setlength{\arrayrulewidth}{1pt} 
\renewcommand{\arraystretch}{1} 
\resizebox{\textwidth}{!}{%
\begin{tabular}{|l|c|c|c|c|c|c|c|c|}
\hline
\textbf{Method} & \textbf{ODE/PDE} & \textbf{Model-} & \textbf{Input} & \textbf{Regime} & \textbf{Self-} & \textbf{Characteristic} & \textbf{Irreducible}& \textbf{Model} \\
 &  & \textbf{Free} & \textbf{Ranking} & \textbf{Detection} & \textbf{Similar} & \textbf{Scales} & \textbf{ Error} &\textbf{Efficiency} \\
\hline
Scaling LAWs \cite{mendez2005scaling} & \xmark & \xmark & \cmark & \xmark & -- & -- & \xmark & \xmark\\
\hline
Active Subspaces \cite{constantine2017data} & \cmark & \xmark & \cmark & \xmark & -- & -- & \xmark & \xmark\\
\hline
AI Feynman \cite{udrescu2020ai} & \xmark & \xmark & \xmark & \xmark & -- & -- & \xmark & \xmark\\
\hline
Clustering \cite{zhang2024clustering} & \cmark & \xmark & \cmark & \cmark & -- & -- & \xmark & \xmark\\
\hline
PyDimension \cite{xie2022data} & \cmark & \xmark & \xmark & \xmark & \cmark & \cmark & \xmark & \xmark\\
\hline
BuckiNet \cite{bakarji2022dimensionally} & \cmark & \xmark & \cmark & \xmark & \cmark & \cmark & \xmark & \xmark\\
\hline
{BSM} \cite{bempedelis2024extracting} & \cmark & \xmark & \xmark & \xmark & \cmark & \cmark & \xmark & \xmark\\
\hline
IT-$\pi$ (Current) & \cmark & \cmark & \cmark & \cmark & \cmark & \cmark & \cmark & \cmark\\
\hline
\end{tabular}%
}
\caption{\textbf{Overview of some data-driven dimensionless learning
    methods and their capabilities.} These capabilities include
  whether each method is applicable to ODEs/PDEs, operates in a
  model-free manner, ranks inputs by predictability, identifies
  distinct physical regimes, uncovers self-similarity, determines
  characteristic scales and dimensionless parameters, provides a bound
  on the irreducible error, and evaluates model efficiency.  Entries
  marked with ``--'' indicate that although the method could
  potentially be extended to infer the corresponding physical property
  after deriving the dimensionless variables, the authors did not
  explicitly perform this step.}
\label{tab:method_comparison}
\end{table}

In this work, we present IT-$\pi$, an information-theoretic,
model-free method for dimensionless learning. An overview of the
method is shown in Fig.~\ref{fig:dim_analysis}.  Our approach is
motivated by a fundamental question: Which dimensionless
variables are best suited for predicting a quantity of interest,
\emph{independent of the modeling approach?} IT-$\pi$ addresses this
question by unifying the Buckingham-$\pi$ theorem with the irreducible
error theorem.  The central idea is that the predictive capability of
a model is bounded by the amount of information shared between the
input (dimensionless variables) and the output (dimensionless quantity
of interest)~\cite{lozano2022information, yuan2024limits}.

This manuscript is organized as follows. We begin by introducing the
information-theoretic irreducible error theorem, which establishes the
minimum achievable error across all possible models. This theorem
serves as the foundation for a generalized formulation of the
Buckingham-$\pi$ theorem, while enabling all capabilities outlined in
Table~\ref{tab:method_comparison}. We then apply IT-$\pi$ to a broad
set of validation and application cases. The validation cases, which
have known analytical solutions, are used to benchmark the performance
of IT-$\pi$. In contrast, the application cases—where no optimal
solution is known—are used to discover new dimensionless variables
governing the underlying physics. Finally, we compare the performance
of IT-$\pi$ with other dimensionless learning methods across all
studied cases.


\subsection*{Results}

\subsubsection*{Dimensionless learning based on information}
%
A physical model (or law) aims to predict a dimensional quantity $q_o$
using a set of $n$ dimensional input variables, $\mathbf{q} =
\left[q_1, q_2, \ldots, q_n\right]$, through the relation $\hat{q}_o =
\mathcal{F}(\mathbf{q})$, where $\hat{q}_o$ is an estimate of
$q_o$. As an example, consider the prediction of the gravitational
force between two objects, $q_o = F_g$, which depends on
$\mathbf{q} = \left[m_1, m_2, r, G\right]$, where $m_1$ and $m_2$
are the masses of the objects, $r$ is the distance between their
centers of mass, and $G$ is the gravitational constant.  According to
the Buckingham-$\pi$ theorem~\cite{buckingham1914physically}, physical
models can be reformulated in a dimensionless form as $\hat{\Pi}_o =
f\bigl(\mathbf{\Pi}\bigr)$, where $\hat{\Pi}_o$ denotes the
predicted dimensionless output, and $\mathbf{\Pi} = \left[\Pi_1,
  \Pi_2, \ldots, \Pi_l\right]$ is the set of dimensionless input
variables. Each dimensionless variable has the form $\Pi_i =
q_1^{a_{i1}}\, q_2^{a_{i2}}\, \cdots\, q_n^{a_{in}}$ and the number of
required dimensionless inputs is upper bounded by $l = n - n_u$, where
$n_u$ is the number of fundamental units involved in the problem
(e.g., length, time, mass, electric current, temperature, amount of
substance, and luminous intensity).  For a given $L_p$-norm, the
success of the model is measured by the error $\epsilon_f = \|\Pi_o -
\hat{\Pi}_o\|_p$.\\

\textbf{Irreducible error as lack of information.}  Our approach is
grounded in the information-theoretic irreducible error theorem [see
  proof in the Supplementary Materials] \cite{lutwak2005crame, cover1999elements}. The key insight is that
prediction accuracy of any model is fundamentally limited by the
amount of information the input contains about the output, where
information here is defined within the framework of information
theory~\cite{shannon1948}. More precisely, the error across all
possible models ${f}$ is lower-bounded by
\begin{equation}
\label{eq:Renyi_lower_bound}
 \epsilon_f
\;\geq\; e^{- I_{\alpha}(\Pi_o ; \mathbf{\Pi})} \cdot c(\alpha,p, h_{\alpha,o})
\;\equiv\; \epsilon_{LB},
\end{equation}
where $I_{\alpha}\bigl(\Pi_o ; \mathbf{\Pi}\bigr)$ is the
R\'enyi mutual information of order $\alpha$~\cite{renyi1961measures},
which measures the shared information between $\Pi_o$ and
$\mathbf{\Pi}$.  The value of $c\ \!\bigl(\alpha, p, h_{\alpha
  ,o}\bigr)$ depends on the $L_p$-norm, $\alpha$ and the information
content of $\Pi_o$, denoted by $h_{\alpha,o}$ [see Methods].
The irreducible (lower bound) error is denoted as $\epsilon_{LB}$.
{The inequality in Eq.~(\ref{eq:Renyi_lower_bound})
  represents a fundamental mathematical constraint that holds
  regardless of the complexity of the statistical relationship between
  input and output, and must therefore be satisfied by any predictive
  model, irrespective of its form.}
When an exact functional relationship exists between the input and the
output, the information measure $I_\alpha\bigl(\Pi_o ;
\mathbf{\Pi}\bigr)$ converges to infinity, indicating that an
exact model is possible ($\epsilon_{LB}=0$). In contrast, if some of
the variables influencing $\Pi_o$ are inaccessible or unmeasurable,
$I_\alpha\bigl(\Pi_o ; \mathbf{\Pi}\bigr)$ remains finite, leading
to an irreducible error ($\epsilon_{LB}>0$) that cannot be eliminated.
It is interesting to note that the inequality in
Eq.~(\ref{eq:Renyi_lower_bound}) holds for a range of values of
$\alpha$. However, the most useful case occurs when $\epsilon_{LB}$ is
maximized, yielding the tightest bound.  { A detailed
  analysis of the role of $\alpha$ in Eq.~(\ref{eq:Renyi_lower_bound})
  is provided in the Supplementary Material.}

The irreducible error $\epsilon_{LB}$ has several useful
properties. First, it is independent of any particular model $f$.
Second, it is invariant under bijective transformations of inputs
when the output is fixed, reflecting the principle that such transformations produce
alternative yet equivalent model formulations. Third, it is sensitive
to the choice of the $L_p$-norm for the error. For example, predicting
extreme events (captured by high $L_p$-norms) may be more challenging
and require different variables than predicting weaker, common events
(captured by low $L_p$-norms) [see example in the Supplementary
  Materials].  Finally, Eq.~(\ref{eq:Renyi_lower_bound}) naturally
leads to the definition of the normalized irreducible error $
\tilde{\epsilon}_{LB} = e^{-I_\alpha(\Pi_o ; \mathbf{\Pi})}, $
which ranges from $0$---when exact predictions are possible---to
$1$---when predictions are essentially random guesses.  Occasionally,
we will refer to the percentage form of $\tilde{\epsilon}_{LB}$,
defined as $\%\tilde{\epsilon}_{LB} = \tilde{\epsilon}_{LB} \times
100$.\\

\begin{figure}
  \vspace{-0.7cm}
    \centering
    \includegraphics[width=1.0\linewidth]{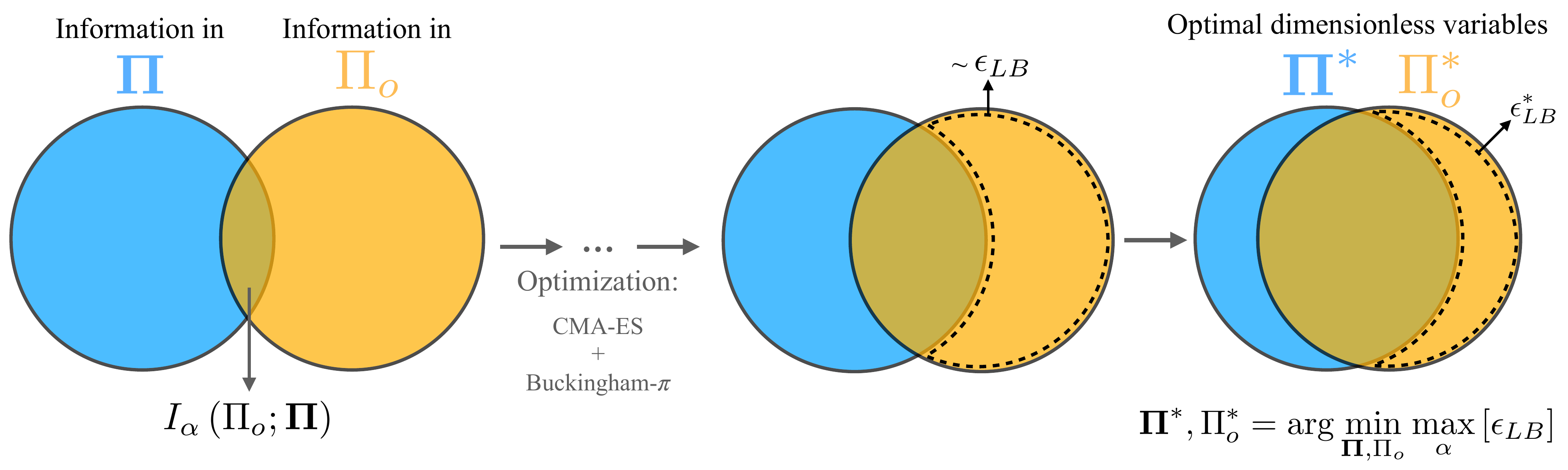}
    \caption{{Schematic of the optimization process in
        IT-$\pi$. Circles represent the information content of the
        dimensionless input ($\mathbf{\Pi}$) and output
        ($\Pi_o$). The overlapping region denotes their mutual
        information of order $\alpha$, $I_\alpha(\Pi_o;
        \mathbf{\Pi})$. The area enclosed by the dashed line
        corresponds to the amount of information in $\Pi_o$ that
        cannot be inferred from $\mathbf{\Pi}$, which is related
        to the irreducible error $\epsilon_{LB}$. The optimization is
        conducted with CMA-ES and proceeds from left to right,
        evaluating candidate pairs of $\mathbf{\Pi}$ and $\Pi_o$
        to minimize $\epsilon_{LB}$ until the optimal value
        $\epsilon_{LB}^*$ is reached. The candidates are constrained
        to be dimensionless by the Buckingham-$\pi$ theorem. The value
        of $\alpha$ determines the sensitivity of IT-$\pi$ to
        different $L_p$-norms and is learned as part of the
        optimization process.  }}
    \label{fig:IT-pi}
\end{figure}
\begin{figure}
    \centering
    \vspace{-0.25cm}
    \includegraphics[width=1.0\linewidth]{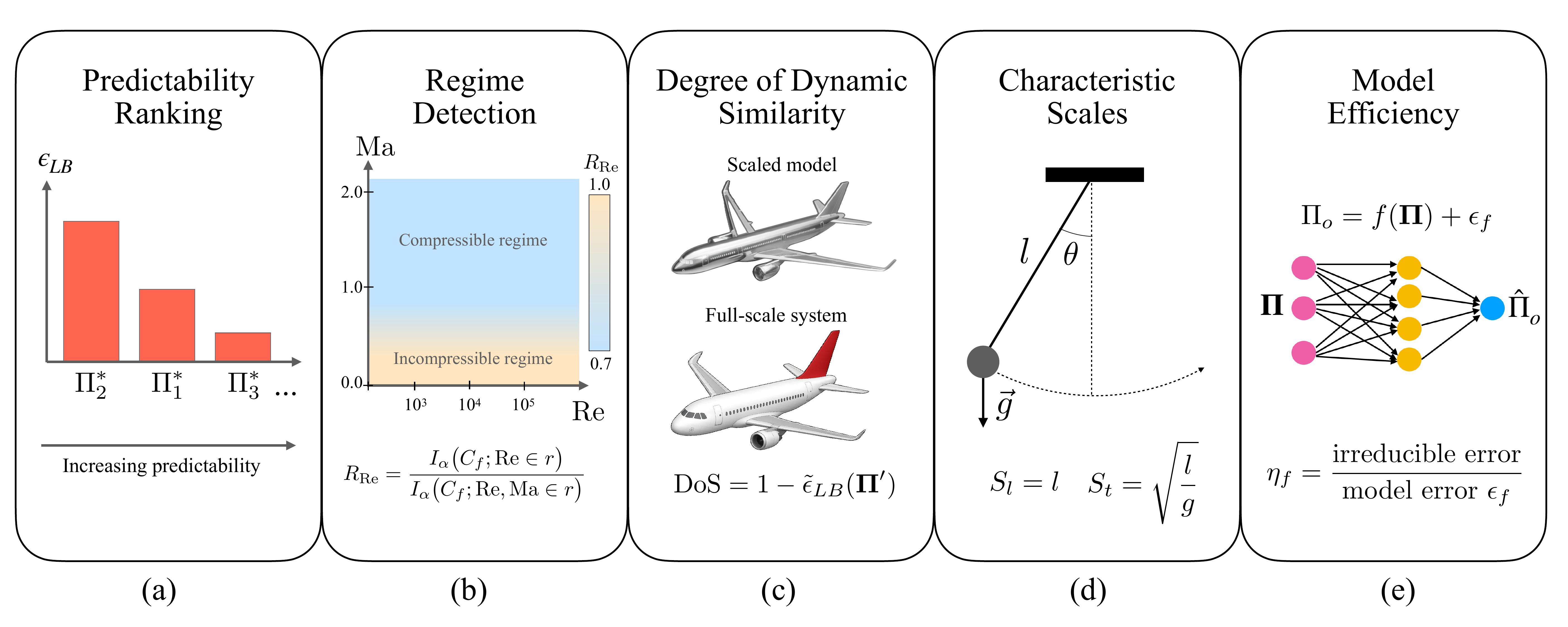}
    \caption{{ Illustration of the dimensionless
        learning capabilities enabled by IT-$\pi$.  (a) Ranking of
        candidate dimensionless variables based on their
        predictability.  (b) Example of regime detection for the skin
        friction coefficient $C_f$ in a wall-bounded turbulent flow,
        with Reynolds number (Re) and Mach number (Ma) as
        dimensionless inputs. The score $R_{\text{Re}}$ quantifies the
        relative importance of Re in predicting $C_f$.  (c) Degree of
        dynamic similarity (DoS) between the lift coefficient of a
        wind tunnel–scaled aircraft model and that of the
        corresponding full-scale vehicle.  (d) Characteristic length
        scale ($S_l$) and time scale ($S_t$) identified for a simple
        pendulum.  (e) Model efficiency in predicting a target
        quantity $\Pi_o$ using a neural network with inputs
        $\mathbf{\Pi}$ and prediction error $\epsilon_f$.
        Additional details and discussion of the examples are provided
        in the text.  }}
    \label{fig:dim_analysis}
\end{figure}

\textbf{Information-theoretic Buckingham-$\pi$ theorem (IT-$\pi$).}
Following Eq.~(\ref{eq:Renyi_lower_bound}), we define the
optimal dimensionless inputs $\mathbf{\Pi}^* =\ \left[\Pi^*_1,
  \Pi^*_2, \ldots, \Pi^*_{l^*}\right]$ and dimensionless output $\Pi_o^*$ 
  for a given $L_p$-norm as those
  satisfying
\begin{equation}
\label{eq:IT_pi}
\mathbf{\Pi}^*, \; \Pi_o^* = \arg \min _{\mathbf{\Pi}, \Pi_o} \max_{\alpha} [ \epsilon_{LB} ].
\end{equation}
{Figure~\ref{fig:IT-pi} illustrates the optimization
  process from Eq.~(\ref{eq:IT_pi})}.  This model-free formulation
ensures that the identified dimensionless variables yield the highest
predictive capabilities irrespective of the modeling approach. If
desired, the output can be fixed in dimensionless form, requiring only
$\mathbf{\Pi}^*$ to be discovered. The irreducible error using the
optimal dimensionless inputs $\mathbf{\Pi}^*$ is denoted by
$\epsilon_{LB}^* = \epsilon_{LB}(\mathbf{\Pi^*})$. It is satisfied
that $\epsilon_f \geq \epsilon_{LB} \geq \epsilon_{LB}^*$.  The
optimization problem from Eq.~(\ref{eq:IT_pi}) can be efficiently
solved by employing the covariance matrix adaptation evolution
strategy (CMA-ES)~\cite{hansen2003reducing} constrained to the
dimensionless candidates for $\mathbf{\Pi}$ and $\Pi_o$ from the
(classical) Buckingham-$\pi$ theorem [see
  Methods]. {Next, we discuss the different
  capabilities enabled by IT-$\pi$, with illustrative examples
  provided in Figure~\ref{fig:dim_analysis}}.\\

\textbf{Ranking of dimensionless variables by predictability.}  The
variables in $\mathbf{\Pi}^*$ can be ranked by predictability
according to $\epsilon_{LB}(\Pi_1^*) \geq \epsilon_{LB}(\Pi_2^*) \geq
\cdots \geq \epsilon_{LB}(\Pi_{l^*}^*)$.  This ranking applies not
only to individual variables but also to pairs of variables, such as
$\epsilon_{LB}([\Pi_1^*, \Pi_2^*]) \geq \epsilon_{LB}([\Pi_1^*,
  \Pi_3^*])$, triplets, and so on. If considering additional variables
no longer provides new information about the output, then the error
does not decrease further, i.e., $\epsilon_{LB}([\Pi_1^*, \Pi_2^*,
  \ldots, \Pi_{l^*}^*]) = \epsilon_{LB}([\Pi_1^*, \Pi_2^*, \ldots,
  \Pi_{l^*+1}^*]) =\epsilon_{LB}^{*} $, where $l^*$ is the minimum
number of dimensionless variables to maximize predictability of the
output.  {Note that the number of dimensionless
  variables provided by the original Buckingham-$\pi$ theorem, $l = n
  - n_u$~\cite{buckingham1914physically}, represents an upper
  bound---i.e., the actual number of dimensionless variables required
  for a predictive model may be smaller. In contrast, the value of
  $l^*$ is tight, improving upon both Buckingham’s original bound, $l
  \geq l^*$, and the subsequent refinement proposed by
  Sonin~\cite{sonin2004generalization}}.  Identifying the exact number
of required inputs has several advantages.  For instance, it provides
a clear guideline for selecting the optimal set of inputs to balance
model complexity with prediction accuracy. {In the
  Applications section, we present a case in which Buckingham-$\pi$
  yields $l = 7$ dimensionless variables, whereas IT-$\pi$ identifies
  a significantly smaller set with $l^* = 2$}.\\


\textbf{Detection of physical regimes.}  Physical regimes are distinct
operating conditions of a system, each governed by a particular set of
dimensionless variables. As these variables vary and fall within
specific intervals, the system transitions to a different regime,
where new  effects become dominant. For instance, in fluid
mechanics, incompressible and compressible flow represent two distinct
physical regimes, each governed by unique flow characteristics.  In
the incompressible regime, the flow physics are governed solely by the
dimensionless Reynolds number. In contrast, compressible flows require
both the Reynolds and Mach numbers to accurately characterize the
dynamics.

IT-$\pi$ identifies physical regimes by evaluating the predictive
significance of each dimensionless input, $\Pi_i^*$, within specific
regions of the dimensionless space.  First, $\mathbf{\Pi}^*$ is
divided into $M$ regions, labeled as $r_1, r_2, \ldots, r_M$. In each
region $r_k$, a prediction score for $\Pi_i^*$ is computed as
$R_i\bigl(r_k\bigr) = I_{\alpha}\bigl(\Pi_o^* ; \Pi_i^* \in r_k
\bigr)/I_{\alpha}\bigl(\Pi_o^*; \mathbf{\Pi}^* \in r_k \bigr) \in
     [0,1]$.  The score $R_i\bigl(r_k\bigr)$ represents the relative
     importance of $\Pi_i^*$ in predicting the output $\Pi_o^*$ within
     the region $r_k$. By comparing these scores across regions, one
     can categorize dimensionless inputs into distinct physical
     regimes.  {Consider the example of predicting the
       skin friction of a turbulent flow in a pipe, the Reynolds
       number (Re) dominates in the incompressible flow regime without
       the influence of additional physical effects (e.g., buoyancy,
       magnetic forces, etc), as it would be indicated by a prediction
       score $R_{\mathrm{Re}}\bigl(r_{\text{incompressible}}\bigr)
       \approx 1$}.  However, in the compressible flow regime,
     $R_{\mathrm{Re}}\bigl(r_{\text{compressible}}\bigr) < 1$,
     indicating the need for an additional dimensionless number to
     fully determined the skin friction---in this case, the Mach
     number.\\

\textbf{Degree of dynamic similarity}.  According to classical
dimensional analysis, dynamic similarity is achieved when all
dimensionless inputs governing a physical system are exactly matched
between the scaled model and full-scale system. IT-$\pi$ generalizes
this concept by requiring similarity only for the optimal subset of
$l^*$ dimensionless variables, relaxing the conservative requirement
of matching all $l$ variables prescribed by classical
theory. Furthermore, the quantity $\text{DoS}=1-\tilde{\epsilon}_{LB}
\in [0,1]$ measures the \emph{degree of dynamic similarity} that can
be achieved.  Consider, for example, a wind tunnel experiment of a
scaled model of an aircraft where technical limitations restrict the
control to only a few dimensionless variables $\mathbf{\Pi}'$. In
this scenario, the value of
$\text{DoS}=1-\tilde{\epsilon}_{LB}(\mathbf{\Pi}')$ quantifies the
degree of dynamic similarity attainable matching only those
variables. This contrasts with traditional theory, which merely
indicates whether dynamic similarity is or is not attained without
offering insight into the extent of similarity when it is not
perfectly achieved.\\

\textbf{Characteristic scales.}  The characteristic scales of a
physical problem refer to the length, time, mass, and other
fundamental quantities that can be constructed from the parameters
involved in the system under study. These are essential not only for
non-dimensionalization, but also for understanding the order of
magnitude of the variables controlling the system.  To define these
scales, we can divide the dimensional inputs into two sets
$\mathbf{q}=\left[\mathbf{q}_{v}, \mathbf{q}_{p}\right]$,
where $\mathbf{q}_{v}$ consists of variables that vary in each
simulation/experiment (i.e., dependent and independent variables), and
$\mathbf{q}_{p}$ consists of variables that remain fixed for a
given simulation/experiment but change across problem configurations
(i.e., parameters).  The characteristic scales are constructed from
$\mathbf{q}_{p}$. For example, for a pendulum with the governing
equation $\mathrm{d}/\mathrm{d} t[\theta,
  \dot{\theta}]=\left[\dot{\theta},-g/l \sin \theta\right]$, the
variables are time ($t$), angular displacement ($\theta$), and angular
velocity ($\dot{\theta}$), yielding $\mathbf{q}_{v} = [t, \theta,
  \dot{\theta}]$, whereas the parameters include the pendulum length
($l$) and gravitational acceleration ($g$), giving
$\mathbf{q}_{p}=[l, g]$. As such, the characteristic length and
time scales of the pendulum are obtained from $\mathbf{q}_{p}$ as
$[S_l,S_t] = [l,\sqrt{l/g}]$.

IT-$\pi$ extracts the characteristic scales, $\mathbf{S} =
\left[S_1, S_2, \ldots, S_{n_u}\right]$, from $\mathbf{\Pi}^*$ by
identifying the combination of quantities in $\mathbf{q}_{p}$
required to non-dimensionalize the variables in $\mathbf{q}_{v}$
[see the Supplementary Materials for the theory and algorithm]. In the
previous example of the pendulum, IT-$\pi$ will identify the optimal
variable $\Pi^* = \dot{\theta} S_t$ with characteristic time scale
$S_t= \sqrt{l/g}$. If the dimensional group $\Pi_i$ depends solely on
quantities from $\mathbf{q}_{p}$, then it represents a
\emph{dimensionless parameter} (rather than a dimensionless variable),
as it encapsulates a relationship only between characteristic
scales. One example of dimensionless parameter is the Reynolds number,
that can be expressed as a ratio of two length scales and does not
change for a given flow setup.\\

\textbf{Self-similarity.} Another capability of IT-$\pi$ is the
detection of self-similar variables---those that cannot be made
dimensionless using only the parameters in $\mathbf{q}_{p}$. In
such instances, IT-$\pi$ identifies the need to incorporate additional
variables from $\mathbf{q}_{v}$ to non-dimensionalize $\Pi^*$. The
latter variable is then classified as self-similar, as it reveals an
invariance between the ratios of the dependent and/or independent
variables that govern the system.\\

\textbf{Model efficiency.}  A foundational property of IT-$\pi$ is its
model-free formulation.  This naturally leads to a definition of model
performance relative to the theoretical optimum.  Specifically, we
introduce the model efficiency $\eta(f) = \epsilon_{LB}^*/\epsilon_f
\in [0,1]$, which quantifies how closely the predictions of the model,
\(\hat{\Pi}_o = f(\mathbf{\Pi})\), approach the theoretical limit.
A low value of $\eta$ indicates that the model underperforms relative
to the optimal model. This underperformance may stem from inadequate
inputs or insufficient model complexity (e.g., too few layers or
neurons in an artificial neural network). Conversely, a value of
$\eta$ close to 1 implies that the model is extracting all the useful
information from the inputs, and further improvements are not
possible. An interesting interpretation of this efficiency is its
analogy to the Carnot cycle in thermodynamics~\cite{carnot1824}; in
this context, it serves as the \emph{Carnot cycle of physical laws},
setting a theoretical benchmark for the limits of predictive model
performance. {A diagnostic tool for assessing whether
  a model is suboptimal, optimal, or overfitting under finite-sample
  conditions is provided in the Supplementary Materials.}

\subsubsection*{Validation}

%
\begin{figure}
    \centering
    \includegraphics[width=0.95\linewidth]{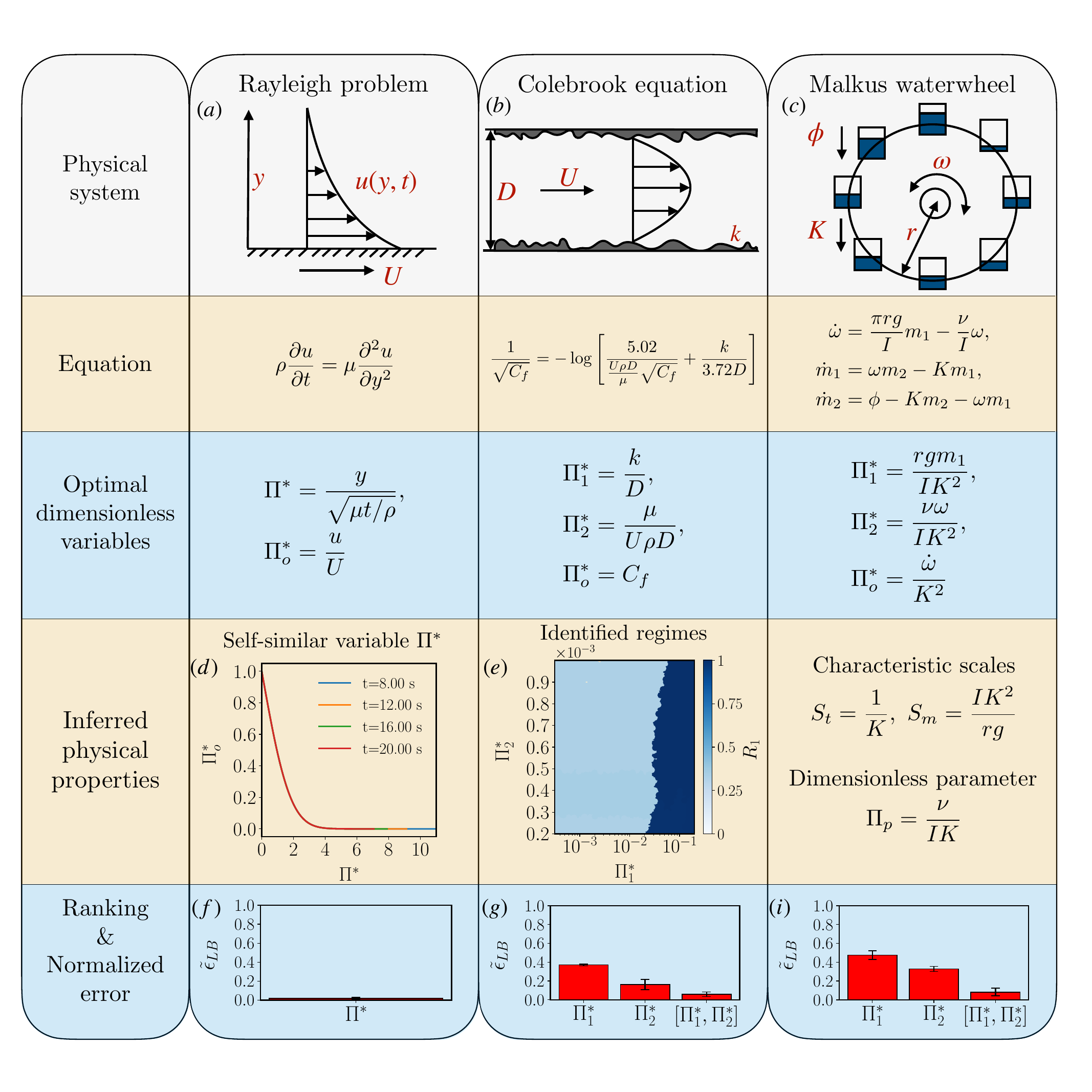}
    \caption{
\textbf{Summary of validation cases.}  
{Schematic representations of the physical systems (a, b, c)}, the corresponding systems of equations, and the optimal dimensionless inputs and outputs discovered by IT-$\pi$, are shown alongside the inferred physical properties, including self-similarity, distinct physical regimes, characteristic scales, relevant dimensionless parameters (where applicable), and the irreducible model error.  
(d) Rayleigh problem: Optimal dimensionless velocity profiles ($\Pi_o^*$) as a function of $\Pi^*$, with colors indicating different times $t$.  
(e) Colebrook equation: Contour plot of the prediction score $R_1$ {(i.e., the relative importance of $\Pi_1^*$)} as a function of the dimensionless inputs.  
{The result is obtained by dividing the dimensionless inputs $\mathbf{\Pi}^*$ into 10 clusters using the KNN method.}  
(f, g, i) Normalized irreducible error ($\tilde{\epsilon}_{LB}$) for individual components of $\mathbf{\Pi}^*$ and for all components combined, for (f) the Rayleigh problem, (g) the Colebrook equation, and (i) the Malkus waterwheel. Error bars indicate the uncertainty in the normalized irreducible error,  
{computed as the difference between the error bound estimated using the full dataset and that estimated using half the data.  
Additional details regarding the symbols and definitions of all variables are provided in the main text.}
}
\label{tab:validation_summary}
\end{figure}

We validate IT-$\pi$ on physical systems with known optimal
dimensionless inputs and physical properties. Our test cases include
the Rayleigh problem, the Colebrook equation, and the
Malkus-Howard-Lorenz water wheel. Figure~\ref{tab:validation_summary}
summarizes these cases by presenting the system equations alongside
the physical properties identified by IT-$\pi$, such as the optimal
dimensionless inputs and outputs, self-similarity, physical regimes,
characteristic scales, input ranking, and the information-theoretic
irreducible error. Although IT-$\pi$ infers a complete set of
properties for each case, the figure highlights only the most relevant
ones for clarity. Additional validation cases—including turbulent
Rayleigh-Bénard convection and the Blasius laminar boundary layer—are
discussed in the Methods. \\

\textbf{The Rayleigh Problem}~\cite{batchelor2000introduction} (see
Figure~\ref{tab:validation_summary}, Column 2) involves an infinitely
long wall that suddenly starts moving with a constant velocity $U$ in
the wall-parallel direction within an initially still, infinite
fluid. In the absence of a pressure gradient, the analytical solution
for the flow velocity is $u = U\operatorname{erfc}\left(\xi/2
\right)$, where $\xi = y/\sqrt{\mu t /\rho }$ is a self-similar
variable that combines the distance from the wall ($y$), viscosity
($\mu$), density ($\rho$), and time $(t)$ such that the flow profile
remains constant when scaled by $U$.

We generated samples of the velocity over time, $q_o = u$, with input
variables $\mathbf{q} = [\mathbf{q}_v , \mathbf{q}_p ]$,
where $\mathbf{q}_v = [y, t]$ and $\mathbf{q}_p = [U, \mu,
  \rho]$, and performed dimensionless learning using IT-$\pi$. The
optimal dimensionless input and output discovered are $\Pi^* = y
\rho^{0.5} /(t^{0.5} \mu^{0.5})$ and $\Pi_o^* = u/U$,
respectively. These dimensionless variables coincide with the
analytical solution and successfully collapse the velocity profiles
across different times, as shown in
Figure~\ref{tab:validation_summary}(d).  IT-$\pi$ further identifies
$\Pi^*$ as a self-similar variable because the characteristic length
and time scales cannot be constructed using only $U$, $\mu$ and
$\rho$. Finally, the near-zero irreducible error reported in
Figure~\ref{tab:validation_summary}(f) indicates that there exists a
model capable of exactly predicting the output.  Consequently, no
additional dimensionless inputs are required. Note that IT-$\pi$
identifies the need of only one dimensionless input ($l^*=1$), which
is less than the number of two inputs ($l=2$) inferred from the
Buckingham-$\pi$ theorem.\\

\textbf{The Colebrook Equation}~\cite{colebrook1939correspondence}
(see Figure~\ref{tab:validation_summary}, Column 3) is a widely used
formula in fluid mechanics for calculating the friction coefficient,
$C_f$, which measures the resistance encountered by turbulent flow
inside a pipe.  Accurately determining $C_f$ is crucial for designing
efficient piping systems and predicting energy losses due to friction
in various engineering applications~\cite{Smits2013}. This coefficient
depends on several factors, including the average roughness in the
interior surface of the pipe ($k$), its diameter ($D$), the flow
velocity ($U$), density ($\rho$), and viscosity ($\mu$).

After generating samples for $q_o = C_f$ and $\mathbf{q} =
[U,\rho,D,k,\mu]$, IT-$\pi$ discovered the optimal dimensionless
inputs $\Pi_1^* = k/D$, and $\Pi_2^* = \mu/(U \rho D)$, both of which
are consistent with the equation. The former represents the relative
roughness height, whereas the latter is related to the Reynolds number
$Re_D\equiv 1/\Pi_2^*$. The ranking in
Figure~\ref{tab:validation_summary}(g) shows that $\Pi_1^*$ and
$\Pi_2^*$ individually yield normalized irreducible errors
($\tilde{\epsilon}_{LB}$) in the output prediction of 40\% and 20\%,
respectively. When both inputs are considered, they reduce the
normalized irreducible error to nearly 0\%. The physical regimes
identified by IT-$\pi$ are illustrated in panel (e) of
Figure~\ref{tab:validation_summary}.  The figure depicts the prediction
score $R_1$ for $\Pi_1^*$ across the dimensionless input space, that
quantifies the importance of the roughness height in predicting the
friction coefficient $C_f$.  The results reveal two flow regimes: one
where the relative roughness height, $\Pi_1^*$, predominantly
determines the friction factor ($R_1 \approx 1$), and a second regime
where both the relative roughness height, $\Pi_1^*$, and the Reynolds
number, $\Pi_2^*$, are needed to explain $C_f$. A similar conclusion
can be drawn from $R_2$, which is omitted here for brevity.  The
regimes identified by IT-$\pi$ are consistent with those from
classical rough-wall turbulence analysis: the fully rough regime,
where pressure drag dominates over viscous drag, and the
transitionally rough regime, where both pressure and viscous drag
influence the total drag~\cite{jimenez2004turbulent,
  chung2021predicting}.\\

\textbf{The Malkus-Howard-Lorenz Water
  Wheel}~\cite{strogatz2018nonlinear} (see
Figure~\ref{tab:validation_summary}, Column 4) is a mechanical system
that exhibits chaotic dynamics. Water flows into compartments on a
rotating wheel, creating complex, unpredictable motion similar to that
observed in the Lorenz system~\cite{lorenz1963deterministic}.  The
dynamics of the system depend on the angular velocity ($\omega$) and
mass distributions ($m_1$ and $m_2$). The key system parameters
include the wheel's radius ($r$), gravitational acceleration ($g$),
moment of inertia ($I$), rotational damping ($\nu$), leakage rate
($K$), and the water influx ($\phi$).

Without loss of generality, we focus on the output $\dot{\omega}$,
although the same approach extends to the other outputs, $\dot {m_1}$
and $\dot{m_2}$.  The optimal dimensionless inputs discovered by
IT-$\pi$ are $\Pi_1^* = r g m_1/(I K^2)$ and $\Pi_2^* = \nu \omega/(I
K^2)$ with the dimensionless output $\Pi_o^*=\dot{\omega}/K^2$, which
recover the analytically derived dimensionless variables. The ranking
in Figure~\ref{tab:validation_summary}(i) reports the predictive
capabilities of the discovered $\mathbf{\Pi}^*$ groups. Using
$\Pi_1^*$ or $\Pi_2^*$ alone as inputs results in
$\%\tilde{\epsilon}_{LB}$ of 50\% and 30\%, respectively, while
considering both of them reduces considerably the normalized irreducible error. Finally, IT-$\pi$ uncovers the characteristic time and
mass scales as $S_t = 1/K$ and $S_m = I K^2/(r g)$, along with the
dimensionless parameter $\Pi_p = \nu / (IK)$. Hence, the dimensionless
input and output can be rewritten as $\Pi_o^* = \dot{\omega} S_t^2$,
$\Pi_1^* = m_1/S_m$, and $\Pi_2^* = \omega S_t \Pi_p$.\\

\subsubsection*{Applications}
\begin{figure}
    \centering
    \includegraphics[width=0.93\linewidth]{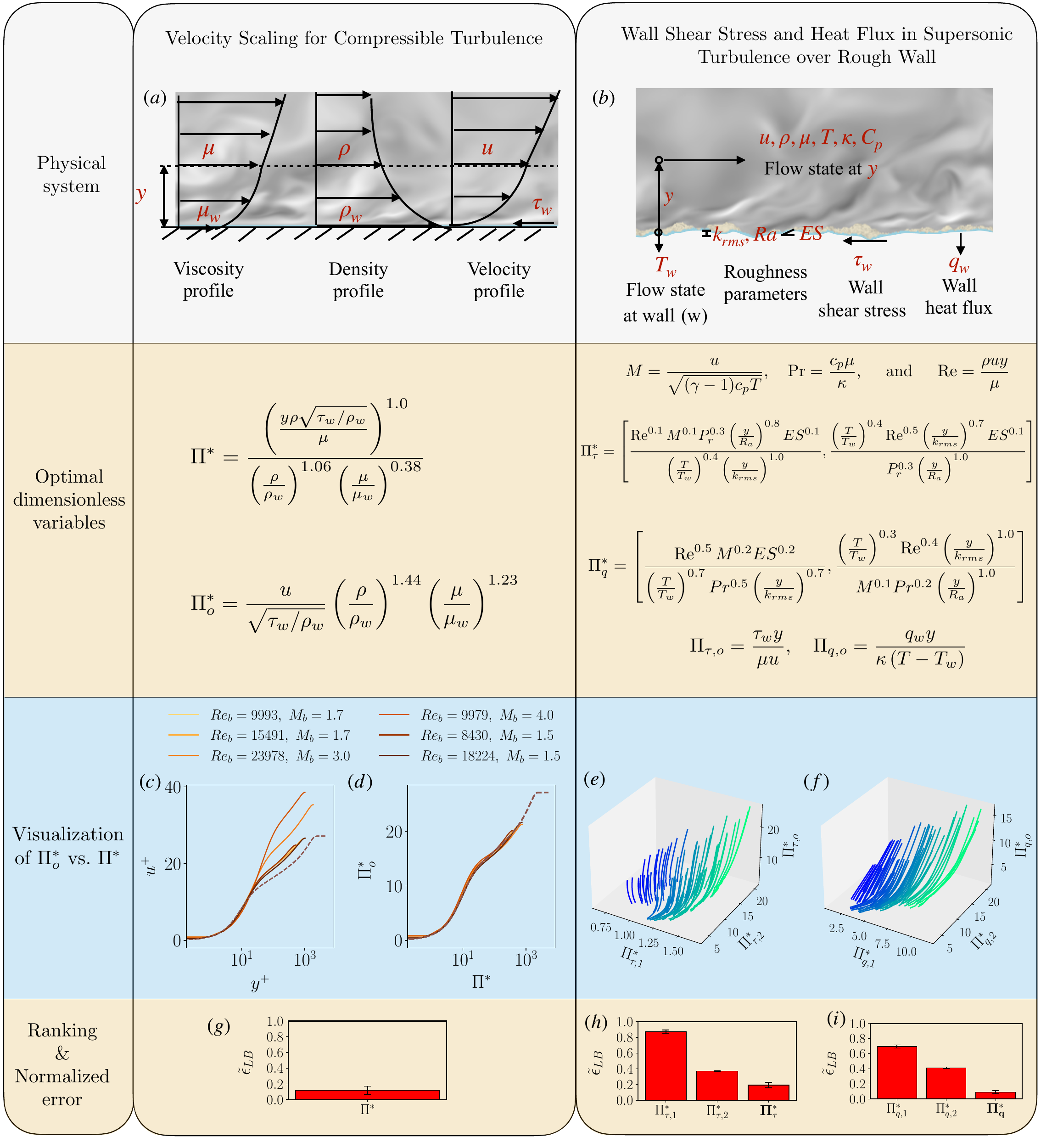}
\caption{ \small{
\textbf{Summary of compressible wall-bounded turbulence applications.}  
Left column: Velocity scaling for compressible turbulence.  
Right column: Wall shear stress and heat flux in supersonic turbulence over rough walls.  
{(a, b)} Schematic representations of the physical systems.  
(c, d): Solid lines of different colors represent velocity profiles at various Mach and Reynolds numbers; the dashed line corresponds to the incompressible turbulence velocity profile.  
(c) Viscous velocity profile $u^{+}$ versus $y^{+}$, where $u^{+} = u / \sqrt{\tau_w / \rho_w}$ and $y^{+} = y \rho \sqrt{\tau_w / \rho_w} / \mu$.  
(d) Dimensionless velocity profile obtained using IT-$\pi$ scaling.  
(e, f): Dimensionless (e) wall shear stress and (f) wall heat flux plotted against the optimal dimensionless inputs.  
{For compactness, the optimal dimensionless variables are expressed in terms of the Mach number ($M$), Prandtl number ($Pr$), and Reynolds number ($Re$), with definitions provided in the figure panels.}  
{The continuous lines represent data from the same simulation, colored by their mean $\Pi_1^*$ to aid visualization.}  
(g, h, i): Normalized irreducible error, with uncertainty shown as error bars {reported in Figure~\ref{tab:validation_summary}}, for (g) dimensionless velocity profile, (h) wall shear stress, and (i) wall heat flux.  
{Additional details regarding the symbols and definitions of all variables are provided in the main text.}
}}
\label{tab:compressible_wall_turbulence_summary}
\end{figure}

We have applied IT-$\pi$ to dimensionless learning across several
challenging problems, including supersonic turbulence, aerodynamic
drag on both smooth and irregular surfaces, magnetohydrodynamic power
generation, and  laser-metal interaction. Here, we focus on the discovery
of previously unknown scaling laws for supersonic flows over smooth
and rough surfaces.  The other applications can be found in the
Methods section.

Accurate prediction of high-speed turbulence near
solid boundaries is essential for advancing both commercial aviation
and space exploration~\cite{spina1994physics}. However, significant
challenges arise due to the complex interplay of the variables within
these systems. The challenges are twofold. From a fundamental physics
perspective, it is necessary to determine the scaling laws that govern
key quantities of interest, such as mean velocity and wall
fluxes. From a computational modeling standpoint, developing
parsimonious models is needed for achieving accurate predictions. We
leverage IT-$\pi$ to tackle both challenges. We also demonstrate the
use of the model efficiency in guiding the complexity of artificial
neural network (ANN) to predict wall heat flux.\\

{\bf Dimensionless learning for mean velocity}. Firstly, we discover a
local scaling for the mean velocity profile in compressible turbulent
channels using high-fidelity simulation data from existing
literature~\cite{trettel2016mean,modesti2019direct}. The dataset,
which spans different Reynolds and Mach numbers, includes the mean
velocity $q_o = u$ and the flow state $\mathbf{q} = \left[y, \rho,
  \mu, \rho_w, \mu_w, \tau_w\right]$, where $y$ is the wall-normal
distance, $\rho$ and $\mu$ are the local density and viscosity,
$\rho_w$ and $\mu_w$ are the density and viscosity at the wall, and
$\tau_w$ is the wall shear stress. By limiting the number of inputs to
one, IT-$\pi$ identifies the optimal dimensionless variable with the
highest predictive capabilities.  The dimensionless inputs and outputs
discovered by IT-$\pi$ are summarized in
Figure~\ref{tab:compressible_wall_turbulence_summary} (Column 2, Row
2). Panels (c) and (d) demonstrate that the scaling identified by
IT-$\pi$ improves the collapse of the compressible velocity profiles
across the range of Mach and Reynolds numbers considered compared to
the classic viscous scaling~\cite{karman1930mechanische}. A closer
inspection of the dimensionless input and output variables reveals
that this improvement is accomplished by accounting for local
variations in density and viscosity.\\

{\bf Dimensionless learning for wall fluxes}. Next, we identify the
optimal dimensionless variables for predicting wall fluxes in
compressible turbulence over rough walls~\cite{ma2024building}. The
output variables are the wall stress and heat flux, $q_o = [\tau_w,
  q_w]$, while the input variables are $\mathbf{q} = [y, u, \rho,
  T, T_w, \mu, \kappa, c_p, k_{rms}, R_a, ES]$. Here, $y$ is the
wall-normal distance; $u$, $\rho$, $T$, and $\mu$ represent the
velocity, density, temperature, and viscosity, respectively; $T_w$ is
the wall temperature; $\kappa$ is the thermal conductivity; and $c_p$
is the specific heat capacity. The last three inputs ($k_{rms}, R_a,
ES$) characterize the geometric properties of the surface
roughness. These include the root-mean-square roughness height
($k_{rms}$), the first-order roughness height fluctuations ($R_a$),
and the effective slope ($ES$)~\cite{ma2024building}.

Figure~\ref{tab:compressible_wall_turbulence_summary} (Column 3, Row 2)
summarizes the dimensionless forms of the optimal inputs and outputs
discovered by IT-$\pi$. These forms combine the local Reynolds, Mach,
Prandtl, and roughness numbers. The dimensionless wall shear stress
and heat flux are presented in
Figure~\ref{tab:compressible_wall_turbulence_summary}(e,f) as functions
of the identified dimensionless inputs. For both wall shear stress and
heat flux, two dimensionless inputs were sufficient to achieve
$\tilde{\epsilon}_{LB}^*\approx 0.1$, while the addition of further
variables resulted in only marginal improvements in the irreducible
error. Note that this number of variables is considerably smaller than the seven
dimensionless variables anticipated by the Buckingham-$\pi$
theorem. \\

{\bf Artificial neural network model for wall heat flux}. To
illustrate the application of the model efficiency $\eta$ in guiding
model complexity, we train three separate ANNs to predict the wall
heat flux using the optimal dimensionless inputs from IT-$\pi$. The
models are denoted by ANN$_1$, ANN$_2$ and ANN$_3$.  Each model
exhibits a different degree of complexity: $\mathrm{ANN}_1$ has $9$
tunable parameters (i.e., weights and biases), $\mathrm{ANN}_2$ has
$120$, while $\mathrm{ANN}_2$ has $781$. The simplest model,
$\mathrm{ANN}_1$, achieves an efficiency of $\eta_{1} = 30\%$,
indicating the need for additional layers and neurons to better
capture the underlying input-output relationships.  The second model,
$\mathrm{ANN}_2$, improves upon this with an efficiency of $\eta_{2} =
65\%$. The third model, $\mathrm{ANN}_3$, attains an efficiency of
$\eta_{3} = 98\%$, essentially matching the information-theoretic
limit in predictability.  As a result, we can conclude that no
additional model complexity is needed beyond $\mathrm{ANN}_3$.  We
show in the Supplementary Materials that training an ANN of similar
complexity to $\mathrm{ANN}_3$ using four suboptimal inputs from the
Buckingham-$\pi$ theorem results in a reduced efficiency of 82\%
despite using four inputs instead of two.

\subsubsection*{Comparison of IT-$\pi$ with previous dimensionless learning methods}

We compare IT-$\pi$ against four dimensionless learning methods:
Active Subspaces~\cite{constantine2017data},
PyDimension~\cite{xie2022data},
BuckiNet~\cite{bakarji2022dimensionally}, and
{BSM}~\cite{bempedelis2024extracting}. {
  BuckiNet specifically refers to the constrained optimization
  approach via kernel ridge regression, which is one of the three
  approaches proposed by Bakarji \emph{et al.} (2022).}  The
comparison spans all validation and application cases discussed
above. A summary of each method's capabilities was provided in
Table~\ref{tab:method_comparison}, and further details on their
formulations are available in the Supplementary Materials.

The results are summarized in Table~\ref{tab:compa_performance}. The
specific dimensionless variables identified by each method, along with
implementation details and model parameters, are provided in the
Methods section. Here, we offer an overview of the performance. In the
validation cases, success is measured by the ability to recover the
analytical optimal dimensionless variables. For the application
cases---where ground-truth solutions are unknown---performance is
quantified by the normalized irreducible error $\tilde{\epsilon}_{LB}$
associated with the input and output variables identified by each
method, with lower values indicating better performance.  The results
clearly demonstrate that IT-$\pi$ consistently outperforms or matches the other
methods across both validation and application cases, particularly in
the latter. It is worth noting that even in scenarios where existing
methods successfully identify the optimal dimensionless variables,
only IT-$\pi$ is capable of simultaneously inferring key physical
properties such as self-similar variables, distinct physical regimes,
characteristic scales, and governing dimensionless
parameters. Moreover, none of the other methods can provide a lower
error bound that is independent of specific modeling assumptions.

In terms of computational cost, all methods generate solutions within
seconds to minutes for the cases considered [see
  Table~\ref{tab:running time} in Methods]. Therefore, the
predictability of the discovered dimensionless inputs and outputs is
more important than the sheer computational cost of the method. This
situation may change when dealing with a large number of samples. In
such scenarios, IT-$\pi$ offers efficient linear scaling with respect
to the number of samples, performing similarly to or better than other
methods. {Beyond scaling, computational efficiency
  also depends on the structural design of the algorithm itself. Some
  of the previous approaches rely on a two-level optimization process,
  involving an outer loop that searches over candidate input
  combinations (e.g., powers of dimensionless groups), and an inner
  loop that fits model parameters (e.g., regression coefficients) to
  evaluate the performance of each candidate.  In contrast, IT-$\pi$
  eliminates this overhead by directly evaluating the irreducible
  model error, bypassing the need to fit any model.}
\begin{table}[h!]
\centering
\setlength{\arrayrulewidth}{1 pt} 
\resizebox{0.9 \textwidth}{!}{%
\begin{tabular}{|l|c|c|c|c|c|}
\hline
\textbf{Method} & \textbf{Active} & \textbf{PyDim-} & \textbf{Bucki-} & \textbf{{BSM}} & \textbf{IT-$\pi$}\\  & \textbf{Subspaces} & \textbf{ension} &  \textbf{Net} & & \textbf{(Current)}\\
\hline
Rayleigh problem  & \cmark & \cmark & \cmark & \cmark &\cmark \\
\hline
Colebrook equation & \cmark & \xmark & \cmark & \xmark &\cmark \\
\hline
Malkus waterwheel & \xmark & \xmark & \xmark & \xmark &\cmark \\
\hline
Rayleight-Bénard convection & \xmark  & \cmark & \xmark & \xmark &  \cmark\\
\hline
Blasius boundary layer& \cmark  & \cmark &\cmark  & \cmark  &  \cmark\\
\hline
Velocity scaling & $72\%$ & $21\%$ & $50\%$ & N/A & $12\%$ \\
\hline
Wall shear stress & $78\%$  & $54\%$  & $74\%$  & N/A & $19\%$ \\
\hline
Wall heat flux & $62\%$  & $44\%$ & $37\%$ & N/A & $9\%$ \\
\hline
Skin friction  & $19\%$  & $27\%$ & $75\%$ & N/A & $17\%$ \\
\hline
MHD generator & $7\%$  & $7\%$ & $98\%$ & N/A & $5\%$ \\
\hline
Laser–metal interaction & $94\%$& $25\%$ &$100\%$ & N/A & $23\%$\\
\hline
\end{tabular}%
}
\caption{Comparison of Active Subspaces, PyDimension, BuckiNet, and
  {BSM} across validation and application cases:
  Rayleigh problem, Colebrook equation, Malkus waterwheel,
  Rayleight-Bénard convection, Blasius boundary layer, Velocity
  scaling, Wall flux, Skin friction, MHD generator, Laser-metal
  interaction.  For the validation cases, the table presents whether
  the methods could identify the correct dimensionless variables for
  validations cases (\cmark or \xmark). For the application cases, the
  table shows the normalized irreducible error $\tilde{\epsilon}_{LB}$
  associated to the dimensionless input and output variables
  identified by each method with lower values indicating better
  performance.}
      \label{tab:compa_performance}

\end{table}
\section*{Discussion}


The concept of dimensional homogeneity---i.e., the invariance of
physical laws under transformation of units---is arguably one of the
most fundamental principles in physics. This simple yet powerful idea
gave rise to the field of dimensional analysis, which is widely used
across multiple disciplines.  In this work, we have introduced
IT-$\pi$, a formulation of dimensional analysis based on
information. Our approach is rooted in the information-theoretic
irreducible error, which allows us to identify the most predictive
dimensionless numbers with respect to a quantity of interest. The idea
goes beyond merely identifying a unique set of variables; it is the
realization that the information content in the variables of a system
is fundamental to understanding the governing physical laws and their
inherent limitations~\cite{landauer1991information,
  landauer1996physical, lozano2022information}. One can view IT-$\pi$
as the \emph{Carnot cycle of physical laws}: just as the thermodynamic
Carnot cycle sets an upper limit on the work extractable from two
thermal reservoirs---irrespective of the engine's technology---IT-$\pi$
extends this principle to predictive models irrespective of the
modeling approach. In this interpretation, the predictive power of a
set of variables is fundamentally constrained by the amount of
information they share with the quantity to be predicted, regardless
of whether the relationships are modeled through linear regression,
sophisticated neural networks, or analytical equations.

We have shown that IT-$\pi$ offers a complete set of dimensionless
learning tools, including ranking inputs by predictability,
identifying distinct physical regimes, uncovering self-similar
variables, and extracting characteristic scales and dimensionless
parameters.  IT-$\pi$ is also sensitive to the norm used to quantify
errors and the optimal set of dimensionless variables may vary
depending on the error metric of interest (e.g., prediction of ordinary
versus rare events).  Although some of these features are available
through other methods, none encompass them all. Even in cases where
alternative methods apply, IT-$\pi$ distinguishes itself by being
grounded in a theorem rather than relying on heuristic
reasoning. This makes IT-$\pi$ independent of specific modeling
assumptions.

In additional to its model-free nature, IT-$\pi$ offers unique
capabilities that other methods do not, such as establishing bounds on
the irreducible error and evaluating model efficiency. The former
allows us to precisely determine the actual number of relevant
dimensionless variables, $l^*$, which is typically overestimated by
the Buckingham-$\pi$ theorem.  Moreover, IT-$\pi$ quantifies the
degree of dynamic similarity achievable with the optimal variables,
rather than providing merely a binary yes-or-no answer as classical
dimensional analysis does. This feature can be decisive in designing
laboratory experiments for extrapolation to real-world
applications. For example, consider predicting the heat flux over a
rough surface as discussed in the application above. According to
Buckingham-$\pi$, seven dimensionless variables would be required. If
three different values must be measured to capture the scaling
behaviour of each variable, that would entail approximately $3^7=
2,187$ experiments. In contrast, IT-$\pi$ determined that only two
dimensionless variables are necessary to achieve a dynamic similarity
of 92\% (i.e., an 8\% normalized irreducible error). This entails a
significantly reduced effort of only $3^2=9$ experiments.  The same
reasoning applies to the construction of predictive modeling: models
with fewer inputs require orders of magnitude less training data
compared to those with high-dimensional inputs. In the previous
example, this factor would be of the order of 1,000.

Model efficiency is another distinctive feature of IT-$\pi$ that can
guide the structural complexity in model design. For instance,
machine-learning models are typically built with various architectures
and tunable parameters (e.g., weights and biases). In this context,
the model efficiency can determine whether a model operates near its
theoretical optimum---eliminating the need to explore alternative
architectures---or if there is potential for further improvement. We
have applied this concept to determine the optimal number of tunable
parameters for developing an ANN model for wall heat prediction. Our
results have shown that ANNs with only a few tens of parameters fail
to fully leverage the available input information, whereas nearly
1,000 parameters are necessary to extract that information
efficiently.

We have successfully validated IT-$\pi$ using cases with established
optimal dimensionless variables. These include classic problems in
fluid dynamics and dynamical systems, such as the Rayleigh problem,
the Colebrook equation, the Malkus-Howard-Lorenz water wheel, the
Rayleigh-Bénard convection, and the Blasius laminar boundary
layer. Moreover, IT-$\pi$ was applied to conduct dimensionless
learning for supersonic turbulence, aerodynamic drag on both smooth
and irregular surfaces, MHD power generation, and high-energy material
processing. In all cases, IT-$\pi$ has been shown to outperform or
match existing methods for dimensionless learning.

It is also important to acknowledge some shortcomings of the
approach. The first relates to its model-free nature. As mentioned
above, one of the key strengths of IT-$\pi$ is that its results do not
depend on any underlying model. However, some may view this as a
weakness, as it leaves the task of identifying the optimal model to
the practitioner. A more evident challenge is the amount of data
required. When many variables are involved, IT-$\pi$ necessitates the
estimation of mutual information in high dimensions. Although advanced
tools exist for high-dimensional
estimation~\cite{kraskov2004estimating, suzuki2008relative,
  singh2014nonparametric, gao2017demystifying, belghazi2018mine}, the
curse of dimensionality can render results inconclusive in certain
scenarios. Therefore, estimating the uncertainty in the normalized
irreducible error is crucial to determine whether the conclusions
drawn from IT-$\pi$ are statistically significant or merely reflect
insufficient data. {In all the results presented
  above, we have quantified the statistical uncertainty in the
  irreducible error, $\Delta \epsilon_{LB}$, which is represented by
  the error bars in the plots.  The methodology used to estimate
  uncertainty in IT-$\pi$ under a finite sample regime is detailed in
  the Supplementary Materials, along with a further evaluation of the
  sensitivity of IT-$\pi$ to the amount of available data.}

In conclusion, IT-$\pi$ offers a new perspective to dimensional
analysis rooted in information. Its broad applicability makes it a
useful tool across diverse disciplines---from fluid dynamics and
thermodynamics to electromagnetism, astrophysics, materials science,
and plasma physics. By effectively addressing challenges in scaling
laws, similarity solutions, and the identification of governing
dimensionless parameters, IT-$\pi$ provides a powerful tool for
dimensionless learning of complex physical systems.

\section*{Methods}

\subsection*{Constructing dimensionless variables using the Buckingham-$\pi$ theorem}
\label{sec:A_Buck_PI}

The Buckingham-$\pi$ theorem is used to construct dimensionless
candidates $\mathbf{\Pi}$ and $\Pi_o$.  The $i$-th dimensionless
variable has the form $$\Pi_i = q_1^{a_{i 1}} \cdot q_2^{a_{i 2}}
\cdots q_n^{a_{i n}}\equiv\mathbf{q}^{\mathbf{a}_i},$$ where
$\mathbf{a}_i=\left[a_{i 1}, a_{i 2}, \ldots, a_{i n}\right]^T$ is
the vector of exponents for $\Pi_i$.  The input candidate
$\mathbf{\Pi}$ is then obtained from the solution to
$\mathbf{D} \mathbf{a}_i=0$, where $\textit{\textbf{D}}$ is
the dimension matrix containing the powers of the fundamental units
for $\mathbf{q}$, $\mathbf{D} = \left[\mathbf{d}_1,
  \mathbf{d}_2, \ldots, \mathbf{d}_n\right],$ and
$\mathbf{d}_i$ is the dimensional vector for the physical quantity
$q_i$.  For example, the velocity $q_1=u= \mathrm{[length]}^1
\mathrm{[time]}^{-1} $ has $\mathbf{d}_1=[1,-1,0,0,0,0,0]^T$ and
so on. The solution $\mathbf{a}_i$ can be expressed as $
\mathbf{a}_i= \sum_{j=1}^{n-n_u} c_{i j} \mathbf{w}_j =
\mathbf{W} \mathbf{c}_i,$ where
$\mathbf{W}=\left[\mathbf{w}_1, \mathbf{w}_2, \ldots,
  \mathbf{w}_{n-n_u}\right]$ is the matrix of basis vectors of the
null space of $\mathbf{D}$, and $\mathbf{c}_i=\left[c_{i 1},
  c_{i 2}, \ldots, c_{i (n-n_u)}\right]^T$ is the coefficient vector
corresponding to $\mathbf{a}_i$.  In conclusion, non-dimensional
variables are obtained by
$\mathbf{\Pi}=\mathbf{q}^{\mathbf{W
    C}}=\left[\mathbf{q}^{\mathbf{W} \mathbf{c}_1},
  \mathbf{q}^{\mathbf{W} \mathbf{c}_2}, \ldots,
  \mathbf{q}^{\mathbf{W} \mathbf{c}_l}\right],$ where
$\mathbf{C}=\left[\mathbf{c}_1, \mathbf{c}_2, \ldots,
  \mathbf{c}_l\right]$.  The dimensionless output $\Pi_o$ is
constructed similarly with the matrix of basis vectors
$\mathbf{W}_o$ and coefficients $\mathbf{c}_o$. An important
consideration when some variables $q_i$ may be negative is to
generalize the formulation to avoid imaginary numbers. Specifically,
we define $ \Pi_i = \text{sgn}(q_1) \, |q_1|^{a_{i1}} \,
\text{sgn}(q_2) \, |q_2|^{a_{i2}} \, \cdots \, \text{sgn}(q_n) \,
|q_n|^{a_{in}}, $ where $|\cdot|$ denotes the absolute value and
$\text{sgn}(\cdot)$ is the sign function. This approach preserves the
sign information of each $q_i$ while ensuring that the resulting
dimensionless variables remain real-valued.

\subsection*{Information content of variables}
\label{sec:Info_content}

Consider the random variables $\Pi_o$ and $\mathbf{\Pi}$, whose
realizations are denoted by $\pi_o$ and $\mathbf{\pi}$,
respectively. They are characterized by the joint probability
distribution $\rho_{\Pi_o, \mathbf{\Pi}}(\pi_o, \mathbf{\pi})$
with corresponding marginal distributions $\rho_{\Pi_o}(\pi_o)$ and
$\rho_{\mathbf{\Pi}}(\mathbf{\pi})$.
The R\'enyi mutual information of order
$\alpha>0$~\cite{renyi1961measures} between $\Pi_o$ and
$\mathbf{\mathbf{\Pi}}$ is
$$
I_\alpha(\Pi_o ; \mathbf{\mathbf{\Pi}}) = h_\alpha(\Pi_o )- h_\alpha(\Pi_o \mid \mathbf{\mathbf{\Pi}}),
$$
where $h_\alpha(\Pi_o )$ and $h_\alpha(\Pi_o | \mathbf{\Pi})$ are
the R\'enyi entropy and conditional R\'enyi entropy, respectively,
which are given by
\begin{align*}
    h_{\alpha,o}  &\equiv h_\alpha(\Pi_o) = \lim _{\alpha' \rightarrow \alpha} \frac{1}{1-\alpha'} \log \left(\int \rho_{\Pi_o}^{\alpha'}(\pi_o) \, \mathrm{d} \pi_o\right), \\
    h_\alpha\left(\Pi_o| \mathbf{\Pi}\right)
    &=\lim _{\alpha^{\prime} \rightarrow \alpha}
    \frac{1}{1-\alpha^{\prime}} \int \rho_{\mathbf{\Pi}}(\mathbf{\pi}) \log \left(\int \rho_{\Pi_o \mid \mathbf{\Pi}}^{\alpha^{\prime}}\left(\pi_o \mid \mathbf{\pi}\right) d \pi_o\right) d \mathbf{\pi}.
\end{align*}
%

The Rényi mutual information between $\Pi_o$ and $\mathbf{\Pi}$
quantifies the amount of information about $\Pi_o$ that can be
extracted from $\mathbf{\Pi}$. It generalizes the Shannon mutual
information~\cite{shannon1948} by introducing the order parameter
$\alpha$, which is particularly valuable in situations where emphasis
on tail distributions is critical. IT-$\pi$ leverages the parameter
$\alpha$ to adjust sensitivity with respect to the $L_p$-norm,
balancing the influence of high-probability events against that of
low-probability events. When the value of $\alpha$ is equal to one,
the R\'enyi entropy corresponds to the Shannon
entropy~\cite{shannon1948}.\\

\subsection*{Optimization with CMA-ES}
\label{sec:M_IT-PI}

 The optimal dimensionless input and output variables are identified
 by solving Eq.~(\ref{eq:IT_pi}), where the candidate sets
 $\mathbf{\Pi}$ and $\Pi_o$ are constructed using the previously
 defined coefficient matrix $\mathbf{C}$. The optimization is
 carried out over the entries of $\mathbf{C}$. Specifically, the
 problem {$\arg \min_{\mathbf{\Pi}, \Pi_o}
   \max_{\alpha} [\epsilon_{LB}]$} is solved using the Covariance
 Matrix Adaptation Evolution Strategy
 (CMA-ES)~\cite{hansen2003reducing}, a stochastic, derivative-free
 algorithm designed for non-linear and non-convex continuous
 optimization problems. CMA-ES generates candidate solutions by
 sampling from a multivariate Gaussian distribution and iteratively
 updates its mean and covariance matrix to efficiently explore the
 search space and converge toward an optimal solution that maximizes
 the mutual information.  {To determine the value of
   $\alpha$ that yields the tightest possible bound, $\max_{\alpha}
   [\epsilon_{LB}]$, we perform a golden-section search over the
   interval $\alpha \in \left(1/(1+p),
     10\right]$~\cite{press2007numerical}. A schematic overview of the
   IT-$\pi$ workflow for discovering optimal dimensionless variables
   is shown in Fig.~\ref{fig:dim_learning_IT_PI_detailed}, and the
   corresponding pseudocode is provided in the Supplementary
   Materials.}

In the main text, the error norm is set to $p = 2$, which is the
standard choice for measuring prediction errors. An illustrative
example demonstrating the use of alternative $L_p$ norms is provided
in the Supplementary Materials. The CMA-ES algorithm is executed with
a population size of 300, coefficient bounds $c_{ij} \in [-2, 2]$, a
maximum of 50,000 iterations, and an initial standard deviation of
0.5. {For all cases presented above, the exponents
  $c_{ij}$ are rounded to a finite number of significant digits
  without compromising the value of $\epsilon_{LB}$.}  When $\Pi_o^*$,
$p$, and $\alpha$ are fixed, the function $c(\alpha, p, h_{\alpha,o})$
becomes identical, and the task of identifying the most predictive
non-dimensional input simplifies to maximizing mutual information: $
\mathbf{\Pi}^* = \arg \max_{\mathbf{\Pi}}
I_{\alpha}\left(\Pi_o^*; \mathbf{\Pi}\right). $
\begin{figure}[h!]
    \centering
    \includegraphics[width=1\linewidth]{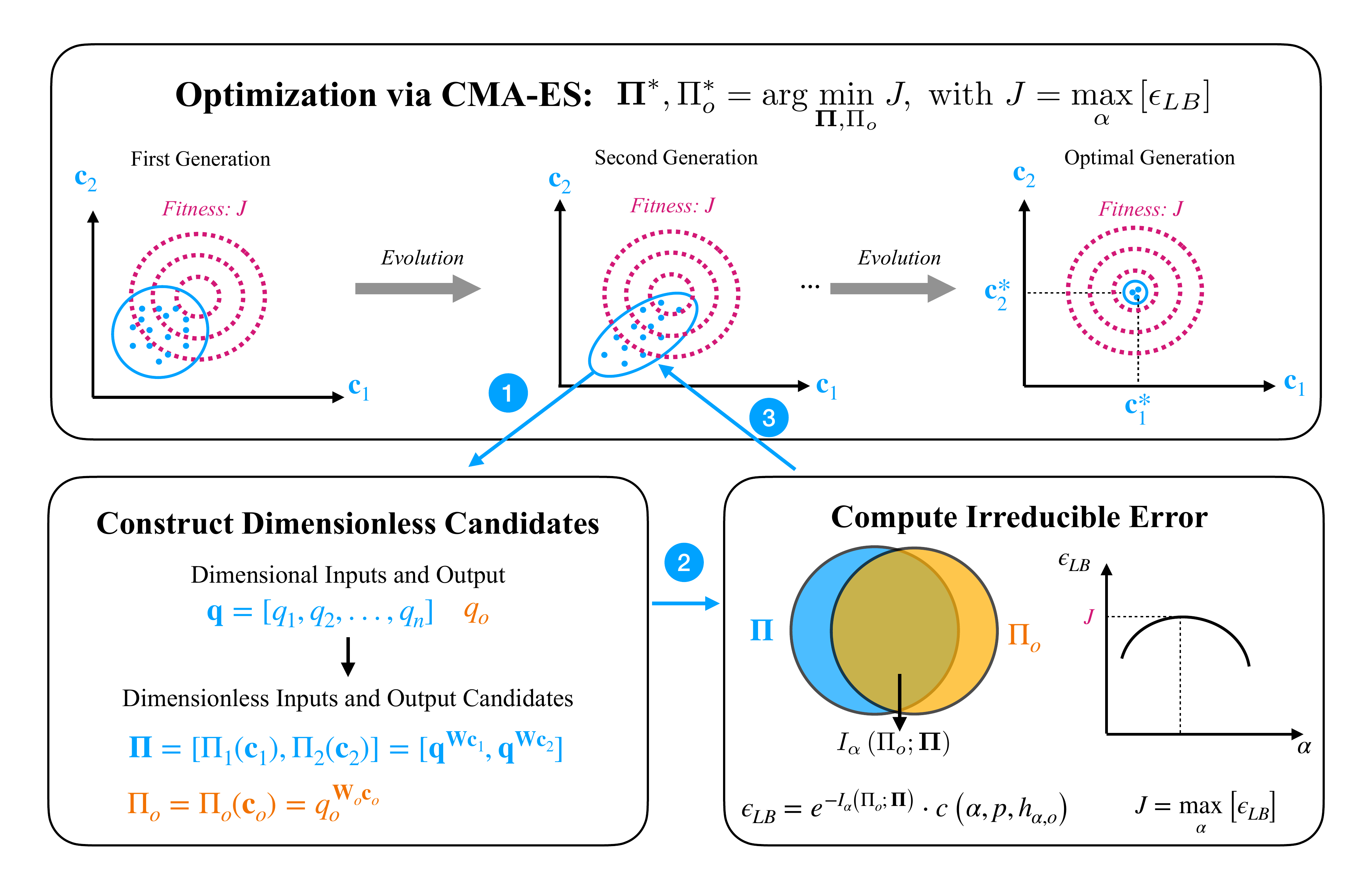}
    \caption{{ Workflow of IT-$\pi$ for discovering
        optimal dimensionless variables.  This schematic illustrates a
        specific example in which the input dimensionless variables
        $\mathbf{\Pi} = [\Pi_1, \Pi_2]$ are optimized. Each
        generation of the CMA-ES algorithm maintains a population of
        candidate solutions (represented by blue dots), where each
        individual encodes a set of exponents $(\mathbf{c}_1,
        \mathbf{c}_2)$ that define the candidate dimensionless
        variables.  For each individual, the variables are constructed
        as $\Pi_1 = \mathbf{q}^{\mathbf{W}\mathbf{c}_1}$
        and $\Pi_2 = \mathbf{q}^{\mathbf{W}\mathbf{c}_2}$,
        where $\mathbf{q}$ represents the set of dimensional
        quantities and $\mathbf{W}$ is the matrix of basis
        vectors. The fitness of each candidate is evaluated using the
        irreducible error across Rényi orders, $ J =
        \max_{\alpha}[\epsilon_{LB}] = \max_{\alpha}
        \left[e^{-I_\alpha(\Pi_o ; \mathbf{\Pi})} \cdot c(\alpha,
          p, h_{\alpha, o})\right], $ where $I_\alpha$ denotes the
        Rényi mutual information.  The optimal $\alpha$ is found
        through golden-section search in the range $(1/(1+p),10]$.
        The algorithm evolves the population across generations by
        updating the mean and covariance of the sampling distribution,
        navigating the fitness landscape (whose constant-value
        contours are shown as dark purple dashed lines) to minimize
        the irreducible error. The final generation yields the
        maximally predictive dimensionless variables
        $\mathbf{\Pi}^* = [\Pi_1^*, \Pi_2^*] =
        [\mathbf{q}^{\mathbf{W}\mathbf{c}_1^*},
          \mathbf{q}^{\mathbf{W}\mathbf{c}_2^*}]$.  The
        dimensionless output $\Pi_o$ can either be included in the
        optimization or treated as fixed. For clarity, this schematic
        illustrates only the optimization of $\mathbf{c}_1$ and
        $\mathbf{c}_2$ associated with the input variables.  }}
    \label{fig:dim_learning_IT_PI_detailed}
\end{figure}

\subsection*{Details about validations cases}

The Rayleigh problem dataset consists of samples uniformly generated
over $y \in[0,1] \mathrm{m}$, $t \in[0.01,5] \mathrm{s}$, $U
\in[0.5,1.0] \mathrm{m/s}$, $\mu \in\left[10^{-3}, 10^{-2}\right]
\mathrm{kg/m/s}$ and $\rho =1 \mathrm{kg/m^3}$. For the Colebrook
dataset, $\log _{10} R e_D$ is uniformly sampled in [3,5], yielding
Reynolds numbers in $\left[10^3, 10^5\right]$, and $\log _{10}(k / D)$
is uniformly sampled in $[-5,-0.7]$. The discovered dimensionless
inputs $\left[\Pi_1^*, \Pi_2^*\right.$] are divided into 10 clusters
using the K-Nearest Neighbors (KNN) clustering algorithm
\cite{cover1967nearest}.  For the Malkus waterwheel dataset, the
physical variables are uniformly sampled within the following ranges:
radius $r \in[0.3,0.7] \mathrm{m}$, water influx rate $q
\in[0.0001,0.0005] \mathrm{kg} / \mathrm{s}$, moment of inertia $I
\in[0.05,0.2] \mathrm{kg} \cdot \mathrm{m}^2$, rotational damping $\nu
\in[0.01,0.1]\mathrm{kg} \cdot\mathrm{m^2} /\mathrm{s}$, and water
leakage rate $K \in[0.01,0.1] \mathrm{s}^{-1}$, gravitational
acceleration $g = 9.8 \mathrm{m/s^2}$. The system is simulated over a
time span of $t \in[0,50]$s with 500 evaluation points. 

\subsection*{Details about application cases}

\indent {\bf Data for compressible wall-bounded turbulence application}

\noindent  The dataset for mean-velocty transformation comprises mean flow
profiles from direct numerical simulation (DNS) of four compressible
channel flows and two compressible pipe flows
\cite{trettel2016mean,modesti2019direct}, characterized by bulk Mach
numbers $\left(M_b = {U_b}/{\sqrt{\gamma R T_w}} \right)$ between 1.5
and 4.0, and bulk Reynolds numbers $\left(Re_b = {\rho_b U_b
  \delta}/{\mu_w} \right)$ from 8430.2 to 23977.6, where
$\rho_b={1}/{\delta} \int_0^{\delta} \rho d y$ and $ U_b={1}/{\delta}
\int_0^{\delta} u d y$ are the bulk density and velocity,
respectively; $T_w$ and $\mu_w$ are the mean temperature and dynamic
viscosity at the wall; and $\delta$ is the channel half-height.
The dataset for wall shear stress and heat flux in supersonic
turbulence over rough wall includes DNS of turbulent channel flows
over and rough surfaces \cite{ma2024building}. Ten irregular,
multiscale rough surfaces were generated using Gaussian probability
density functions. Simulations were driven with uniform momentum and
energy sources to achieve $M_c={U_c}/{\sqrt{\gamma R T_w}}=0.5, 1, 2,
4$ and $Re_c={\rho_c U_c \delta}/{\mu_w}=4000, 8000, 16000$, where
$\rho_c$ and $U_c$ are the mean density and velocity at the channel
centerline, respectively; $T_w$ and $\mu_w$ are the mean temperature
and dynamic viscosity at the wall; and $\delta$ is the channel
half-height.\\

\noindent {\bf Neural networks for predicting the wall flux}

\noindent  The data is split into training (70\%), validation (15\%), and testing
(15\%) sets, with $L_2$ regularization (factor 0.9) used to control
overfitting.  Each network follows a feedforward architecture.  The
simplest network ANN$_1$ consists of 1 hidden layers with 2 neurons,
ANN$_2$ have 2 hidden layers with 10 neurons per layer, while ANN$_3$
have 4 hidden layers with 15 neurons per layer. All are trained using
gradient descent with momentum and an adaptive learning rate. The
training process employs a learning rate of $10^{-5}$, with a maximum
of 50000 iterations and a validation tolerance of 40,000 epochs without
improvement before stopping.

\subsection*{Additional validation cases}

\begin{figure}
    \centering
    \includegraphics[width=1.0\linewidth]{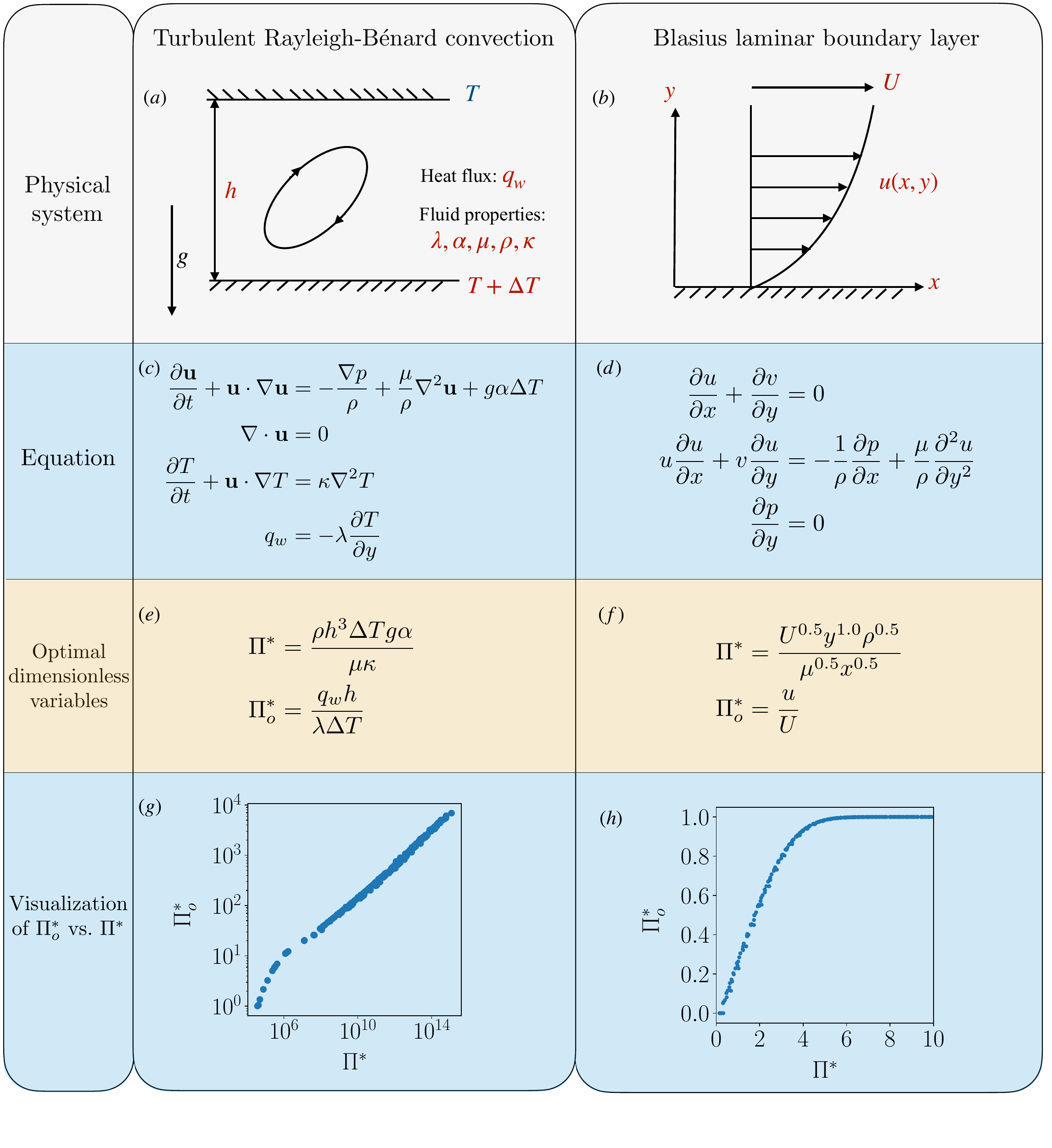}
\caption{\textbf{Summary of additional validation cases}. System
  equations, optimal dimensionless inputs and outputs discovered from
  IT-$\pi$, visualization of $\Pi_o^*$ as a function of $\Pi^*$.  {Additional details regarding the symbols and definitions are provided in the main text.}}
\label{tab:additional_validation_summary}
\end{figure}

We validate IT-$\pi$ using datasets from previous studies
\cite{xie2022data, bakarji2022dimensionally, bempedelis2024extracting}
with known optimal dimensionless inputs: the turbulent Rayleigh–Bénard
convection and the Blasius laminar boundary layer. Figure
\ref{tab:additional_validation_summary} summarizes each case,
detailing the system equations, optimal dimensionless inputs and
outputs discovered from IT-$\pi$. The figure also shows a visualization of $\Pi_o^*$ as a
function of $\mathbf{\Pi}^*$. \\

\noindent {\bf The Rayleigh–Bénard convection system} (Figure
\ref{tab:additional_validation_summary}, Column 2) describes
convection occurring in a planar horizontal layer of fluid heated from
below in a container with height $h$. The system is governed by the
equations in Figure \ref{tab:additional_validation_summary} (Column 2,
Row 2),
with parameters include viscosity ($\mu$), density ($\rho$), temperature
differences between the top and the bottom plane ($\Delta T$), thermal
expansion coefficient ($\alpha$), and thermal diffusivity ($\kappa$). The
dimensionless output is set to the Nusselt number $\Pi_o^* = {q_w
  h}/\left({\lambda \Delta T}\right)$, where $q_w$ is the heat flux,
$\lambda$ is the thermal conductivity. The data\cite{xie2022data}
include samples of the output $q_o = q_w$, inputs $\mathbf{q} =
[h,\Delta T, \lambda,g,\alpha,\mu, \rho,\kappa]$.  IT-$\pi$ discovered
the optimal dimensionless input $\Pi^*={\rho h^3 \Delta T g \alpha
}/\left({\mu \kappa}\right)$, which is consistent with the 
Rayleigh number \cite{rayleigh1916lix}. \\

\noindent {\bf The Blasius laminar boundary layer} (Figure
\ref{tab:additional_validation_summary}, Column 3) describes the
two-dimensional laminar boundary layer that forms on a semi-infinite
plate which is held parallel to a constant unidirectional flow. The
system is governed by the equations in Figure
\ref{tab:additional_validation_summary} (Column 3, Row 2), with
variables including the streamwise velocity ($u$), wall-normal
velocity ($v$), free-stream velocity ($U$), pressure ($p$), viscosity
($\mu$), density ($\rho$), streamwise distance ($x$) and wall normal
distance ($y$).  We focus on the output $u$. The
data~\cite{bakarji2022dimensionally} include samples of the output
$q_o=u$, and inputs $\mathbf{q}=\left[U, \mu, \rho, x,y\right]$.
IT-$\pi$ discovers the optimal, self-similar, dimensionless input
$\Pi^*={U^{0.5} y^{1.0}}\rho^{0.5}/\left({\mu^{0.5} x^{0.5}}\right)$,
which is equivalent to the analytical Blasius similarity variable
\cite{blasius1907grenzschichten}.\\

\subsection*{Additional application cases}

{\bf Skin friction under pressure gradient effects}

\noindent  We apply IT-$\pi$ to identify the most predictive model for wall
friction in turbulent flow over smooth surfaces under different mean
pressure gradients. Friction scaling and predictive modeling in
smooth-wall turbulence have been extensively studied for over a
century, owing to their crucial role in reducing operational costs in
engineering applications such as pipeline transport and aviation.
%
We use the data compiled by Dixit et \emph{al.} \cite{dixit2024generalized},
which includes experimental measurements and simulation results for
various flow conditions: mean zero-pressure-gradient (ZPG) flows in
channels, pipes, and turbulent boundary layers; mean
adverse-pressure-gradient (APG) turbulent boundary layers; mean
favorable-pressure-gradient (FPG) turbulent boundary layers; and
turbulent boundary layers on the pressure side of an airfoil. For a
detailed description of the data, please refer to Dixit et \emph{al.}
\cite{dixit2024generalized} and the references therein.

\begin{figure}
    \centering
    \includegraphics[width=1\linewidth]{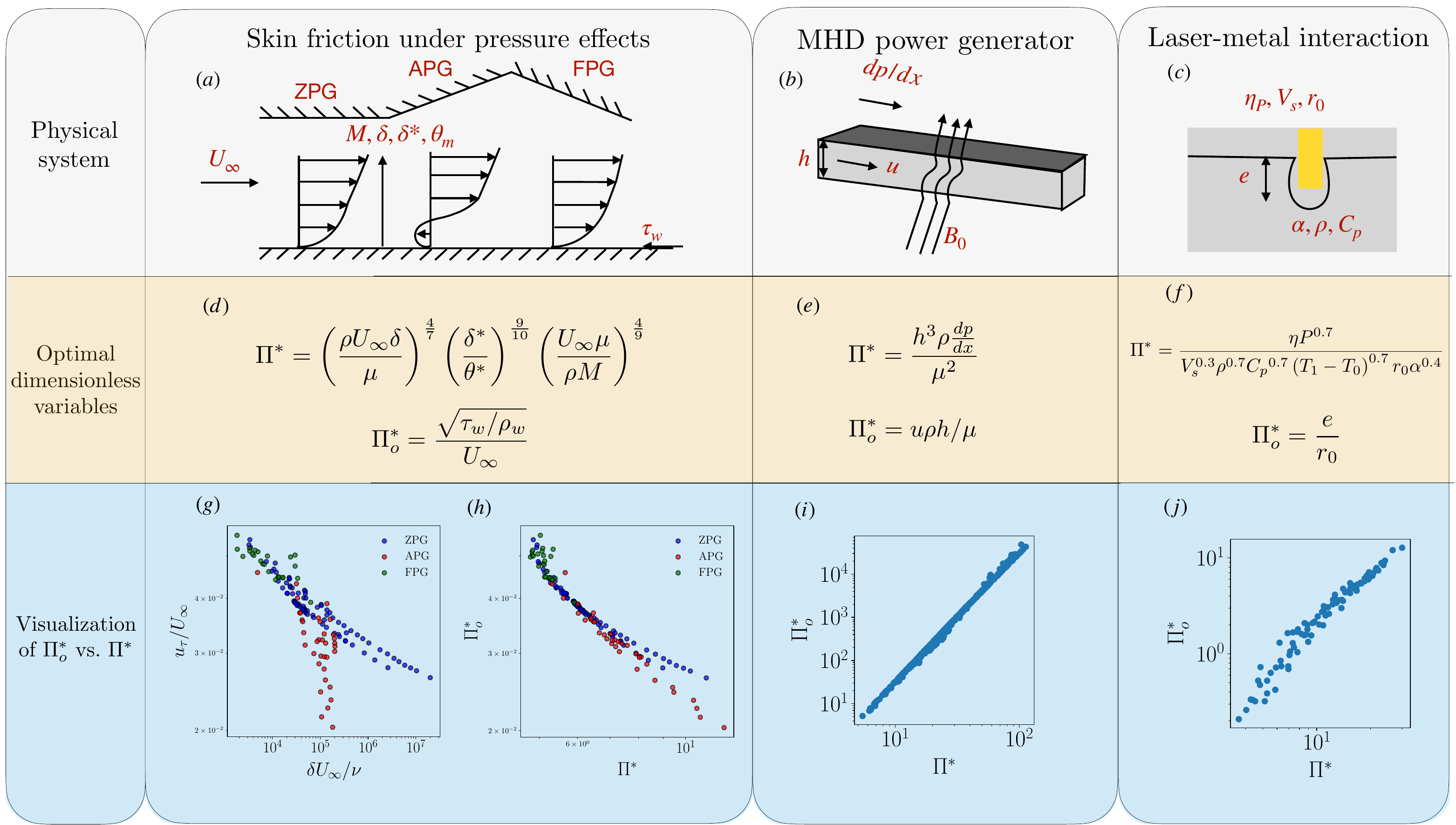}
\caption{\textbf{Summary of additional application cases}. System
  schematic, optimal dimensionless inputs and outputs discovered from
  IT-$\pi$, visualization of $\Pi_o^*$ as a function of $\Pi^*$.  {Additional details regarding the symbols and definitions are provided in the main text.}}
\label{tab:additional_application_summary}
\end{figure}

The dimensional input variables include $q=\left[U_{\infty}, \mu,
  \rho, M, \delta, \delta^*, \theta_m\right]$, where $U_{\infty}$ is
the free-stream velocity, $\mu$ is the viscosity, $\rho$ is the
density, $M=$ $\int_0^\delta u^2 d y$ is the total mean-flow kinetic
energy, $\delta$ is the boundary layer thickness at 99\% of the
free-stream, $\delta^*=\int_0^\delta\left(1-u / U_{\infty}\right) d y$
is the boundary layer displacement thickness, and
$\theta_m=\int_0^\delta {u}/{U_{\infty}}\left(1-u / U_{\infty}\right)
d y$ is the boundary layer momentum thickness, where $y$ is the
wall-normal distance.  The output variable $q_o=u_\tau =
\sqrt{\tau_w/\rho_w}$ is the friction velocity, with $\tau_w$ the wall
shear stress and $\rho_w$ the flow density at the wall.
%
We define the dimensionless output as $\Pi_o^* = u_{\tau}/U_{\infty}$,
where $u_{\tau}$ is the friction velocity, as this is a common form
for modeling skin friction.  For simplicity, we restrict the number of
input variables to one.  Under these conditions, IT-$\pi$ identifies
the most predictive single variable as
$$
\Pi^* = \left(\frac{\rho U_{\infty}\,\delta}{\mu}\right)^{\frac{4}{7}}
\left(\frac{\delta^*}{\theta_m}\right)^{\frac{9}{10}}
\left(\frac{U_{\infty}\,\mu}{\rho M}\right)^{\frac{4}{9}},
$$ with the exponents constrained to be rational
numbers. Figure~\ref{tab:additional_application_summary}(g),(h)
demonstrates that the scaling identified by IT-$\pi$ significantly
improves the collapse of the friction velocity data compared to the
classic approach~\cite{karman1930mechanische}. $\Pi^*$ is the product
of three dimensionless groups: the first two correspond to the classic
free-stream Reynolds number and the shape factor.  The third term is
more interesting: it represents the {transfer of kinetic energy from
  mean flow to large eddies of turbulence, which is derived from the
  momentum integral equation by Dixit et
  \emph{al.}~\cite{dixit2022universal} }.\\
%

\noindent {\bf Magnetohydrodynamics power generator}

\noindent Magnetohydrodynamic (MHD) generators represent an innovative solution
for sustainable and clean energy production
\cite{thring1965magnetohydrodynamic,Hammond1972}. Unlike conventional
generators that rely on moving mechanical components—such as
turbines—MHD generators convert thermal energy directly into
electrical power. This direct conversion not only minimizes mechanical
losses but also allows these systems to operate efficiently at
extremely high temperatures \cite{Hammond1972}. Moreover, owing to their unique
operational characteristics, MHD generators offer the highest
theoretical thermodynamic efficiency among all established methods of
electricity generation. In this section, we employ IT-$\pi$ to
identify the critical dimensionless input variables governing the flow velocity 
within the generator. 

The dataset used is obtained from numerical simulations of
steady-state MHD duct flow reported by Glaws \emph{et al.}
\cite{glaws2016dimension}. In this MHD generator configuration, an
electric current is induced by propelling a conducting fluid through
a square cross-sectional duct at a specified flow rate while
subjecting it to an externally applied vertical magnetic field. The
interaction between the moving fluid and the magnetic field causes the
field lines to bend, thereby producing a horizontal electric
current. The set of dimensional input variables is defined as
$
\mathbf{q} = \left[h,\mu,\rho,\frac{dp}{dx},\eta,B_0\right],
$
where $h$ denotes the side length of the square duct, $\mu$ and $\rho$
represent the viscosity and density of the conducting fluid,
respectively, $\frac{dp}{dx}$ is the applied pressure gradient, $\eta$
is the magnetic resistivity of the fluid, and $B_0$ is the magnitude
of the applied magnetic field. 

The quantity to predict is the average flow velocity, $u$, and we are
interested in identifying the single dimensionless input with the
highest predictive capability.  Using the dimensionless output
$\Pi_{o}^* = u\,\rho\,h/{\mu}$, IT-$\pi$ identifies the most
predictive dimensionless input as $\Pi^* =
h^3\,\rho\,\frac{dp}{dx}/\mu^2$. This result is consistent with
physical intuition: the average flow velocity is fundamentally
governed by the balance between the driving force (represented by the
pressure gradient) and the resisting force (arising from
viscosity). Hence, the dimensionless group $\Pi^*$ encapsulates the
interplay between these competing effects.\\

\noindent {\bf Laser-metal interaction}

\noindent Quantifying laser–metal interactions is critical for improving precision in advanced manufacturing processes such as additive manufacturing, laser cutting, and welding in aerospace applications~\cite{gu2012laser}. We employ IT-$\pi$ to identify the single most predictive dimensionless input governing the formation of a keyhole in a puddle of liquid metal melted by the laser.

The dataset used comes from high-speed X-ray imaging experiments of keyhole dynamics reported by Xie \emph{et al.}~\cite{xie2022data}. 
The set of dimensional input variables is defined as
$$\mathbf{q} = \left[
\eta P, V_s, r_0, \alpha, \rho, C_p, T_1 - T_0
\right],
$$
where $\eta P$ denotes the effective laser power, $V_s$ represents the laser scan speed, $r_0$ is the laser beam radius, and $\alpha$, $\rho$, and $C_p$ are the thermal diffusivity, density, and heat capacity of the material, respectively. $T_1 - T_0$ is the temperature difference between melting and ambient conditions. The quantity of interest is the normalized keyhole depth, defined as $\Pi_{o}^* = e / r_0$. IT-$\pi$ identifies the most predictive dimensionless input as
$
\Pi^* =
\frac{{\eta P}^{0.7}}{{V_s}^{0.3}  \rho^{0.7}  {C_p}^{0.7}  (T_1 - T_0)^{0.7}  {r_0}  \alpha^{0.4}}.
$

\subsection*{Details about comparison with other dimensionless learning methods}

The dimensionless input discovered by other methods using the same
output $\Pi_o^*$, is summarized in
the Supplementary Material.
Active Subspaces employs
Gaussian Process Regression with a radial basis function (RBF) kernel,
which is initialized with a width of $1$ and optimized using $5$
restarts. The gradients of the response surface are estimated using
finite differences.  PyDimension uses a 10th-order polynomial
regression model, optimizing the basis coefficients with a pattern
search method that is initialized on a grid ranging from $-1$ to $1$
in intervals of $0.1$.  BuckiNet utilizes Kernel Ridge Regression with
an RBF kernel (width = 1) and a regularization parameter of
$1\times10^{-4}$. Its optimization includes an $L_1$ regularization
term of $1\times10^{-3}$.  For {BSM}, the optimization
minimizes a distance-based loss over $40$ iterations, balancing
prediction accuracy and dimensional consistency with a weight of
$1\times10^{-3}$. It is also worth noting that the results were found
to be sensitive to the model parameters for each method.
{For all validation cases where the optimal solution
  is known, we tune the hyperparameters of all baseline methods to
  best match the known results. In application cases (where the true
  solution is not known), we evaluate the performance of existing
  methods using the standard or recommended hyperparameter settings
  provided by their original implementations. This approach reflects
  practical usage scenarios, where domain users often rely on default
  or minimally tuned settings. A discussion on the influence of
  hyperparameters and the sensitivity of each method is provided in
  the Supplementary Material.}

\begin{table}[h!]
\centering
\setlength{\arrayrulewidth}{1 pt} 
\resizebox{0.85 \textwidth}{!}{%
\begin{tabular}{|l|c|c|c|c|c|}
\hline
\textbf{Method} & \textbf{Active} & \textbf{PyDim-} & \textbf{Bucki-} & \textbf{{BSM}} & \textbf{IT-$\pi$}\\  & \textbf{Subspaces} & \textbf{ension} &  \textbf{Net} & & \textbf{(Current)}\\
\hline
Rayleigh problem  & $10.1$ & $0.2$ & $7.1$ & $1.4$ & $8.8$ \\
\hline
Colebrook equation & $8.6$ & $0.1$ & $14.3$ & N/A &$2.5$ \\
\hline
Malkus waterwheel & $36.5$ & $1.2$ & $269$ & N/A & $214$\\
\hline
Rayleight-Bénard convection &  $2.6$ & $2.7$ & $1.8$ & N/A &  7.0 \\
\hline
Blasius boundary layer& $0.6$  & $0.4$ & $1.7$ & $1.6$  & $1.2$ \\
\hline
Velocity scaling & $23.9$ & $4.8$ & $9.5$ & N/A & $10.6$ \\
\hline
Wall shear stress & $522$  & $109$  & $236$  & N/A & $29.4$ \\
\hline
Wall heat flux & $688$ & $288$ & $ 894 $ & N/A & $28.6$
\\
\hline
Skin friction  & $0.2$  & $3.5$ & $0.5$ & N/A & $1.5$ \\
\hline
MHD generator & $1.5$  & $1.4$ & $6.3$ & N/A & $3.0$ \\
\hline
Laser–metal interaction & $1.6$& $0.9$ &$0.2$ & N/A & $0.8$\\
\hline
\end{tabular}%
}
\caption{Comparison of running times (in seconds) to compute the
  dimensionless variables for Active Subspaces, PyDimension, BuckiNet,
  {BSM}, and IT-$\pi$ across different validation and
  application cases.}
      \label{tab:running time}

\end{table}

The corresponding running times for methods across various cases are
summarized in Table~\ref{tab:running time}. Assuming identical
optimization methods and candidate solutions across all approaches,
the primary cost differences arise from the function evaluation of a
single solution.  Active Subspaces and BuckiNet require kernel matrix
inversions with a computational cost of $O(N_{\text{samples}}^3)$,
where $N_{\text{samples}}$ is the number of data samples. For
PyDimension, fitting an $m$-th order polynomial model incurs a cost of
$O(N_{\text{samples}}^2 m^2)$. In the case of {BSM},
the cost is $O\left(N_{\text{samples}} \cdot n_t^2\right)$, where
$n_t$ is the number of independent variables. For IT-$\pi$, when using
the histogram method with $N_{\text{bins}}$ bins to estimate the
probability distribution, the computational cost is
$O\left(N_{\text{samples}} + N_{\text{bins}}^{l+1}\right)$.

\section*{Data availability}
The data generated in this study as well as the analysis code have been deposited in a Zenodo database \cite{yuan2025dimensionless} under identifier \url{https://doi.org/10.5281/zenodo.17080657}.

\section*{Code availability}
The code for this work is available at \url{https://github.com/ALD-Lab/IT_PI}.
\clearpage 





%
\bibliography{reference} 
\bibliographystyle{sciencemag}

%
%
%
%
%
%


\section*{Acknowledgments}
A.L.-D. and Y.Y. acknowledge support from the National Science Foundation under grant
No. 2140775 and grant No. 2317254, Early Career Faculty grant from NASA’s Space
Technology Research Grants Program (grant No. 80NSSC23K1498) and MISTI Global Seed Funds.
\section*{Author contributions statement}
Writing, visualization, validation, methodology, and conceptualization: Y.Y; 
Writing, project administration, methodology, investigation, funding acquisition, and conceptualization: A.L.-D.

\section*{Competing interests statement}
There are no conflicts of interest.

\clearpage
\clearpage
\renewcommand{\thefigure}{S\arabic{figure}}
\renewcommand{\thetable}{S\arabic{table}}
\renewcommand{\theequation}{S\arabic{equation}}
\renewcommand{\thepage}{S\arabic{page}}
\setcounter{figure}{0}
\setcounter{table}{0}
\setcounter{equation}{0}
\setcounter{page}{1} 

\renewcommand{\thesection}{S.\arabic{section}}
\setcounter{section}{0}
\begin{center}
\section*{Supplementary Materials for\\ Dimensionless learning based on information}
Yuan Yuan$^{\ast}$ \&
Adri\'an Lozano-Dur\'{a}n\\
\small$^\ast$Corresponding author. Email: yuany999@mit.com\\
\end{center}
\tableofcontents
\newpage

\section{{~Challenges and Methods for Dimensionless Learning}}


A brief overview of dimensionless learning methods and their primary
limitations and challenges was provided in the Introduction. In this
section, we offer a more in-depth analysis of these challenges,
examining both methodological and practical aspects, with particular
attention to how they are addressed by IT-$\pi$. We also review the
key technical strategies adopted by various dimensionless learning
approaches, summarizing their key underlying assumptions and modeling
frameworks.

\subsection{\textcolor{black}{Challenges}}
\label{sec:challenges}

Table~\ref{table:challenges} summarizes the key challenges associated
with dimensionless learning methods and outlines how each is addressed
by IT-$\pi$. These challenges can be grouped into the following main
categories:
\begin{itemize}

\item \textbf{Non-uniqueness of dimensionless variables.} Traditional
  dimensional analysis does not yield unique dimensionless
  variables. According to the Buckingham-$\pi$ theorem, the derived
  dimensionless variables span a linear subspace of exponents, and any
  basis within this subspace is equally valid. This lack of uniqueness
  can complicate the identification of the most relevant dimensionless
  parameters for a given physical system, potentially leading to
  ambiguity in model interpretation and application. For instance,
  different sets of dimensionless variables may describe the same
  system, making it challenging to select the most physically
  meaningful set without additional context or expertise. Most
  dimensionless learning methods, including IT-$\pi$ (see
  \S\ref{sec:proof}), address this challenge by identifying
  dimensionless variables that are optimal from a predictive
  standpoint. However, alternative objectives---such as simplicity of
  the dimensionless variables or ease of physical
  interpretability---are also important and have been comparatively
  less explored in the literature.

\item \textbf{Assumed model structure.} Most existing dimensionless
  learning methods rely on a predefined model structure to relate
  input and output variables. For example, some approaches assume that
  the relationship between dimensionless groups follows a power-law
  form, while others adopt polynomial expansions or, in more
  sophisticated cases, leverage machine learning models such as neural
  networks. Although these assumptions can simplify model training and
  interpretation, they inherently constrain the space of admissible
  functional relationships. In practice, the true underlying dynamics
  of a physical system may not conform to the assumed model
  structure. This mismatch can result in suboptimal or biased
  identification of dimensionless variables, especially in systems
  governed by highly nonlinear, discontinuous, or multiscale
  behavior. Consequently, methods that depend on rigid model forms may
  exhibit limited generalizability across diverse physical domains and
  may require additional validation or reconfiguration when applied to
  unfamiliar systems. Model-free approaches, such as IT-$\pi$, aim to
  address this issue by avoiding assumptions about the functional
  form, instead relying on the information-theoretic irreducible error
  theorem (see \S\ref{sec:proof}).

\item \textbf{Applicability to noisy real-world data.} Many
  dimensionless learning methods are primarily designed for idealized
  settings in which experiments can be freely conducted and
  measurement noise is minimal. In contrast, real-world
  applications—especially in engineering and experimental
  sciences—often involve data that are noisy, sparse, or
  incomplete. Such conditions pose significant challenges for
  dimensionless learning, as noise can obscure the underlying physical
  relationships and degrade the accuracy of the identified
  dimensionless variables. 
  IT-$\pi$ addresses this challenge through its theoretical robustness
  to noise. This resilience arises from the properties of the
  information-theoretic irreducible error, whose minimum remains
  invariant under additive noise (see
  \S\ref{sec:sensitivity_noise}). We have shown that IT-$\pi$ can
  tolerate noise levels of up to 20\% without significant degradation
  in performance, consistently recovering the correct dimensionless
  relationships across a wide range of test cases.

\item \textbf{Computational efficiency.} As the complexity of physical
  systems increases, so too does the computational burden associated
  with dimensionless learning. The optimization procedures required to
  identify optimal dimensionless groups can be susceptible to local
  minima, particularly when employing zero-order methods such as grid
  search or pattern search. Although increasing the number of initial
  samples or using global optimization strategies can mitigate this
  issue, such approaches come at the cost of increased computational
  expense—posing scalability challenges in high-dimensional parameter
  spaces or large-scale applications. These computational demands can
  become a significant bottleneck, especially when repeated
  evaluations of complex models are involved.  Despite these
  challenges, IT-$\pi$ demonstrates favorable computational
  performance. Its cost is generally low, and in worst-case scenarios,
  it remains comparable to that of other state-of-the-art
  dimensionless learning methods. Additional details supporting this
  comparison are provided in the ``Methods'' section under ``Details
  about comparison with other dimensionless learning methods.''

\item \textbf{Data requirements.}  Dimensionless learning methods
  often require substantial amounts of data to reliably identify
  predictive dimensionless variables, particularly in high-dimensional
  settings. As the number of input variables increases, the volume of
  the associated parameter space grows exponentially, leading to the
  well-known ``curse of dimensionality.'' In such regimes, data
  sparsity can undermine the reliability of statistical
  estimates—especially for metrics such as mutual information or
  regression error—and may result in inconclusive or misleading
  outcomes. This challenge is further exacerbated when the underlying
  relationships between variables are nonlinear or exhibit
  multi-regime behavior, which requires even denser sampling to
  capture accurately.  IT-$\pi$ partially mitigates this limitation
  through its model-free formulation and information-theoretic
  optimization strategy. As demonstrated in
  \S\ref{sec:sensitivity_stat}, IT-$\pi$ remains effective across a
  range of dimensionalities, provided that the available dataset
  sufficiently covers the relevant subspaces of interest.
  Nevertheless, when data are extremely limited or the input space is
  poorly sampled, the statistical uncertainty in the mutual
  information estimates can remain large, and the discovered variables
  should be interpreted with caution. In such cases, IT-$\pi$ is
  built-in with uncertainty quantification as described in the next
  point.

\item \textbf{Need for uncertainty quantification.} Reliable
  application of dimensionless learning requires rigorous
  quantification of uncertainty in the identified dimensionless
  variables. Without such analysis, it is difficult to assess whether
  the discovered relationships are statistically significant or merely
  artifacts of limited sample size, noisy measurements, or
  overfitting. In particular, when mutual information or
  regression-based metrics are estimated from finite datasets, the
  resulting values can exhibit high variance---especially in
  high-dimensional or poorly sampled regimes---leading to potentially
  spurious conclusions.  While several methods exist for estimating
  statistical uncertainty, including bootstrapping, subsampling, and
  analytical approximations, their application requires careful
  implementation and interpretation. These procedures often increase
  computational cost and introduce additional methodological
  complexity, which can deter widespread adoption. IT-$\pi$ addresses
  this need directly by incorporating uncertainty quantification into
  its framework. As described in \S\ref{sec:sensitivity_stat}, the
  results produced by IT-$\pi$ are accompanied by error bars on the
  estimated irreducible error $\epsilon_{LB}$, enabling users to
  assess the robustness and statistical reliability of the discovered
  dimensionless variables.

\item \textbf{Sensitivity to error norm.} Methods often implicitly
  assume a specific error norm---typically the $L_2$ norm---when
  evaluating model performance or identifying optimal dimensionless
  variables. However, the choice of error norm directly influences
  which aspects of the data are prioritized. For example, the $L_2$
  norm emphasizes frequent, moderate-magnitude events; higher-order
  norms (e.g., $L_4$ or $L_\infty$) place greater weight on extreme
  events; and lower-order norms (e.g., $L_1$) emphasize weaker
  events. This sensitivity can result in substantially different sets
  of dimensionless variables depending on the norm used, potentially
  yielding models that are well-suited to certain tasks (e.g., average
  prediction) but poorly suited to others (e.g., extreme event
  forecasting).  IT-$\pi$ explicitly addresses this sensitivity
  through its formulation of the irreducible error bound, which
  depends on the chosen Rényi entropy order $\alpha$. This enables the
  framework to accommodate different $L_p$ norms for quantifying
  prediction error (see \S\ref{sec:norm}).

\item \textbf{Dependency on the list of variables.} The effectiveness
  of any dimensionless learning method critically depends on the
  completeness of the input list of dimensional variables. If relevant
  variables are omitted, the resulting dimensionless variables may
  fail to capture key physical mechanisms, leading to suboptimal or
  misleading representations. Conversely, including irrelevant or
  redundant variables can obscure meaningful relationships and
  increase model complexity unnecessarily.  An additional challenge
  arises with discrete-valued variables, which many prediction-based
  methods are not well suited to handle due to their reliance on
  differentiability for gradient-based optimization.  IT-$\pi$
  addresses these issues through several mechanisms. It can diagnose
  missing variables by monitoring the normalized irreducible error,
  and detect redundancy by systematically analyzing subsets of input
  combinations (see \S\ref{sec:sensitivity_input}). Moreover, IT-$\pi$
  ranks input variables by their predictive contribution, enabling
  users to identify and exclude superfluous inputs that do not
  significantly decrease the irreducible error. Finally,
  discrete-valued variables are naturally supported through the
  discrete formulation of Rényi entropy, making IT-$\pi$ applicable to
  mixed-type input spaces.

\item \textbf{Sensitivity to hyperparameter choices.}  Dimensionless
  learning methods rely on hyperparameters that may influence their
  performance and the quality of the identified dimensionless
  variables. These hyperparameters include the degree of polynomial
  expansions, regularization weights, the number of terms in sparse
  regression libraries, or architectural choices in neural networks
  such as depth, width, and activation functions. Improper tuning can
  lead to overfitting, underfitting, or instability in the discovered
  dimensionless relationships, particularly when data are limited or
  noisy. As a result, these methods often require careful calibration
  or cross-validation, which adds to their computational and
  methodological complexity.  In contrast, IT-$\pi$ exhibits
  comparatively low sensitivity to hyperparameter choices. As
  demonstrated in \S\ref{sec:sensitivity_hyperparameter}, its
  performance remains robust across a wide range of hyperparameter
  settings, provided a sufficient number of samples are
  available. This robustness stems from the model-free nature of
  IT-$\pi$, which avoids fitting a specific function to the data and
  instead relies on information-theoretic criteria to evaluate
  candidate dimensionless variables.

\item \textbf{Identification of regimes, self-similarity, and more.}
  One of the major promises of dimensionless learning is its ability
  not only to reduce the number of variables but also to uncover
  deeper physical structures such as distinct regimes of behavior,
  characteristic scales, and self-similar solutions. However, most
  existing methods are limited in their ability to systematically
  detect such features. For instance, identifying regime
  boundaries---where dominant physical effects change---typically
  requires prior knowledge or manual inspection of the
  data. Similarly, discovering self-similar variables or invariant
  scalings often depends on assumed model structures or predefined
  transformation groups, which can bias results or limit
  applicability.  IT-$\pi$ addresses these challenges through its
  unified framework that is able to tackle identification of regimes
  and characteristic scales, self-similarity, a ranking of variables,
  and model efficiency (see Results).

\item \textbf{Model overfitting.} Overfitting is a significant concern
  in model-based dimensionless learning methods, particularly when
  function approximators such as high-degree polynomials, symbolic
  regressors, or neural networks are used. These methods may achieve
  near-perfect agreement with the training data by fitting spurious
  patterns or noise rather than capturing the underlying physical
  relationships. This is especially problematic in regimes with
  limited data, where high-capacity models can memorize the training
  set and generalize poorly to new observations. In such cases, the
  resulting dimensionless variables may appear predictive during
  training but fail to retain their relevance under data perturbations
  or when applied to unseen conditions.  IT-$\pi$ offers a principled
  approach to diagnose and mitigate overfitting by separating the
  intrinsic information content of the input-output relationship from
  the artifacts of a particular model fit. As detailed in
  \S\ref{sec:diagnostics_overfitting}, IT-$\pi$ can also quantify
  model overfitting using the statistical uncertainty-aware
  irreducible error, allowing practitioners to assess whether apparent
  improvements in predictability are statistically significant or
  likely due to overfitting.

\item \textbf{Generalization and transferability.} A central challenge
  in dimensionless learning is ensuring that the identified models or
  dimensionless variables generalize well to new, unseen data or to
  different physical systems. This issue is particularly important
  when the underlying assumptions, boundary conditions, or parameter
  distributions shift. In such cases, dimensionless variables
  discovered from one dataset may fail to capture the dominant
  physical mechanisms in another, resulting in reduced predictive
  accuracy or incorrect physical interpretation.  This lack of
  generalization is often not a limitation of the dimensionless
  learning method itself, but rather a consequence of the training
  data used. If the data fail to span the relevant physical regimes or
  omit key phenomena, the learned model will naturally exhibit poor
  transferability.  IT-$\pi$ provides a partial solution to this issue
  by offering a model-free formulation that explicitly quantifies the
  amount of information shared between the selected inputs and the
  target output. As shown in \S\ref{sec:data_partitioning}, IT-$\pi$
  can be used to evaluate the generalization properties of learned
  dimensionless variables by testing them across independently
  partitioned datasets. In the example of the ANN model for wall heat
  flux, we demonstrate that dimensionless variables discovered by
  IT-$\pi$ retain their predictive relevance even when applied to data
  drawn from different subsets of the original parameter space.
  Nonetheless, ensuring generalizability in entirely new or
  extrapolative contexts remains a challenging and open problem for
  all dimensionless learning methods.

\end{itemize}
  
\begin{table}[h]
  \centering
  \small

\begin{tabular}{p{4cm} p{8cm} p{2cm}}
\toprule
\textbf{Challenge} & \textbf{IT-\(\pi\) Strategy to tackle the challenge} & \textbf{Section} \\
\midrule
Non-uniqueness 
  & Identification dimensionless variables  optimal for predictive performance
  & \S\ref{sec:proof} \\

Assumed model structure
  & Model-free approach based on the information-theoretic irreducible error theorem
  & \S\ref{sec:proof} \\

Applicability to noisy  data
  & Theoretical robustness: irreducible error bound invariant under additive noise
  & \S\ref{sec:sensitivity_noise} \\

Computational efficiency
  & Low computational cost; comparable worst-case performance to other methods
  & Methods  \\

Data requirements
  & Low sensitivity of the discovered dimensionless variable to sample size
  & \S\ref{sec:sensitivity_stat} \\

Uncertainty quantification
  & Error bars on the estimated irreducible error for statistical significance
  & \S\ref{sec:sensitivity_stat} \\

Sensitivity to error norm
  & Irreducible error bound depending on Rényi entropy order and $L_p$ norm
  & \S\ref{sec:norm} \\

List of variables
  & Diagnose missing/superfluous inputs; handle discrete-valued inputs; rank variables by predictive contribution
  & \S\ref{sec:sensitivity_input} \\

Method hyperparameters
  & Robust performance across a wide range of hyperparameter settings
  & \S\ref{sec:sensitivity_hyperparameter} \\

Identification of regimes, self-similarity, and more
  & Unified framework for discovering regimes, characteristic scales, and self-similar solutions
  & Results \\

Model overfitting
  & Overfitting detected via uncertainty-aware irreducible error
  & \S\ref{sec:diagnostics_overfitting} \\

Generalization 
  & Test learned variables across independently partitioned datasets
  & \S\ref{sec:data_partitioning} \\
\bottomrule
\end{tabular}
\caption{\textcolor{black}{ Summary of key challenges in dimensionless
    learning, the IT-$\pi$ strategies developed to address each
    challenge, and the manuscript sections where these strategies are
    presented.}\label{table:challenges}}
\end{table}



\subsection{Overview of methods}
\label{sec:methods}

We provide an overview of the methodologies and distinguishing
features of various data-driven approaches for dimensionless
learning. Several of these methods are summarized in
Table~
1 of the main text.
\begin{itemize}
    \item \textbf{Scaling Laws} \cite{mendez2005scaling} combines
      dimensional analysis with backward elimination in multivariate
      regression under the assumption of a power-law relationship
      between input and output variables,
    $$ {\Pi_o}=e^{{\beta}_0} q_1^{\beta_1} \cdots q_n^{\beta_n} +
      \epsilon,$$ where $\epsilon$ is the model error.  The optimal
      coefficient vector $\mathbf{\beta}=[\beta_0,\cdots,\beta_n]$
      is determined by minimizing the linear regression model error,
    $$\min_{\mathbf{\beta}} \Big|\Big| \log \Pi_o
      -\beta_0-\sum_{j=1}^n \beta_j \log q_j \Big|\Big|_2.$$

    \item \textbf{Active Subspaces} \cite{constantine2017data}
      integrates dimension reduction techniques with global
      sensitivity analysis to construct unique dimensionless groups
      via two algorithms: the response surface-based approach and the
      finite difference-based approach.  The method constructs a
      response surface that relates the dimensionless output $\Pi_o$
      to the dimensionless input
      $\mathbf{\Pi}=\mathbf{q}^{\mathbf{W^T}}$, where
      $\mathbf{W}$ denotes the null space of the dimension matrix
      $\mathbf{D}$. The log of the input is
      $\mathbf{\gamma}=\log (\mathbf{\Pi})=\mathbf{W}^T
      \log (\mathbf{q})$.  A model is then used to
      approximate the output using the log of the input, $
      \Pi_o=\hat{g}(\mathbf{\gamma})+\epsilon $ where
      $\hat{g}(\mathbf{\gamma})$ is the predicted response
      function, and $\epsilon$ represents the modeling error. 
    In practice, $\hat{g}$ may be realized through various function approximators, such as kernel regression~\cite{constantine2017data} or neural networks~\cite{xu2022artificial}.
      The
      active subspace is identified through eigen-decomposition,
    $$ \int \nabla \hat{g} \nabla \hat{g}^T \sigma(\mathbf{\gamma}) d \mathbf{\gamma}
    \approx {\mathbf{U}} {\mathbf{\Lambda}} \mathbf{{U}}^T,
    $$ where $\nabla \hat{g}$ is the gradient of the regression model,
    $\sigma(\mathbf{\gamma})$ is the weight function,
    $\mathbf{U}$ is the eigenvector matrix, $\mathbf{\Lambda}$
    is a diagonal matrix of eigenvalues.  
    The discovered dimensionless
    groups are given by $\mathbf{\Pi}^* = [\Pi_1^*,\Pi_2^*,
      \cdots, \Pi_n^*]$, where
    $$ {\Pi}^*_i(\mathbf{q})=\exp \left(\mathbf{z}_i^T \log
    (\mathbf{q})\right),
    $$ where ${\mathbf{z}}_i=\mathbf{W} {\mathbf{u}}_i$
    and $\mathbf{u}_i$ is the $i$-th eigenvector from the
    eigenvector matrix $\mathbf{U}$.

    \item \textbf{AI Feynman} \cite{udrescu2020ai} discovers symbolic
      expressions from data by integrating dimensionless
      principles. The dimensionless input and output are constructed
      as
$$\Pi_i^{\prime} \equiv q_1^{W_{i 1}} q_2^{W_{i 2}} \cdots
      q_n^{W_{\text {in }}}, \quad \Pi_o^{\prime} \equiv
      \frac{q_o}{q_o^*}, \quad q_o^* \equiv q_1^{p_1} q_2^{p_2} \cdots
      q_n^{p_n},$$ where $\mathbf{D p}=\mathbf{b}$ with
      $\mathbf{p}=[p_1,p_2,\cdots]$, and $\mathbf{D
        W}=\mathbf{0}$, with $\mathbf{D}$ the dimension
      matrix, \(\mathbf{b}=[b_1,b_2,\cdots]\) is the dimension
      vector of the output. The method then employs a decision tree to
      systematically explore candidate models and identify a symbolic
      relationship between \(\Pi_o'\) and \(\mathbf{\Pi'}\).

    \item \textbf{Clustering} \cite{zhang2024clustering} introduces
      clustering-based methods to identify different physical regimes
      and discover dominant dimensionless parameters following the
      active subspace method. Data points \(\gamma\) are clustered
      into \(K\) groups, \(\Omega^1, \Omega^2, \ldots, \Omega^K\),
      using a clustering algorithm based on gradients \(\nabla
      \hat g(\gamma)\).

    \item \textbf{PyDimension} \cite{xie2022data} embeds the principle
      of dimensional invariance into a two-level polynomial regression
      scheme to discover dimensionless numbers by solving the
      optimization problem,
    \[
    \min_{\mathbf{C},\mathbf{\beta}}\|\Pi_o-f(\mathbf{\Pi}(\mathbf{C}),
    \mathbf{\beta})\|_2,
    \]
    where \(f\) is a polynomial regression model, $\mathbf{\beta}$ represents
    its coefficients, and $\mathbf{C}$ is the basis coefficients
    for constructing \(\mathbf{\Pi}\).

    \item \textbf{BuckiNet} \cite{bakarji2022dimensionally} uses
      machine learning and Sparse Identification of Nonlinear Dynamics
      (SINDy) to collapse data into dimensionless groups by solving a
      constrained optimization problem,
    \[
   \min _{\mathbf{C},\psi}\left\|\Pi_o-\psi\left(\mathbf{\Pi}\left(\mathbf{C}\right)\right)\right\|_2,
    \]
    where the function \(\psi\) serves as an approximation model,
    which can be implemented using ridge regression, a neural network,
    or a SINDy-based approach, and $\mathbf{C}$ is the basis
    coefficients for constructing \(\mathbf{\Pi}\).

    \item \textbf{{BSM}}
      \cite{bempedelis2024extracting} combines optimization and
      symbolic regression to extract similarity variables by first
      identifying them and then discovering the analytic form of their
      transformations. Given a set of similarity variables \(\xi\) and
      \(\tilde{q}\), the similarity transformation follows
    \[
    q_o(s, t) \rightarrow  \tilde{q_o}(\xi),
    \]
    with 
    \[
    \xi=\alpha(t) s+\beta(s, t), \quad \tilde{q}_o=\gamma(t) q_o+\delta(s, t),
    \]
    where $q_o$ is the quantity of interest, $s$ and $t$ are the
    independent variables, $\xi$ is the self similar variable, $\tilde
    q_o$ is the self-similar output, $[\alpha, \beta]$, and $[\gamma,
      \delta]$ are elementary dilation/translation groups.  The
    similarity variables are determined by solving the optimization
    problem,
    \[
    {\arg \min }_{\alpha, \beta,\gamma, \delta}  \sum_{i=1}^{n_t} \sum_{j=1}^{n_t}\left\|\tilde{q}_o\left(\xi, t_i\right)-\tilde{q}_o\left(\xi, t_j\right)\right\|_2^2,
    \]
    where $i$ and $j$ denotes the $i$-th and $j$-th value of the
    independent variable $t$.  Symbolic regression is then employed to
    extract the analytic form of transformation variables $\alpha,
    \beta,\gamma, \delta$.
  \item { \textbf{Information-based methods} have also
    been proposed in the context of dimensional analysis. To the best
    of our knowledge, the first attempt to apply information-theoretic
    principles to dimensionless learning was developed by our
    group~\cite{ma2024building}, where mutual information was used to
    extract dimensionless variables in hypersonic flows. 
   \[
\mathbf{\Pi}(\mathbf{C}) = \arg\max_{\mathbf{C}} I\left( \Pi_o ; \mathbf{\Pi}(\mathbf{C}) \right)
\]
    A subsequent method, referred to as MIDL~\cite{zhang2025}, also
    leveraged mutual information for learning dimensionless variables,
    with an emphasis on invariance under reparameterizations and
    robustness to outliers.  There are, nonetheless, several key
    conceptual differences between MIDL and IT-$\pi$ that are worth
    highlighting. First, IT-$\pi$ is rigorously grounded in the
    irreducible error theorem, while MIDL does not provide a
    theoretical justification linking mutual information to model
    error. As shown in \S\ref{sec:proof}, mutual information alone is
    insufficient to establish this connection; it is the conditional
    differential entropy of the output given the input that governs
    the irreducible error.  Furthermore, IT-$\pi$ is formulated using
    Rényi entropies, allowing the framework to accommodate different
    error norms. In contrast, MIDL is based solely on Shannon
    entropy. As demonstrated in \S\ref{sec:norm}, relying exclusively
    on Shannon mutual information can lead to failure in a variety of
    scenarios.  Finally, IT-$\pi$ offers additional capabilities
    beyond identifying dimensionless variables. These include
    identifying physical regimes, discovering self-similar structures
    and characteristic scales, and quantifying model efficiency.}
\end{itemize}

\section{~IT-$\pi$ Formulation}

\subsection{Proof of information-theoretic lower bound to irreducible error}
\label{sec:proof}

{\bf Definitions}. Consider the physical law or model $$\hat{\Pi}_o=
f(\mathbf{\mathbf{\Pi}}).$$ The irreducible error for a given
$L_p$-norm is defined as $$ \min_ { f} \|\Pi_o - \hat \Pi_o\|_p =
\min_ { f}
\left[\mathbb{E}\left(|\Pi_o-\hat{\Pi}_o|^p\right)\right]^{\frac{1}{p}},$$
where $\mathbb{E}( \, )$ is the expectation operator.  The specific
conditional Rényi entropy of $\Pi_o$ given $\Pi=\pi$ is
$$
h_\alpha(\Pi_o \mid \mathbf{\Pi}=\mathbf{\pi}) = \lim _{\alpha' \rightarrow \alpha} \frac{1}{1-\alpha'} \log \left(\int \rho_{\Pi_o \mid \mathbf{\Pi}}^{\alpha^{\prime}}\left(\pi_o \mid \mathbf{\pi}\right) d \pi_o\right), 
$$
such that
$$
 h_\alpha\left(\Pi_o| \mathbf{\Pi}\right) = \int \rho_{\mathbf{\Pi}}(\mathbf{\pi}) h_\alpha(\Pi_o \mid \mathbf{\Pi}=\mathbf{\pi}) d \mathbf{\pi}. 
$$


\noindent {\bf Theorem.} The irreducible error is lower bounded by
\begin{equation}
\label{eq:a_Renyi_lower_bound}
\min_ { f} \|\Pi_o - \hat \Pi_o\|_p 
\geq e^{- I_{\alpha}(\Pi_o ;\mathbf{\Pi})}\cdot c\ \!\bigl(\alpha, p, h_{\alpha ,o}\bigr),
\end{equation}
for  $\alpha>1 /(1+p)$. The function $c\ \!\bigl(\alpha, p, h_{\alpha ,o}\bigr)$ is given by  
$$
c\ \!\bigl(\alpha, p, h_{\alpha ,o}\bigr) = e^{h_{\alpha}(\Pi_o )} \cdot c_1(\alpha, p),
$$

$$
c_1(\alpha, p)= \begin{cases}(p \alpha+\alpha-1)^{-\frac{1}{p}+\frac{1}{1-\alpha}}(p \alpha)^{\frac{1}{\alpha-1}} c_2(\alpha, p) & \text { if } \alpha \neq 1, \\  \left({pe}\right)^{-\frac{1}{p}} c_2(1, p) & \text { if } \alpha=1,\end{cases}
$$
and
$$
c_2(\alpha, p)= \begin{cases}\frac{p(1-\alpha)^{\frac{1}{p}}}{2 \beta\left(\frac{1}{p}, \frac{1}{1-\alpha}-\frac{1}{p}\right)} & \text { if } \alpha<1, \\ \frac{p}{2 \Gamma\left(\frac{1}{p}\right)} & \text { if } \alpha=1, \\ \frac{p(\alpha-1)^{\frac{1}{p}}}{2 \beta\left(\frac{1}{p}, \frac{\alpha}{\alpha-1}\right)} & \text { if } \alpha>1 ,\end{cases}
$$
where $\beta$ is the Beta function and $\Gamma$ is the Gamma
function.\\

\begin{proof}
Applying the law of total expectation, the $L_p$-norm of the error is
decomposed into a weighted sum of the conditional error given each
state of the input $\mathbf{\Pi} = \mathbf{\pi}$,

$$
\begin{aligned}
\left\|\Pi_o-\hat{\Pi}_o\right\|_p & =\left[\mathbb{E}\left(\left|\Pi_o-\hat{\Pi}_o\right|^p\right)\right]^{\frac{1}{p}} \\
& =\left[\int \rho_{\Pi}(\mathbf{\pi}) \mathbb{E}\left(\left|\Pi_o-\hat{\Pi}_o\right|^p \mid \mathbf{\Pi}=\mathbf{\pi}\right) d \mathbf{\pi}\right]^{\frac{1}{p}},
\end{aligned}
$$
where the $L_p$-norm of conditional error given
$\mathbf{\Pi}=\mathbf{\pi}$ is denoted as,
$$
\begin{aligned}
\mathbb{E}\left(\left|\Pi_o-\hat{\Pi}_o\right|^p \mid \mathbf{\Pi} =\mathbf{\pi}\right)
&=\int \rho_{\Pi_o-\hat\Pi_o \mid \mathbf{\Pi}  }\left(\pi_o -\hat{\pi}_o \mid \mathbf{\pi}\right)\left|\pi_o-\hat{\pi}_o\right|^p d \pi_o
 \\
 & = \int \rho_{\mathcal{E} \mid \mathbf{\Pi}  }\left(\epsilon \mid  \mathbf{\pi} \right)|\epsilon|^p d \epsilon\\
 & = \|\mathcal{E} \mid \mathbf{\Pi}=\mathbf{\pi}\|_p^p,
\end{aligned}
$$
where $\mathcal{E}=\Pi_o-\hat{\Pi}_o$ is the error random variable,
and $\epsilon=\pi_o-\hat{\pi}_o$ is its realization.

The distribution that maximizes the Rényi entropy of a continuous
random variable with a fixed $p$-th moment is a generalized Gaussian
distribution of order $\alpha$ \cite{lutwak2005crame} of the form
$$
\rho^G_{\mathcal{E}_G \mid  \mathbf{\Pi} }(\epsilon \mid \mathbf{\pi}) = G\left(\frac{\epsilon}{t}\right)/t,
$$
when $\alpha>1 /(1+p)$. The scaling factor $t$ is $$t =
\frac{\|\mathcal{E} \mid
  \mathbf{\Pi}=\mathbf{\pi}\|_p}{(p\alpha+\alpha-1)^{-1/p}},$$
and the standard generalized Gaussian distribution of order $\alpha$
is
\begin{equation}
 G(x)= \begin{cases}c_2(p, \alpha)\left(1+(1-\alpha)|x|^p\right)^{\frac{1}{\alpha-1}} & \text { if } \alpha \neq 1, \\ c_2(p, 1) e^{-|x|^p} & \text { if } \alpha=1. \end{cases}
\end{equation}
The latter implies that the entropy of the conditional error is
bounded by the entropy of the conditional error with this follows the
generalized Gaussian ($\mathcal{E}_G$) with the same $p$-th moment for
$\alpha>1 /(1+p)$,
\begin{equation}
\begin{aligned} 
h_\alpha(\mathcal{E} \mid \mathbf{\Pi} =
\mathbf{\pi}) 
&\leq h_\alpha(\mathcal{E}_G \mid \mathbf{\Pi} 
= \mathbf{\pi}) \\
&= \log \| \mathcal{E} \mid \mathbf{\Pi}=\mathbf{\pi} \|_p - \log  c_1(\alpha,p).
\end{aligned}
\end{equation}

This leads to the moment-entropy inequality \cite{lutwak2005crame}
implies that the $L_p$-norm of the conditional error $\|\mathcal{E}
\mid \mathbf{\Pi}=\mathbf{\pi}\|_p $ can be bounded using the
entropy of the irreducible error $h_{\alpha}(\mathcal{E} \mid
\mathbf{\Pi}=\mathbf{\pi})$ as
$$
\|\mathcal{E} \mid \mathbf{\Pi}=\mathbf{\pi}\|_p \geq  e^{h_{\alpha}(\mathcal{E} \mid \mathbf{\Pi}=\mathbf{\pi})}\cdot c_1(\alpha,p).
$$
Therefore, the $L_p$-norm of the error is bounded as
$$
\begin{aligned}
\left\|\Pi_o-\hat{\Pi}_o\right\|_p 
&= 
\left[\int \rho_{\mathbf{\Pi}}(\mathbf{\pi}) \mathbb{E}\left(\left|\Pi_o-\hat{\Pi}_o\right|^p \mid \mathbf{\Pi}=\mathbf{\pi}\right) d \mathbf{\pi}\right]^{\frac{1}{p}} \\
&=\left[\int \rho_{\mathbf{\Pi}}(\mathbf{\pi}) 
\|\mathcal{E} \mid \mathbf{\Pi}=\mathbf{\pi}\|_p^p
d \mathbf{\pi}\right]^{\frac{1}{p}}\\
& \geq \left[\int \rho_{\mathbf{\Pi}}(\mathbf{\pi}) 
\left[e^{h_{\alpha}(\mathcal{E} \mid \mathbf{\mathbf{\Pi}}=\mathbf{\mathbf{\pi}})}\cdot c_1(\alpha,p)\right]^p
d \mathbf{\pi}\right]^{\frac{1}{p}}.
\end{aligned}
$$
Applying Jensen's inequality, $\mathbb{E}[\phi(X)]\geq
\phi(\mathbb{E}[X])$, to the convex function $\phi(x) = e^{p x}$, we
obtain,
$$
\begin{aligned}
\int \rho_{\mathbf{\Pi}}(\mathbf{\pi}) e^{p h_\alpha(\mathcal{E} \mid \mathbf{\Pi}=\mathbf{\pi})} d \mathbf{\pi} 
&\geq e^{p \int \rho_{\mathbf{\Pi}}(\mathbf{\pi}) h_\alpha(\mathcal{E} \mid \mathbf{\Pi}=\mathbf{\pi}) d\mathbf{\pi}}\\
&= e^{p \int \rho_{\mathbf{\Pi}}(\mathbf{\pi}) h_\alpha(\Pi_o-f(\mathbf{\mathbf{\pi}}) \mid \mathbf{\Pi}=\mathbf{\pi}) d\mathbf{\pi}}\\
&\overset{(1)}{=} 
 e^{p \int \rho_{\mathbf{\Pi}}(\mathbf{\pi}) h_\alpha(\Pi_o \mid \mathbf{\Pi}=\mathbf{\mathbf{\pi}}) d\mathbf{\mathbf{\pi}}} \\
& = e^{p h_\alpha(\Pi_o \mid \mathbf{\Pi})
},
\end{aligned}
$$
where $(1)$ follows from the fact that adding a constant
$f(\mathbf{\pi})$ does not change the conditional Renyi entropy.
Substituting this bound into the previous inequality,
$$
\begin{aligned}
 ~\forall f, \quad\left\|\Pi_o-\hat{{\Pi}}_o\right\|_p & \geq\left[c_1^p(\alpha, p) e^{p h_\alpha(\Pi_o \mid \mathbf{\Pi})}\right]^{\frac{1}{p}} \\
& =c_1(\alpha, p) e^{h_\alpha(\Pi_o \mid \mathbf{\Pi})}.
\end{aligned}
$$
and therefore, 
$$\min _f\left\|\Pi_o-\hat{\Pi}_o\right\|_p \geq e^{-I_\alpha\left(\Pi_o ; \mathbf{\Pi}\right)} \cdot c\ \!\bigl(\alpha, p, h_{\alpha ,o}\bigr).$$
We refer to 
$$
\epsilon_{LB} \equiv \max_\alpha \left[ e^{-I_\alpha\left(\Pi_o ; \mathbf{\Pi}\right)} \cdot c\ \!\bigl(\alpha, p, h_{\alpha ,o}\bigr)\right],
$$
as the information-theoretic irreducible error.  

\end{proof}

{ The main assumption in the derivation of the
  information-theoretic irreducible error, $\epsilon_{LB}$, is the
  existence of the conditional differential entropy, $h(\cdot \mid
  \cdot)$, which is a fairly general condition. Under this assumption,
  $\epsilon_{LB}$ represents a fundamental mathematical constraint
  that any predictive model must satisfy. This constraint remains
  valid even in the presence of strong nonlinearities and large
  prediction errors.  Another important observation is that
  $\epsilon_{LB}$ does not depend solely on the mutual information
  between input and output. It also depends on the error norm $p$ and
  the information content of the output, quantified by $h_{\alpha,o}$,
  through the scaling function $c(\alpha, p, h_{\alpha,o})$.  This
  implies that even when the mutual information between input and
  output is non-zero, $\epsilon_{LB}$ may still be large if the output
  information $h_{\alpha,o}$ is high, resulting in a large value of
  $c(\alpha, p, h_{\alpha,o})$.  In such scenarios, models are
  fundamentally limited in their ability to exploit the available
  mutual information between input and output, and large prediction
  errors become unavoidable.}

\newpage
\subsection{{Pseudocode of IT-$\pi$}}
\begin{algorithm}[H]
  \caption{IT-$\pi$: Dimensionless learning based on information}
  \begin{algorithmic}[1]
\Require Dimensional input $\mathbf{q}$, output $q_o$,  null space of the dimension matrix $\mathbf{W}$.
\Ensure Optimal dimensionless input $\mathbf{\Pi^*}$, dimensionless output ${\Pi_o^*}$.
\State \textbf{Define} exponent coefficients $\mathbf{C}$ and $\mathbf{c}_o$:
  $$
    \mathbf{\Pi}(\mathbf{C}) = \mathbf{q}^{\mathbf{W}\mathbf{C}}, 
    \quad
    \Pi_o(\mathbf{c}_o) = q_o^{\mathbf{W}\mathbf{c}_o}.
  $$

  \State \textbf{Define} the $L_p$-norm exponent when measuring the error : $p$.

\State \textbf{Define} the population size in CMA-ES (the number of candidate solutions sampled at each iteration): $M$.

\State \textbf{Objective:} Identify the optimal dimensionless input and output to minimize the irreducible model error
$$\mathbf{\Pi}^*, \Pi_o^*=\arg \min _{\mathbf{\Pi}, \Pi_o} \max _\alpha\left[\epsilon_{L B}\right] = \arg \min _{\mathbf{C}, \mathbf{c}_o} J,$$ 
where the fitness is
$$
J = \max_{\alpha}
\left[
  e^{-I_\alpha\bigl(\Pi_o(\mathbf{c}_o);\mathbf{\Pi}(\mathbf{C})\bigr)}
  c(\alpha,p,h_{\alpha,o})
\right].
$$
\State Initialize CMA-ES population over $(\mathbf{C},\mathbf{c_o})$ by generating a Gaussian distribution with an initial mean vector and covariance matrix.
\While{not converged}
  \State \(\{\mathbf{C}_i,\mathbf{c}_{o,i}\}_{i=1}^M \gets\) sample new population
  \For{$i = 1$ to $M$}
    \State Compute dimensionless inputs  
      $\mathbf{\Pi} = \mathbf{q}^{\mathbf{W
    C}_i} $
    \State Compute the dimensionless output 
       $\Pi_o ={q_o}^{\mathbf{W
    } \mathbf{c}_{o,i}}$
    \State Estimate the irreducible error $\epsilon_{LB} = e^{-I_\alpha\bigl(\Pi_o(\mathbf{c}_o);\mathbf{\Pi}(\mathbf{C})\bigr)} c(\alpha,p,h_{\alpha,o})$
    \State Evaluate the fitness $J_i = \max_{\alpha}\left[\epsilon_{LB}\right]$ using golden-section search over $\alpha \in \left(\frac{1}{1+p}, 10\right]$
  \EndFor
  \State Update CMA-ES distribution using fitnesses $\{J_i\}$
\EndWhile
\State \Return Optimal $\mathbf{\Pi^*},{\Pi_o^*}$
\end{algorithmic}
\end{algorithm}

\newpage
\subsection{Infinite equivalent solutions of IT-$\pi$ under bijective transformations}

The Rényi mutual information $I_{\alpha}(\Pi_o; \mathbf{\Pi})$ is
invariant under bijective transformations applied to
$\mathbf{\Pi}$.  Specifically, if $T$ is the bijective
transformations applied to $\mathbf{\Pi}$,
$$
\tilde{\mathbf{\Pi}}=T(\mathbf{\Pi}),
$$ then the Rényi mutual information remains invariant under bijective
transformations,
$$
I_\alpha\left({\Pi}_o ; \tilde{\mathbf{\Pi}}\right)=I_\alpha\left(\Pi_o ; \mathbf{\Pi}\right).
$$ This invariance implies the existence of infinitely many equivalent
optimal solutions for IT-$\pi$, as any bijective transformation of the
optimal inputs preserves an equivalent mutual information. This is
consistent with the intuition that when building a model, applying a
bijective transformation to the input does not alter the amount of
information available about the output. Consequently, models
constructed with bijectively transformed inputs are equivalent in
terms of their predictive capacity.

For example, for the Rayleigh problem, the mutual information remains
the same for $$I_{\alpha}({\Pi}^*;\Pi^o) =
I_{\alpha}\left(\frac{1}{\Pi^*};\Pi^o\right),$$ under the bijective
transformation $T(x) =1/x$. Therefore, $$T(\Pi^*) = t^{0.5} \mu^{0.5}
/(y \rho^{0.5})$$ is an equivalent optimal solution as $$\Pi^* = y
\rho^{0.5} /\left(t^{0.5} \mu^{0.5}\right).$$ \\

\begin{proof}
Under the bijective transformation $T$, the marginal probability
distribution function $\rho_{\tilde{\mathbf{\Pi}}}$ transforms
according to the change-of-variables
formula, $$\rho_{\tilde{\mathbf{\Pi}}}\left(\tilde{\mathbf{\pi}}\right)=\rho_{\mathbf{\Pi}}\left(\mathbf{\pi}\right)
\cdot\left|J_T\left(\mathbf{\pi}\right)\right|^{-1},$$
where $$J_T\left(\mathbf{\pi}\right)=\left|\operatorname{det}\left(\frac{\partial
  T}{\partial \mathbf{\pi}}\right)\right|$$ is the Jacobian
determinant of the transformation $T$.

Similarly, the joint probability distribution function transforms
as $$\rho_{{\Pi}_o, \bar{\mathbf{\Pi}}}\left({\pi}_o,
\tilde{\mathbf{\pi}}\right)=\rho_{\Pi_o,
  \mathbf{\Pi}}\left(\pi_o, \mathbf{\pi}\right)
\cdot\left|1\right|^{-1}
\cdot\left|J_T(\mathbf{\pi})\right|^{-1}.$$

Therefore, the conditional probability distribution function
transforms as $$\rho_{{\Pi}_o \mid
  \tilde{\mathbf{\Pi}}}\left({\pi}_o \mid
\tilde{\mathbf{\pi}}\right) =\frac{\rho_{{\Pi}_o,
    \bar{\mathbf{\Pi}}}\left({\pi}_o,
  \tilde{\mathbf{\pi}}\right)}{\rho_{\tilde{\mathbf{\Pi}}}\left(\mathbf{\pi}\right)}
=\rho_{\Pi_o \mid \mathbf{\Pi}}\left(\pi_o \mid
\mathbf{\pi}\right), $$ which remains unchanged under the
transformation since the Jacobian terms cancel out.

The conditional Renyi entropy under the transformation is
$$
\begin{aligned}
h_\alpha\left({\Pi}_o \mid \tilde{\mathbf{\Pi}}\right)
& =\lim _{\alpha^{\prime} \rightarrow \alpha} \frac{1}{1-\alpha^{\prime}} \int \rho_{\tilde {\mathbf{\Pi}}}(\tilde {\mathbf{\pi}}) \log \left(\int \rho_{\Pi_o \mid \tilde {\mathbf {\Pi}}}^{\alpha^{\prime}}\left(\pi_o \mid \tilde {\mathbf{\pi}}\right) \cdot d \pi_o\right) d \tilde {\mathbf{\pi}} \\
& = \lim _{\alpha^{\prime} \rightarrow \alpha} \frac{1}{1-\alpha^{\prime}} \int \rho_{\mathbf{\Pi}}\left({\mathbf{\pi}}\right) \cdot\left|J_T\left({\mathbf{\pi}}\right)\right|^{-1} \log \left(\int \rho_{\Pi_o \mid \mathbf{\Pi}}^{\alpha^{\prime}}\left(\pi_o \mid  {\mathbf{\pi}}\right) \cdot d \pi_o\right) \left|J_T\left({\mathbf{\pi}}\right)\right| d{\mathbf{\pi}}
\\
& = \lim _{\alpha^{\prime} \rightarrow \alpha} \frac{1}{1-\alpha^{\prime}} \int \rho_{\mathbf{\Pi}}({\mathbf{\pi}}) \log \left(\int \rho_{\Pi_o \mid  \mathbf{\Pi}}^{\alpha^{\prime}}\left(\pi_o \mid {\mathbf{\pi}}\right) \cdot d \pi_o\right) d {\mathbf{\pi}} \\
& = h_\alpha\left({\Pi}_o \mid {\mathbf{\Pi}}\right).
\end{aligned}
$$
Therefore, the Renyi mutual information remains unchanged,
$$
\begin{aligned}
 I_\alpha\left({\Pi}_o ; \tilde{\mathbf{\Pi}}\right)&=h_\alpha\left({\Pi}_o\right)-h_\alpha\left({\Pi}_o \mid \tilde{\mathbf{\Pi}}\right) \\
& = h_\alpha\left({\Pi}_o\right) - h_\alpha\left({\Pi}_o \mid {\mathbf{\Pi}}\right) \\
& = I_\alpha\left({\Pi}_o ; {\mathbf{\Pi}}\right)
\end{aligned}
$$
\end{proof}

\subsection{Discovery of optimal characteristic scales}
\label{sec:Alg_char_scale}


Consider the dimensional inputs 
$\mathbf{q} = \left[ \mathbf{q}_v,\mathbf{q}_p \right]$,
where 
$\mathbf{q}_v = \left[ q_{v_1}, q_{v_2}, \dots,q_{v_{n_v}} \right]$
consists of $n_v$ variables that change during each simulation or experiment, and 
$\mathbf{q}_p = \left[ q_{p_1}, q_{p_2}, \dots,q_{p_{n_p}} \right]$
consists of $n_p$ parameters that remain fixed within a given simulation or experiment but may change across different  cases. After applying IT-$\pi$, the dimensionless variable $\Pi_i^*$ is constructed by non-dimensionalizing each $q_{v_k}$ using the characteristic scales 
\[
\mathbf{S} = \left[ S_1, S_2, \dots, {S_{n_u}} \right],
\]
and the dimensionless parameter $$\mathbf{\Pi}_p=[\Pi_{p_1},\Pi_{p_2},\cdots,\Pi_{p_{n_u}}],$$ both of which depend on the components of $\mathbf{q}_p$.
The characteristic scales correspond to one of the fundamental units (length, time, mass, electric current, temperature, amount of substance, or luminous intensity) and can be expressed as
\[
S_i = q_{p_1}^{a_{s,i1}}\,q_{p_2}^{a_{s,i2}} \cdots q_{p_{n_p}}^{a_{s,i\,n_p}} \equiv \mathbf{q}_p^{\mathbf{a}_{s,i}},
\]
with the exponent vector 
\[
\mathbf{a}_{s,i} = \left[ a_{s,i1}, a_{s,i2}, \dots, a_{s,i\,n_p} \right].
\]
The exponent vectors are arranged in the characteristic-scales coefficient matrix $\mathbf{A}_s$ (with size $n_u \times n_p$) and are constrained to one of the fundamental units. This is enforced  by the equation
\begin{equation}
\label{eq:CS_dimen_cont}
\mathbf{D}_{\mathbf{q}_p} \mathbf{A}_s^T = \mathbf{I},
\end{equation}
where $\mathbf{D}_{\mathbf{q}_p}$ is the dimension matrix for the parameters and  $\mathbf{I}$ is the identity matrix. 

For example, consider the Malkus water wheel characterized by the vector of parameters
\[
\mathbf{q}_p = [r,\, \phi,\, I,\, \nu,\, K,\, g],
\]
the dimension matrix for parameters is $$\mathbf{D}_{\mathbf{q}_p} =
\begin{array}{c|cccccc}
   & r & \phi & I & \nu & K & g \\ \hline
[\mathrm{time}] & 0  & -1 & 0 & -1 & -1 & -2\\[1mm]
[\mathrm{mass}] & 0 & 1 & 1 & 1 & 0  & 0
\end{array},$$
and a possible characteristic-scales coefficient matrix is 
\[
\mathbf{A}_s =
\begin{array}{c|cccccc}
   & r & \phi & I & \nu & K & g \\ \hline
S_t & 0  & 0 & 0 & 0 & -1 & 0\\[1mm]
S_m & -1 & 0 & 1 & 0 & 2  & -1
\end{array}
\]
to guarantee $$\mathbf{D}_{\mathbf{q}_p} \mathbf{A}_s^T =\begin{bmatrix}
    1 \ 0\\
    0 \ 1
\end{bmatrix},$$
i.e., $S_t$ has units of time and $S_m$ has the units of mass.

Similarly, each component of the dimensionless parameter $\mathbf{\Pi}_p$ is expressed as
\[
\Pi_{p_i} = q_{p_1}^{a_{p,i1}}\,q_{p_2}^{a_{p,i2}} \cdots q_{p_{n_p}}^{a_{p,i\,n_p}} \equiv \mathbf{q}_p^{\mathbf{a}_{p,i}},
\]
where 
\[
\mathbf{a}_{p,i} = \left[ a_{p,i1}, a_{p,i2}, \dots, a_{p,i\,n_p} \right],
\]
is chosen to ensure that $\Pi_{p_i}$ is dimensionless.
After appling IT-$\pi$,
each dimensionless variable $\Pi_i^*$ can be reformulated as
\begin{equation}
\label{eq:charac_scales}
\Pi_i^* = \left(\frac{q_{v_1}}{\mathbf{S}^{\mathbf{d}_{i,1}}}\right)^{\beta_{i,1}} \left(\frac{q_{v_2}}{\mathbf{S}^{\mathbf{d}_{i,2}}}\right)^{\beta_{i,2}} \cdots \left(\frac{{q_{v_j}}}{\mathbf{S}^{\mathbf{d}_{i,j}}}\right)^{{\beta_{i,j}}} \cdot \Pi_{p_i},
\end{equation}
where
\[
\mathbf{S}^{\mathbf{d}_1} \equiv S_1^{d_{11}}\,S_2^{d_{21}} \cdots S_{n_u}^{d_{n_u\,1}},
\]
and similarly for \(\mathbf{S}^{\mathbf{d}_j}\) with \( j=2, \dots, n_v \). 
Here, \(\mathbf{d}_{i,j} = \left[ d_{i,1j}, d_{i,2j}, \dots, d_{i,n_u\,j} \right]\) denotes the dimensional vector associated with the variable \( q_{v_j} \), and \(\beta_{i,j}\) is the exponent of \( q_{v_j} \) in \(\Pi_i^*\).

To determine the characteristic scales $\mathbf{S}$ and the dimensionless parameter $\mathbf{\Pi}_p$, 
we need to express 
the optimal solution $\mathbf{\Pi}^*$ as a combination of the characteristic scales
constructed using $\mathbf{q}_p$. The reminder of the units represent the (residual) dimensionless parameter.
Mathematically, this is formulated as the solution to the linear equation obtained by taking the log on both sides of Eq.~(\ref{eq:charac_scales}), while simultaneously satisfying the constraints in Eq~(\ref{eq:CS_dimen_cont}),
\begin{equation}
\label{eq:CS}
\mathbf{A}_p  = -\mathbf{D}_{\mathbf{q}_v}\,\mathbf{A}_s + \mathbf{A}_{\Pi_p}, \quad s.t. \quad \mathbf{D}_{\mathbf{q}_p} \mathbf{A}_s^T=\mathbf{I},
\end{equation}
where the matrix $\mathbf{A}_p$ (of size $l^* \times n_p$) 
represents the contribution of $\mathbf{q}_p$ to each dimensionless variable $\Pi_i^*$ and its rows are given by
\[
{
\mathbf{a}_{p,i} = \left[ a_{p,i1}, a_{p,i2}, \dots, a_{p,i\,n_p} \right].}
\]
The matrix
$\mathbf{D}_{\mathbf{q}_v}$   
 represents the {the fundamental units} of the variables $\mathbf{q}_v$ in $\Pi_i^*$, such that the
$i$-th row of  $\mathbf{D}_{\mathbf{q}_v}$ is a $1 \times n_u$ vector given by the { $\sum_{k=1}^{j} \mathbf{d}_{i,k}\,\beta_{i,k}$} from Eq.~\eqref{eq:charac_scales} for the $i$-th dimensionless variable $\Pi_i^*$. 
The matrix $\mathbf{A}_s$ is the characteristic-scales coefficient matrix to be determined (with size $n_u \times n_p$), and each row is denoted by $\mathbf{a}_{s,i}$. The residual matrix $\mathbf{A}_{\Pi_p}$ corresponds to the coefficients of the dimensionless parameters $\mathbf{\Pi}_p$, also to be determined.

Considering again the example of the Malkus water wheel, 
the vector of variables is  characterized by
\[
\mathbf{q}_v = [\omega,\, m_1,\, m_2].
\]
Using IT-$\pi$, we identify the following optimal dimensionless variables:
\begin{equation*}
\Pi_1^* =  \frac{r\,g\,m_1}{I\,K^2},\quad
\Pi_2^* = \frac{v\,\omega}{I\,K^2},\quad
\Pi_o^* =  \frac{\dot{\omega}}{K^2}.
\end{equation*}
The dimension matrix [mass, time] for these variables is given by 
\[
\mathbf{D}_{q_v} =
\begin{array}{c|cc}
     & {S_t} & {S_m} \\ \hline
{\Pi_1^*} & 0  & 1 \\
{\Pi_2^*} & -1 & 0 \\
{\Pi_o^*} & -2 & 0
\end{array}
\]
and the coefficient matrix (in the order $[r,\, \phi,\, I,\, \nu,\, K,\, g]$) for the optimal variables is
\[
\mathbf{A}_p =
\begin{array}{c|cccccc}
   & r & \phi & I & \nu & K & g \\ \hline
\Pi_1^* & 1 & 0 & -1 & 0 & -2 & 1 \\
\Pi_2^* & 0 & 0 & -1 & 1 & -2 & 0 \\
\Pi_o^* & 0 & 0 & 0 & 0 & -2 & 0
\end{array}
\]
Equation (\ref{eq:CS}) is solved using the  algorithm outlined below, which iteratively minimizes $\mathbf{A}_{\Pi_p}$ to identify the characteristic scales.
The solution to the characteristic-scale coefficient matrix is
\[
\mathbf{A}_s =
\begin{array}{c|cccccc}
   & r & \phi & I & \nu & K & g \\ \hline
S_t & 0  & 0 & 0 & 0 & -1 & 0\\[1mm]
S_m & -1 & 0 & 1 & 0 & 2  & -1
\end{array}
\]
which leads to the characteristic time ($S_t$) and mass scales ($S_m$)
\[
S_t = \frac{1}{K} \quad \text{and} \quad S_m = \frac{I\,K^2}{r\,g}.
\]
The corresponding residual matrix is
\[
\mathbf{A}_{\Pi_p} = 
\begin{array}{c|cccccc}
   & r & \phi & I & \nu & K & g \\ \hline
\Pi_{p1} & 0  & 0 & 0 & 0 & 0 & 0\\[1mm]
\Pi_{p2} & 0 & 0 & -1 & 1 & -1  & 0\\[1mm]
\Pi_{p3} & 0  & 0 & 0 & 0 & 0 & 0\\[1mm]
\end{array},
\]
indicating that the additional dimensionless parameter is given by
\[
\mathbf{\Pi}_p = [\Pi_{p1}, \Pi_{p2}, \Pi_{p3}] =\left[1 , \frac{\nu}{I\,k}, 1 \right].
\]
The dimensionless variables in terms of the characteristic time scale and mass scale, and dimensionless parameter are
\begin{equation*}
\Pi_1^* = \frac{m_1}{S_m},\quad
\Pi_2^* = \omega\,S_t\,\Pi_{p2},\quad
\Pi_o^* = \dot{\omega}\,S_t^2.
\end{equation*}
%

%
\begin{algorithm}[H] \caption{Iterative Identification for Characteristic Scales} \label{alg: characteristic} \begin{algorithmic}[1] \Require 
$\mathbf{D}_{\mathbf{q}_v}$ and $\mathbf{A}_{p}$. 
\Ensure $\mathbf{A}_{\Pi_p}$ and $\mathbf{A}_s$

\State \textbf{Initialize:} $\mathbf{D}_{\mathbf{q}v}^{\text{orig}} \leftarrow \mathbf{D}_{\mathbf{q}v}$, $\mathbf{A}_p^{\text{orig}} \leftarrow \mathbf{A}_p$, $n \leftarrow 0$

\For{$n = 1$ to $N_\mathrm{iter}$} 
\State Solve for $\mathbf{A}_s^{(n)} =\arg \min _{\mathbf{A}_s}\left\|\mathbf{D}_{\mathbf{q}_v} \mathbf{A}_s+\mathbf{A}_p\right\|_2 + \| \mathbf{D_{q_p}} \mathbf{A}_s^T - \mathbf{I}\|_2$ 
\State Compute residuals: $\mathbf{A}_{\Pi_p} = \mathbf{D}_{\mathbf{q}_v} \mathbf{A}_s^{(n)}+\mathbf{A}_p$
\If{$|{\mathbf{A}_{\Pi_p}}_i| \leq \epsilon$ for all $i$}
    \State \textbf{break}
\Else
    \State Select a row $i$ with $|{\mathbf{A}_{\Pi_p}}_i| > \epsilon$ and remove it
    \State Update: $\mathbf{D}_{\mathbf{q}_v} \leftarrow \mathbf{D}_{\mathbf{q}_v} \setminus \mathbf{D}_{\mathbf{q}_v,i}$, $\mathbf{A}_p \leftarrow \mathbf{A}_p \setminus \mathbf{A}_{p,i}$
\EndIf
\EndFor

\State \textbf{Final Calculation:} $\mathbf{A}_{\Pi_p}^{\text {final }}= \mathbf{D}_{\mathbf{q}_v}^{\text {orig }} \mathbf{A}_s^{(n)}+\mathbf{A}_p^{\text {orig }}$ \State \textbf{Output:} $\mathbf{A}_{\Pi_p}^{\text{final}}$ and $\mathbf{A}_s^{(n)}$ \end{algorithmic} \end{algorithm}

\newpage
\section{{~Robustness Analysis}}

{ In this section, we present a robustness analysis of
  IT-$\pi$ alongside other dimensionless learning methods. We begin by
  assessing how the amount of training data influences the fidelity of
  the learned dimensionless groups, then examine the impact of key
  hyperparameter settings and the choice of input variables on both
  convergence and predictive accuracy. To detect overfitting, we
  introduce diagnostics that reveal when a model begins to memorize
  rather than generalize. We also show the role of the Rényi entropy
  order $\alpha$ in shaping the balance between information of
  preservation and error norm. Finally, to evaluate resilience under
  realistic conditions, we subject IT-$\pi$ to a noisy-data study that
  incorporates all validation cases from the main text. }

\subsection{{Sensitivity to the amount of training data}}
\label{sec:sensitivity_stat}

{We evaluate the statistical uncertainties in the
  irreducible error, model efficiency, and dimensionless variables
  from IT-$\pi$. Estimating these uncertainties is essential to
  determine whether the conclusions drawn are statistically
  significant or potentially inconclusive due to limited data.}

The {statistical} uncertainty in the normalized
irreducible error, $\Delta\tilde{\epsilon}_{LB}$, arising from the
{finite sample size}, is defined as the difference
between the error bound estimated using the full dataset
($N_{\mathrm{samples}}$) and that estimated using half the data
($N_{\mathrm{samples}}/2$). This difference serves as a proxy for the
expected variation in $\tilde{\epsilon}_{LB}$ if the number of samples
were doubled. This approach is used throughout the manuscript to
compute the error bars for $\tilde{\epsilon}_{LB}$. Since the
normalized irreducible error is upper bounded by 1,
$\Delta\tilde{\epsilon}_{LB}$ can be directly interpreted as a
relative uncertainty.  {A similar procedure is applied
  to estimate the statistical uncertainty in model efficiency, given
  by $\Delta \eta = \Delta \epsilon_{LB}^* / \epsilon_f$. The final
  source of uncertainty concerns the persistence of the identified
  dimensionless variables with increasing sample size.  As shown
  below, even when uncertainties in $\tilde{\epsilon}_{LB}$ and $\eta$
  are non-negligible, the dimensionless variables identified from
  IT-$\pi$ remain remarkably robust.}

\begin{figure}
    \centering
 \includegraphics[width=0.98\linewidth]
    {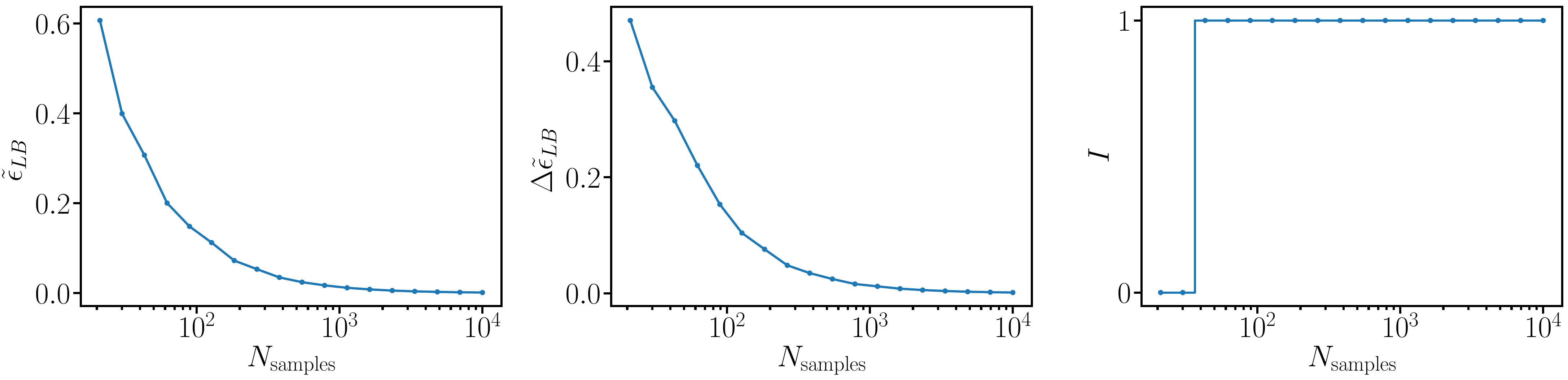}
    \put(-470,100){$(a)$}
    \put(-310,100){$(b)$}
    \put(-155,100){$(c)$}
    \caption{{
        (a) Estimated normalized irreducible error
        $\tilde{\epsilon}_{LB}$, (b) uncertainty in the normalized
        irreducible error $\Delta \tilde{\epsilon}_{LB}$, and (c)
        binary indicator of whether the identified dimensionless
        variable matches the analytical solution (1 = correct, 0 =
        incorrect), all shown as functions of the number of samples
        $N_{\text{samples}}$.}}
    \label{fig:UQ_vs_N}
\end{figure}

{ We evaluate the estimated irreducible model error, its
  associated statistical uncertainty, and the sensitivity of the
  identified dimensionless variable as a function of sample size for
  the Rayleigh problem. The results are presented in
  Figure~\ref{fig:UQ_vs_N}.

\begin{itemize}
    \item \textbf{Irreducible error.} As shown in
      Figure~\ref{fig:UQ_vs_N}(a), the estimated irreducible error
      $\tilde{\epsilon}_{LB}$ is initially overestimated and exhibits
      artificially high values for small sample sizes. However, it
      progressively converges to the correct value (zero) as the
      number of samples increases.

    \item \textbf{Uncertainty in irreducible error.}
      Figure~\ref{fig:UQ_vs_N}(b) displays the uncertainty
      $\Delta\tilde{\epsilon}_{LB}$ associated with the estimated
      irreducible error. As expected, this uncertainty decreases
      monotonically with increasing sample size. This behavior
      provides a useful diagnostic within IT-$\pi$: it can be used to
      (1) assess the statistical reliability of the estimated error in
      Figure~\ref{fig:UQ_vs_N}(a) and (2) extrapolate the number of
      samples required to achieve a desired confidence level when
      working with limited data.

    \item \textbf{Sensitivity of the identified dimensionless
      variable.} Figure~\ref{fig:UQ_vs_N}(c) presents an indicator
      function that takes the value 1 when the identified
      dimensionless variables are correct and 0 otherwise. The results
      demonstrate that the discovered dimensionless variable remains
      remarkably stable across a broad range of sample
      sizes. Deviations occur only when the number of samples is very
      small (a handful), indicating that the method is generally
      robust to moderate data scarcity.
      
   \item \textbf{Sensitivity to extremely limited data.} Finally, we
     assess the impact of extremely limited data on the reliability of
     dimensionless variables identified by Active Subspaces,
     PyDimension, BuckiNet, BSM, and IT-$\pi$ to the Rayleigh problem
     using datasets with 8 and 27 samples.  As shown in
     Table~\ref{tab:limited_data}, when 27 samples are available, only IT-$\pi$ successfully recovers the analytical
     dimensionless variable, whereas the other methods yield spurious
     results. In contrast, with only 8 samples, the discovered
     exponents vary significantly across methods, all deviating from
     the analytical solution. These results indicate that IT-$\pi$
     requires a comparable or smaller amount of data than other
     methods to identify reliable dimensionless variables.     
\end{itemize}
}
\renewcommand{\arraystretch}{1.5}
\begin{table}[h!]
\centering
\setlength{\arrayrulewidth}{1 pt} 
\resizebox{0.95 \textwidth}{!}{%
  {
\begin{tabular}{|l|c|c|c|c|c|}
\hline
{$N_{\text{samples}}$ } & {Active} & {PyDim-} & {Bucki-} & {{BSM}} & {IT-$\pi$}\\  & {Subspaces} & {ension} &  {Net} & & {(Current)}\\
\hline
$ 27$  
& ${y^{1.0}}{t^{-0.4} U^{0.2}(\mu/\rho)^{-0.6}}$ 
& ${y^{1.0}}{t^{-1.0} U ^{-1.0}}$ 
& ${y^{1.0}}{t^{-0.84} U ^{-0.67}} (\mu/\rho)^{-0.16}$ 
& ${y^{1.0}}{U^{-1.0} (\mu/\rho)^{1.0}}$ 
& ${y^{1.0}}{t^{-0.5} (\mu/\rho)^{-0.5}}$ \\
\hline
$ 8$  
& ${y^{1.0}}{  t^{-0.38}  U^{0.23} (\mu/\rho)^{-0.62}}$ 
& ${y^{1.0}}{t^{-1.0} U ^{-1.0}}$ 
& ${y^{1.0}}{t^{-1.0} U ^{-1.0}}$ 
& ${y^{1.0}}{  t^{0.42}  U^{-0.15} (\mu/\rho)^{0.57}}$  
& ${y^{1.0}}{  t^{-0.86}  U^{-0.72} (\mu/\rho)^{-0.14}}$ \\
\hline
\end{tabular}}
}
\caption{{Dimensionless variable $\Pi^*$ identified by
    different methods on the Rayleigh problem with 27 and 8
    samples. The analytical optimal dimensionless input is $y^{1.0}
    t^{-0.5} (\mu/\rho)^{-0.5}$}.}
      \label{tab:limited_data}
\end{table}
\renewcommand{\arraystretch}{1}

\subsection{{Sensitivity to hyperparameter choices}}
\label{sec:sensitivity_hyperparameter}

{We investigate the impact of estimation strategies
  and hyperparameter configurations on various dimensionless-learning
  methods. Results for the Rayleigh problem with 1,000 samples are
  summarized in Table~\ref{tab:estimator_comparison}.}

{For IT-$\pi$, we evaluate two mutual-information
  estimators: the binning method and the KSG estimator \cite{kraskov2004estimating}.  The binning
  method partitions the joint distribution into a fixed number of
  bins; a common rule of thumb is to choose the number of bins
  proportional to the square root of the sample size, yielding
  approximately 30–35 bins for 1,000 samples.  Here, we vary the bin
  count from 10 to 100 to assess its effect.  The KSG estimator, by
  contrast, estimates local densities via $k$–nearest-neighbor
  distances; we test $k$ in the range 5–10, a standard choice that
  balances bias and variance.  As shown in
  Table~\ref{tab:estimator_comparison}, all configurations recover the
  analytically derived dimensionless variable, demonstrating that
  IT-$\pi$ is robust to estimator selection and hyperparameter
  settings when sufficient data are available.}

{Table~\ref{tab:estimator_comparison} also summarizes
  the sensitivity of each method to its hyperparameters.  The results
  show that, except for Active Subspaces, different hyperparameter
  settings can yield markedly different dimensionless variables.  By
  comparison, IT-$\pi$ exhibits similar or lower sensitivity to
  hyperparameter choice for the same number for samples.}
%
 \renewcommand{\arraystretch}{1.5}
\begin{table}[h!]
{
  \centering
\setlength{\arrayrulewidth}{1 pt}
\resizebox{0.98\textwidth}{!}{%
\begin{tabular}{|l|l|l|l|}
\hline
\textbf{Method} & \textbf{Method/Hyperparameters} & \textbf{Hyperparameter Settings} & \textbf{Discovered \(\mathbf{\Pi^*}\)} \\ 
\hline

\multirow{2}{*}{IT-\(\pi\)} 
& \begin{tabular}[t]{@{}l@{}}
Binning estimation:\\ 
Range of number of bins
\end{tabular}
& 10–100 
& $y^{1.0}\,t^{-0.5}\,(\mu/\rho)^{-0.5}$ \\ 
\cline{2-4}

& \begin{tabular}[t]{@{}l@{}}
 KSG Estimator: \\ 
Range of number of neighbors
\end{tabular}
& 5–10 
& $y^{1.0}\,t^{-0.5}\,(\mu/\rho)^{-0.5}$ \\ 
\hline

Active Subspaces 
& Kernel width \(\gamma\)
& 
\begin{tabular}[t]{@{}l@{}}
\(\gamma=1\) \\
\(\gamma=10\)
\end{tabular}
& 
\begin{tabular}[t]{@{}l@{}}
\(y^{1.0}\,t^{-0.5}\,(\mu/\rho)^{-0.5}\) \\
\(y^{1.0}\,t^{-0.5}\,(\mu/\rho)^{-0.5}\)
\end{tabular} \\ 
\hline

PyDimension 
& Polynomial regression order \(m\)
& 
\begin{tabular}[t]{@{}l@{}}
\(m=5\) \\
\(m=10\)
\end{tabular}
& 
\begin{tabular}[t]{@{}l@{}}
\(y^{1.0}\,t^{-0.5}\,(\mu/\rho)^{-0.5}\) \\
\(y^{0.7}\,t^{-0.3}\,(\mu/\rho)^{-0.4}\,U^{0.1}\)
\end{tabular} \\ 
\hline

BuckiNet 
& 
\begin{tabular}[t]{@{}l@{}}
Kernel width \(\gamma\), Ridge reg.\ \(\alpha\), \\ 
\(\ell_1\) reg.\ \(l_1\)
\end{tabular}
& 
\begin{tabular}[t]{@{}l@{}}
\(\gamma=1\), \(l_1=10^{-3}\), \(\alpha=10^{-4}\) \\
\(\gamma=10\), \(l_1=10^{-3}\), \(\alpha=10^{-4}\)
\end{tabular}
& 
\begin{tabular}[t]{@{}l@{}}
\(y^{1.0}\,t^{-0.5}\,(\mu/\rho)^{-0.5}\) \\
\(y^{1.0}\,t^{-0.63}\,(\mu/\rho)^{-0.37}\,U^{-0.25}\)
\end{tabular} \\ 
\hline

BSM 
& Dimension regularization weight \(w\)
& 
\begin{tabular}[t]{@{}l@{}}
\(w=10^{-3}\) \\
\(w=10^{-6}\)
\end{tabular}
& 
\begin{tabular}[t]{@{}l@{}}
\(y^{1.0}\,t^{-0.5}\,(\mu/\rho)^{-0.5}\) \\
\(y^{1.0}\,t^{-0.51}\,(\mu/\rho)^{-0.23}\,U^{-0.54}\)
\end{tabular} \\ 
\hline

\end{tabular}
}
\caption{{Discovered dimensionless variables $\Pi^*$
    for various dimensionless-learning methods under different
    hyperparameter settings on the Rayleigh problem}.}
\label{tab:estimator_comparison}
}
\end{table}
\renewcommand{\arraystretch}{1}

\subsection{{Sensitivity to input variable selection}}
\label{sec:sensitivity_input}

{ This section examines how the choice of input
  variables influences the results of the discovered dimensionless
  variables. In particular, we consider three important scenarios:
  missing variables, redundant variables, and discrete-valued
  variables.
\begin{itemize}
    \item \textbf{Missing variables} arise when relevant dimensional
      variables are omitted from the analysis---either because they
      are inaccessible, unmeasurable, or discarded a priori. In such
      situations, IT-$\pi$ still returns the best possible
      dimensionless variables within the set of available inputs.  We
      illustrate this with the Rayleigh problem, where the
      analytically optimal dimensionless input variable is
    \[
    \Pi^* = \frac{y}{\sqrt{(\mu/\rho)\, t}}
    \]
    given the complete set of dimensional inputs $\mathbf{q} = [U,
      y, t, \mu, \rho]$.  Suppose, however, that the variable $y$ is
    unavailable, and the input list is restricted to $\mathbf{q} =
    [U, t, \mu, \rho]$. In this case, IT-$\pi$ identifies the
    dimensionless variable as
    \[
    \frac{U^{1.0} t^{0.5}}{(\mu/\rho)^{0.5}}.
    \]
    Naturally, this result does not coincide with the analytically
    correct expression, as recovering the latter is impossible without
    $y$. However, the discovered variable is still optimal within the
    limited set of inputs.  A key feature of IT-$\pi$ is its ability
    to detect the presence of missing variables through a non-zero
    irreducible error. In this example, the irreducible error
    increases from $\epsilon = 0\%$ (when $y$ is included) to
    $\epsilon = 100\%$ (when $y$ is omitted), signaling the
    information loss due to the missing variable.
  \item \textbf{Redundant variables} arise when input variables
    contain highly similar or identical information. Consider the
    Rayleigh problem with the set of dimensional inputs
    $\mathbf{q} = [U, y, t, \mu, \rho, \nu]$, where the kinematic
    viscosity is defined as $\nu = \mu / \rho$. In this case, $\nu$
    carries the same information as the ratio $\mu / \rho$, since
    $I_\alpha(\nu ; \mu / \rho) = \infty$. Including both $\nu$ and
    $\mu / \rho$ in the input list introduces redundancy.  Such
    redundancy leads to infinitely many equivalent optimal
    solutions. For example, the dimensionless input
    \[
    \Pi^* = \frac{y}{\sqrt{(\mu/\rho) t}} = \frac{y}{\sqrt{\nu t}} =
    \frac{y}{\sqrt{(\mu/\rho)^{\beta} \nu^{1-\beta} t}}
    \]
    remains unchanged for any value of $\beta \in [0,1]$, since all
    these expressions contain the same information about the
    output. Formally, they yield identical mutual information with the
    dimensionless output:
    \[
    I_{\alpha}\left(\frac{y}{\sqrt{(\mu/\rho) t}} ; \Pi_o \right) =
    I_{\alpha}\left(\frac{y}{\sqrt{(\mu/\rho)^{\beta} \nu^{1-\beta} t}} ; \Pi_o \right).
    \]
    IT-$\pi$ is able to addresses redundancy by exhaustively searching
    over all combinations of input variables---e.g., $[q_i]$, $[q_i,
      q_j]$, $[q_i, q_j, q_k]$, etc.---and selecting the smallest
    subset that minimizes the irreducible error $\epsilon_{LB}$. In
    this example, the minimal non-redundant set would be $[U, y, t,
      \nu]$. This redundancy-aware selection strategy could, in
    principle, be extended to other dimensionless learning frameworks.
      \item \textbf{Discrete-valued variables} refer to input
        variables that take on only a finite set of values (e.g., $x =
        1, 2, 3, 4, \ldots$). The presence of such variables does not
        fundamentally change the operation of IT-$\pi$, as both Rényi
        entropy and mutual information are well defined for discrete,
        continuous, or mixed-type variables.  Therefore, the objective
        of the optimization-maximizing the mutual information between
        the input and the output, or equivalently minimizing the
        irreducible error, remains valid. This enables the
        identification of the most predictable dimensionless
        variables, even when some of the input variables are discrete.
        In contrast, methods based on predictive models, such as those
        employing neural networks, may face challenges in handling
        discrete inputs. These approaches typically assume continuous
        and differentiable variables to compute gradients required for
        optimization during model training, which can limit their
        applicability or performance in settings involving
        discrete-valued features.
\end{itemize}
}

\subsection{{Diagnostics for detecting overfitting}}
\label{sec:diagnostics_overfitting}

{We discuss the contrasting roles of irreducible error
  and model overfitting. When a sufficiently large number of samples
  is available (as discussed in \S\ref{sec:sensitivity_stat}),
  IT-$\pi$ provides a reliable estimate of the irreducible error,
  $\epsilon_{LB}^*$, which represents the fundamental limit below
  which no model---regardless of complexity---can improve its
  predictive performance. In this regime, model performance is best
  assessed by the efficiency metric $\eta$.  However, for finite
  sample sizes, it is well known that an overly complex model may
  achieve perfect prediction on the training data by memorizing it
  (i.e. overfitting).  This phenomenon does not contradict the concept
  of irreducible error, as a model that overfits the training data
  typically fails to generalize and performs poorly on new samples
  drawn from the same underlying probability distribution.  For
  example, fitting ten data points with a model containing a thousand
  parameters may yield excellent agreement with the training data, yet
  fail in new samples following the same distribution. In contrast, as
  the number of samples approaches infinity, the performance of any
  model is ultimately bounded by the irreducible error regardless of
  its complexity. }

{ To evaluate model performance under finite data
  conditions, we compare the model error $\epsilon_f$ against the
  uncertainty-aware irreducible error, expressed as $\epsilon_{LB}^*
  \pm \Delta \epsilon_{LB}^*$. The methodology for computing $\Delta
  \epsilon_{LB}^*$ is described in \S\ref{sec:sensitivity_stat}. Based
  on the relationship between $\epsilon_f$, $\epsilon_{LB}^*$, and
  $\Delta \epsilon_{LB}^*$, we classify model performance into three
  distinct regimes:
\begin{itemize}
    \item \textbf{Suboptimal model within uncertainty:}
    \[
    \epsilon_{LB}^* + \Delta \epsilon_{LB}^* < \epsilon_f.
    \]
    In this regime, the model error exceeds the irreducible error by
    more than the estimated uncertainty, indicating room for
    improvement.
    \item \textbf{Optimal model within uncertainty:}
    \[
    \epsilon_{LB}^* - \Delta \epsilon_{LB}^* \leq \epsilon_f \leq \epsilon_{LB}^* + \Delta \epsilon_{LB}^*.
    \]
    The model error is statistically indistinguishable from the
    irreducible error, suggesting that the model is performing
    optimally given the available data.
    \item \textbf{Overfitting model within uncertainty:}
    \[
    \epsilon_f < \epsilon_{LB}^* - \Delta \epsilon_{LB}^*.
    \]
    In this case, the model achieves an error lower than the estimated
    irreducible bound, which typically indicates overfitting---i.e.,
    the model is capturing noise in the training data rather than
    generalizable structure.
\end{itemize}}

{ To illustrate the procedure described above, we
  construct a synthetic example in which samples of the input variable
  $\Pi$ are drawn uniformly from the interval $[0, 1]$. The output is
  defined as a smooth, nonlinear function of $\Pi$ with added Gaussian
  noise:
\[
\Pi_o = \Pi(1 - \Pi)(1 - 2\Pi)(1 - 3\Pi)(1 - 4\Pi)(1 - 5\Pi) + 0.3\, \mathcal{N}(0,1).
\]
Given access only to the input $\Pi$, the optimal predictive model is
\[
\hat{\Pi}_o^* = \Pi(1 - \Pi)(1 - 2\Pi)(1 - 3\Pi)(1 - 4\Pi)(1 - 5\Pi),
\]
and the irreducible error corresponds to the added Gaussian noise
term, $0.3\, \mathcal{N}(0,1)$. The analytical $L_2$ norm of this
irreducible error is
\[
\epsilon_{LB}^{*\text{analytical}} = \sqrt{\mathbb{E}\left[\left(\Pi_o - \hat{\Pi}_o^*\right)^2\right]} = 0.3.
\]
This value represents the minimum achievable error for any model,
regardless of its complexity, in the limit of infinite data.  In
practice, using a finite sample of $N = 800$ data points, we estimate
the irreducible error bound $\epsilon_{LB}^*$ and its associated
statistical uncertainty $\Delta \epsilon_{LB}^*$ as
\[
\epsilon_{LB}^* = 0.2998, \quad \Delta \epsilon_{LB}^* = 0.0063.
\]
}

\begin{figure}[]
    \centering
    \includegraphics[width=0.8 \textwidth]{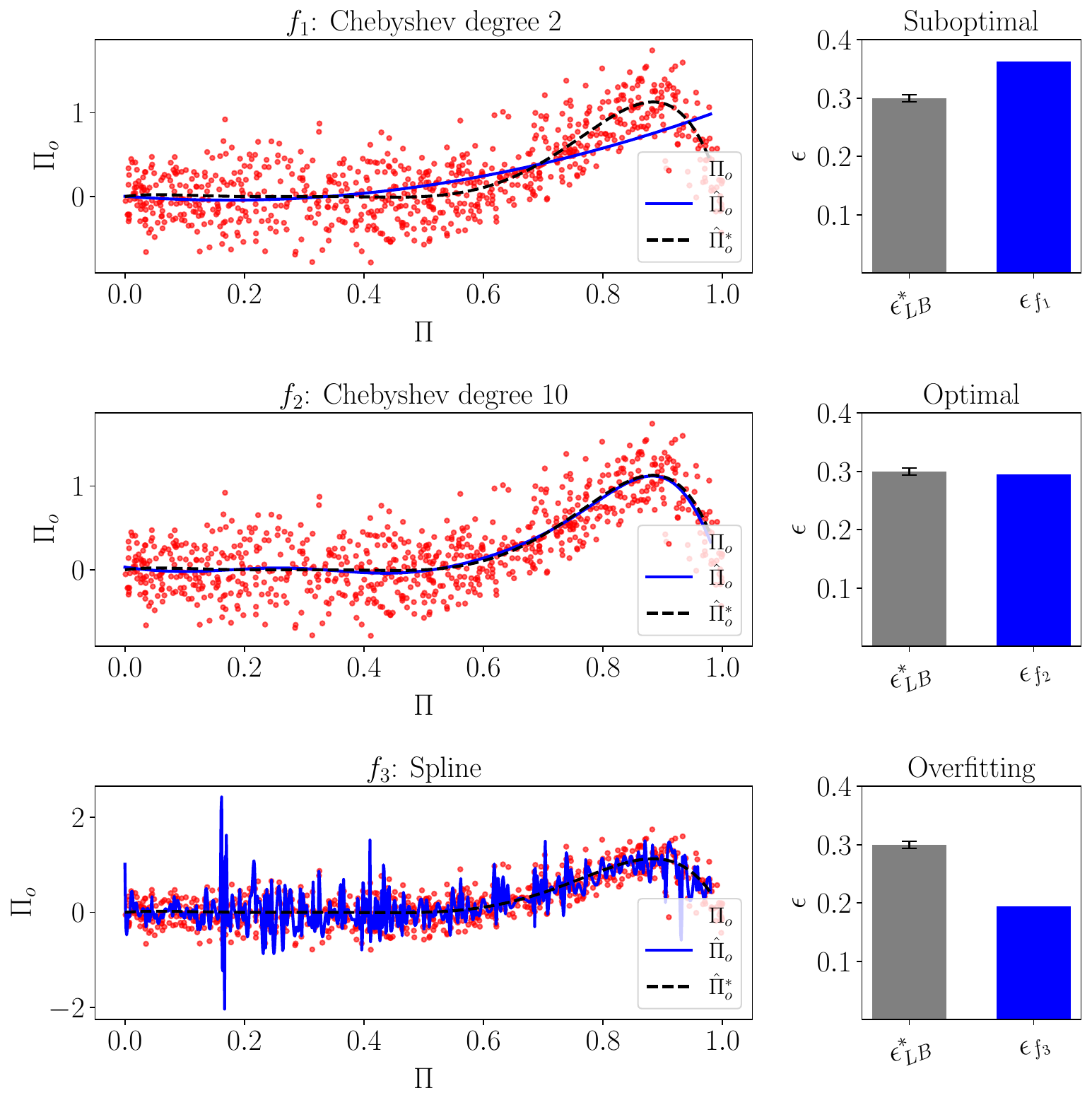}
        \put(-400,380){$(a)$}
        \put(-400,240){$(b)$}
        \put(-400,115){$(c)$}
        \caption{{Example of how statistical
            uncertainty in the estimated irreducible error can be used
            to diagnose model overfitting. Three models, (a), (b), and
            (c), are fitted to the same dataset and evaluated against
            the estimated irreducible error. In each case, the left
            panel shows the analytical optimal model (dashed), the
            fitted model prediction (blue), and the noisy output data
            (red). The corresponding bar plots on the right compare
            the model error $\epsilon_f$ with the irreducible error
            $\epsilon_{LB}^*$ and its uncertainty $\Delta
            \epsilon_{LB}^*$. Based on this comparison, the models are
            classified as: (a) suboptimal, (b) optimal within
            uncertainty, and (c) overfitting. }}
    \label{fig:overfitting_diagnostic}
\end{figure}

{We trained three different models on the
  finite-sample dataset to predict the output $\hat \Pi_o$: a
  low-degree Chebyshev polynomial (degree 2) $f_1$, a higher-degree
  Chebyshev polynomial (degree 10) $f_2$, and a spline $f_3$ with a
  smoothing parameter $30$.  The performance of the models are
  summarized in Figure~\ref{fig:overfitting_diagnostic}.
\begin{itemize}
\item Panel (a) shows the performance of degree 2 Chebyshev polynomial
  $f_1$. The model $\hat \Pi_o$ has too few parameters to approximate
  accurately the analytical optimal model $\hat{\Pi}_o^*$. The model
  error is $\epsilon_{f_1}=0.3627$, which exceeds the upper bound of
  the irreducible error interval $\epsilon_{L B}^*+\Delta \epsilon_{L
    B}^*$. According to the criterion, this model is classified as
  suboptimal, which is consistent with the visible underfitting in the
  plot.
\item Panel (b) shows the performance of the degree 10 Chebyshev
  polynomial $f_2$. This model $\hat{\Pi}_o$ closely matches the
  analytical optimal model $\hat{\Pi}_o^*$ without introducing
  excessive oscillations. The error $\epsilon_{f_2}=0.2945$ falls
  within the uncertainty interval $\left[\epsilon_{LB}^*-\Delta
    \epsilon_{L B}^*, \epsilon_{LB}^*+\Delta \epsilon_{L
      B}^*\right]$. According to the criterion, this model is optimal
  within uncertainty, consistent with the visually good fit.
\item Panel (c) displays the performance of the smoothing spline
  $f_3$. Visually, this model $ \hat \Pi_o$ over-fits the noisy
  fluctuations. The error $\epsilon_{f_3}=0.1937$ is below the lower
  bound of the irreducible error interval $\epsilon_{LB}^*-\Delta
  \epsilon_{LB}^*$. According to our criterion, the model is
  classified as overfitting, which agrees with the observed tendency
  to fit noise rather than only the true predictable function.
\end{itemize}
}

\newpage
\subsection{{Robustness of IT-$\pi$ to noise}}
\label{sec:sensitivity_noise}

\noindent \textbf{Effect of Noise across Benchmark Cases}

\noindent {We systematically evaluate the impact of
  noise on IT-$\pi$ across all benchmark cases. Gaussian noise of
  increasing intensity is added to the dimensional variables, and
  IT-$\pi$ is then reapplied to recover the optimal dimensionless
  variables. The results, summarized in Table~\ref{tab:noise_valid},
  demonstrate that IT-$\pi$ consistently recovers the correct
  dimensionless relationships even under noise levels as high as
  20\%. Below, we provide additional details of the procedure for two
  representative cases: the Rayleigh problem and the Colebrook
  equation.}
\begin{table}[h!]
\centering
\setlength{\arrayrulewidth}{1 pt} 
\resizebox{0.6 \textwidth}{!}{%
{
\begin{tabular}{|l|c|c|c|c|}
\hline
\textbf{Noise Level} & \textbf{1\%} & \textbf{5\%} & \textbf{10\%} & \textbf{20\%} \\
\hline
Rayleigh problem  & \cmark & \cmark & \cmark & \cmark  \\
\hline
Colebrook equation & \cmark & \cmark & \cmark & \cmark \\
\hline
Malkus waterwheel & \cmark & \cmark & \cmark & \xmark \\
\hline
Rayleight-Bénard convection & \cmark  & \cmark & \cmark & \cmark \\
\hline
Blasius boundary layer& \cmark  & \cmark &\cmark  & \cmark  \\
\hline
\end{tabular}%
}}
\caption{{Effect of noise on the discovered dimensionless variable by
  IT-$\pi$ for all validation cases. Four noise levels are tested:
  $1\%, 5\%, 10\%, 20\%$. \cmark\ indicates that the method
  successfully recovered the analytical dimensionless variable}. }
      \label{tab:noise_valid}

\end{table}

\begin{figure}[h!]
    \centering
    \begin{subfigure}[b]{0.24\textwidth}
        \centering
        \includegraphics[width=\textwidth]{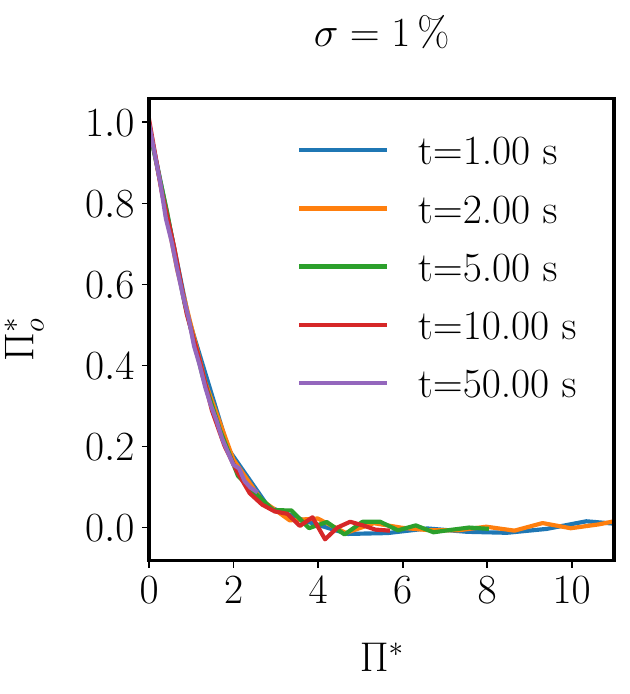}
    \end{subfigure}
    \hfill
    \begin{subfigure}[b]{0.24\textwidth}
        \centering
        \includegraphics[width=\textwidth]{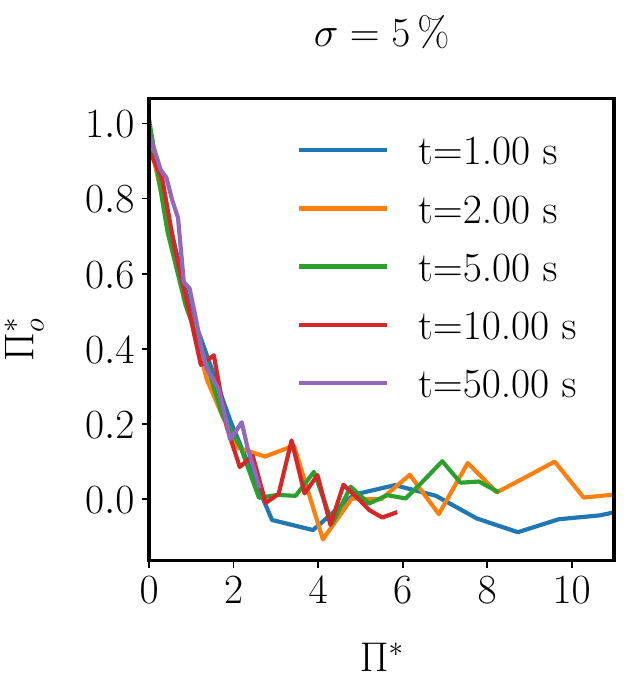}
    \end{subfigure}
    \hfill
    \begin{subfigure}[b]{0.24\textwidth}
        \centering
        \includegraphics[width=\textwidth]{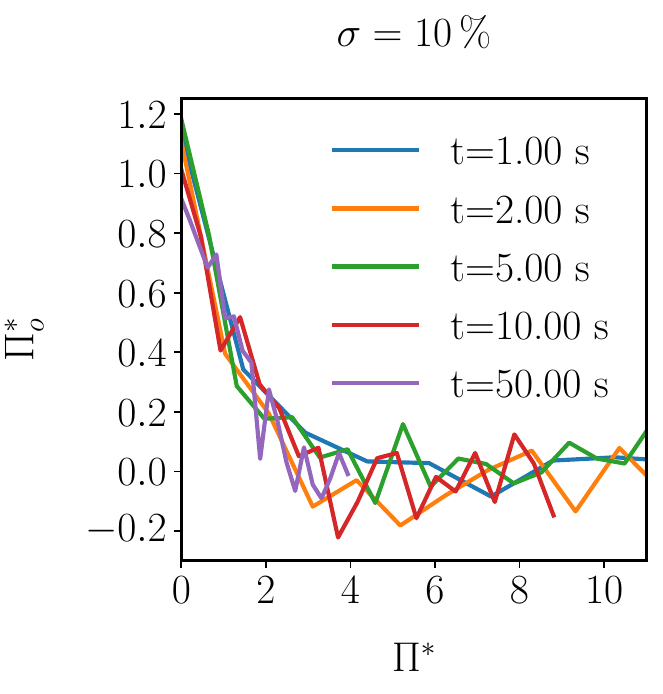}
    \end{subfigure}
    \hfill
    \begin{subfigure}[b]{0.24\textwidth}
        \centering
        \includegraphics[width=\textwidth]{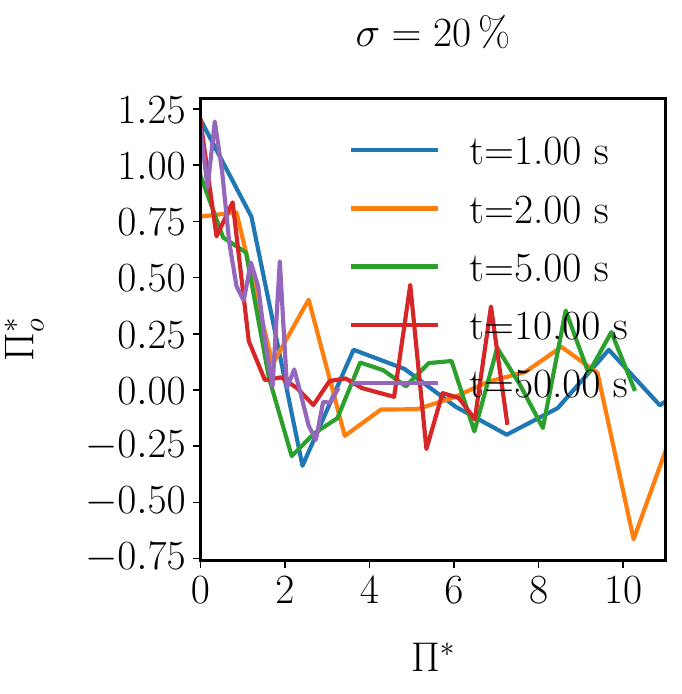}
    \end{subfigure}

    \caption{Robustness of IT-$\pi$ to noise { for
        Rayleigh problem}. Visualization of $\Pi_o^*$ as a function of
      $\Pi^*$ for different noise levels. }
    \label{fig:velocity_profiles_noise}
\end{figure}

\noindent \underline{Rayleigh problem}. We evaluate the robustness of IT-$\pi$
on the Rayleigh problem by introducing noise into the velocity profile
as follows:
\[
u_{\text{noisy}} = u + \epsilon,
\]
where
\[
u = U\, \operatorname{erfc}\!\left(\frac{\xi }{2}\right)
\]
is the noiseless solution, and 
\[
\epsilon \sim \mathcal{N}\Big(0, (\sigma \cdot U)^2\Big)
\]
represents Gaussian noise with zero mean and a standard deviation
proportional to \(U\). Here, \(\sigma\) is a scaling factor that
determines the noise level as a fraction of \(U\). We test four noise
levels: 1\%, 5\%, 10\% and 20\%, corresponding to \(\sigma = 0.01\),
\(0.05\), \(0.1\), and \(0.2\),
respectively. Figure~\ref{fig:velocity_profiles_noise} visualizes the
relationship between \(\Pi_o^*\) and \(\Pi^*\) as discovered by
IT-$\pi$ at different noise levels. For all noise levels, the
resulting dimensionless variable agrees with that obtained from the
noiseless data:
\[
\Pi^* = y^{1.0}\, t^{-0.5}\, \mu^{-0.5}\, \rho^{0.5}.
\]

\begin{figure}[h!]
    \centering
    \begin{subfigure}[b]{0.24\textwidth}
        \centering
        \includegraphics[width=\textwidth]{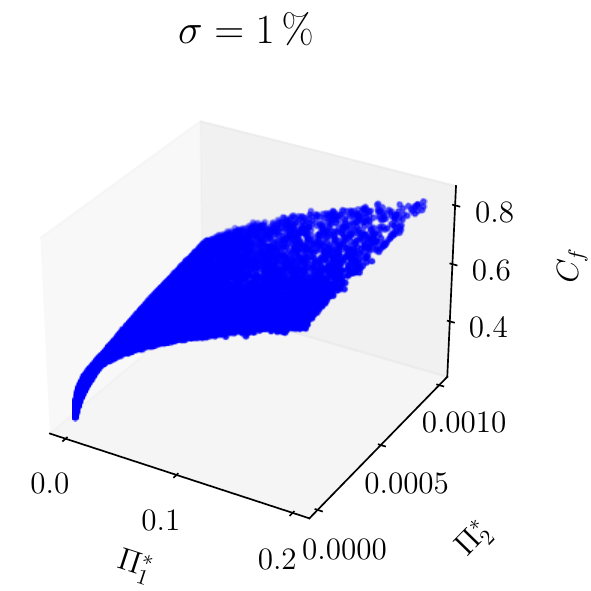}
    \end{subfigure}
    \hfill
    \begin{subfigure}[b]{0.24\textwidth}
        \centering
        \includegraphics[width=\textwidth]{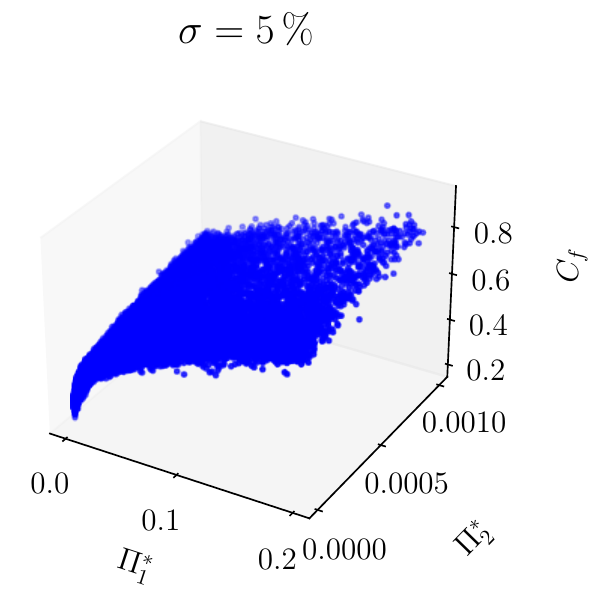}
    \end{subfigure}
    \hfill
    \begin{subfigure}[b]{0.24\textwidth}
        \centering
        \includegraphics[width=\textwidth]{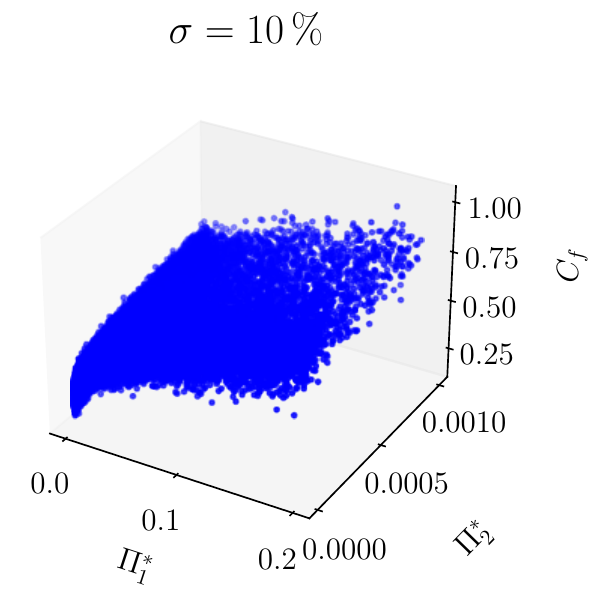}
    \end{subfigure}
    \hfill
    \begin{subfigure}[b]{0.24\textwidth}
        \centering
        \includegraphics[width=\textwidth]{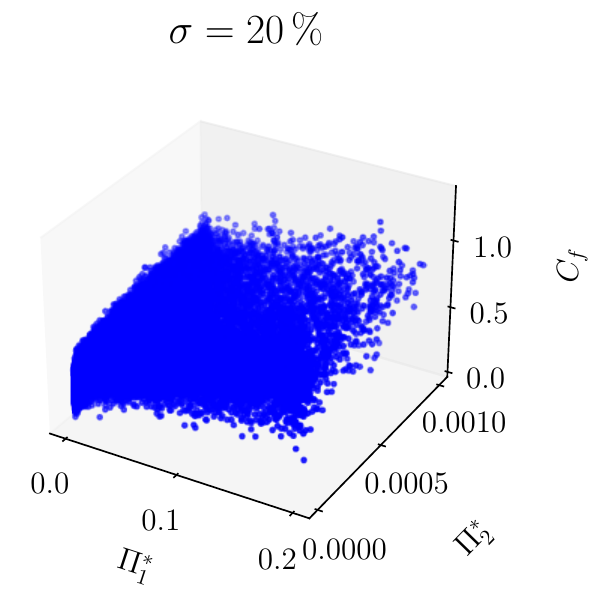}
    \end{subfigure}

    \caption{{Robustness of IT-$\pi$ to noise for
        Colebrook equation. Visualization of $\Pi_o^*$ as a function
        of $\Pi^*$ for different noise levels.} }
    \label{fig:Colebrook_noise}
\end{figure}

\noindent {\underline{Colebrook equation}}
{We assess the robustness of IT-$\pi$ under noisy data
  using the Colebrook equation.  To simulate measurement noise, we
  perturb the friction coefficient as $$C_{f, \text { noisy
  }}=C_f+\epsilon,$$ where $C_f$ is the noiseless solution computed
  from the Colebrook equation and $$\epsilon \sim
  \mathcal{N}\left(0,\left(\sigma \cdot C_f\right)^2\right)$$ is the
  zero-mean Gaussian noise with standard deviation proportional to
  $C_f$. We evaluate four noise levels of 1\%, 5\%, 10\% and 20\%
  ($\sigma=0.01,0.05,0.1$, and $0.2$,
  respectively). Figure~\ref{fig:Colebrook_noise} visualizes the
  relationship between \(\Pi_o^*\) and \(\Pi^*\) as discovered by
  IT-$\pi$ at different noise levels.  Across all noise levels, the
  identified dimensionless variables remain consistent with those from
  the noiseless data,
$$
\Pi_1^*  =\frac{k}{D}, \quad \Pi_2^*  =\frac{\mu}{U \rho D}.
$$}

\noindent \textbf{Theoretical Analysis of IT-$\pi$ under Noise}

\noindent The robustness of IT-$\pi$ to noise can be explained by the
invariance of the optimal mutual information under additive
noise. Consider the assumption of additive noise sources
$\mathbf{W}_I$ and $W_O$, which are independent of each other and
also independent of $\mathbf{\Pi}$ and $\Pi_o$. Let us also assume
that there are no statistical uncertainties in the estimation of the
mutual information. By the data processing inequality
\cite{cover1999elements}, we have
$$ I_\alpha\bigl(\mathbf{\Pi}+\mathbf{W}_I;\, \Pi_o+W_o\bigr)
\leq I_\alpha\bigl(\mathbf{\Pi};\, \Pi_o\bigr),
$$ which implies that the mutual information generally decreases due
to the added noise.

However, while this degradation increases the irreducible error, the
optimal dimensionless variables remain unchanged {if
  $\mathbf{\Pi^*}$, $\Pi_o^*$ are jointly Gaussian distributed,
  $\mathbf{W_I}$ and $W_O$ are independent Gaussian variables},
$$
\mathbf{\Pi}^*,\, \Pi_o^* = \arg\min_{\mathbf{\Pi},\, \Pi_o} \max_{\alpha} [ \epsilon_{LB} ] 
= \arg\min_{\mathbf{\Pi}+\mathbf{W}_I,\, \Pi_o+W_O} \max_{\alpha}  [ \epsilon_{LB} ].
$$ This invariance under noise ensures that IT-$\pi$ consistently
identifies the most predictive dimensionless inputs, despite the
reduction in mutual information caused by noise. This is a key
property that makes IT-$\pi$ robust in real-world applications where
noisy measurements are inevitable.\\

\begin{proof}
Consider the minimizer for the irreducible error
$\mathbf{\Pi}^*,\, \Pi_o^* = \arg\min_{\mathbf{\Pi},\, \Pi_o}
[ \epsilon_{LB} ] $ which is equivalently the minimizer for the
conditional entropy $\mathbf{\Pi}^*,\, \Pi_o^* =
\arg\min_{\mathbf{\Pi},\, \Pi_o} h_\alpha\left(\Pi_o \mid
        {\mathbf{\Pi}}\right)$, which indicates that
$$
\forall \Pi_o,{\mathbf{\Pi}} \quad  h_\alpha\left({{\Pi}}_o \mid {\mathbf{\Pi}}\right) \geq h_\alpha\left(\Pi_o^* \mid {\mathbf{\Pi}}^*\right).
$$ We want to prove that adding independent Gaussian noise to both the
input and output preserves the minimizer of the conditional entropy,
$$
 h_\alpha\left(\Pi_o \mid {\mathbf{\Pi}}\right) \geq h_\alpha\left(\Pi_o^* \mid {\mathbf{\Pi}}^*\right) \Rightarrow h_\alpha\left(\Pi_o+W_O \mid {\mathbf{\Pi}}+{\mathbf{W}_I}\right)\geq h_\alpha\left(\Pi_o^{*}+W_O \mid {\mathbf{\Pi}}^{*}+{\mathbf{W}_I}\right).
$$ To prove that, consider the Rényi conditional differential entropy
 jointly Gaussian random variables is given by
$$
h_\alpha(\Pi_o \mid {\mathbf{\Pi}}) = \frac{1}{2} \log\Bigl(2\pi\,\operatorname{Var}(\Pi_o\mid{\mathbf{\Pi}})\Bigr) - \frac{1}{2(1-\alpha)}\log\alpha.
$$ Since the logarithm is an increasing function, it follows that
$$
h_\alpha\left(\Pi_o\mid{\mathbf{\Pi}}\right) \geq h_\alpha\left(\Pi_o^{*}\mid{\mathbf{\Pi}}^{*}\right) \quad\iff\quad \operatorname{Var}\left(\Pi_o\mid{\mathbf{\Pi}}\right) \geq \operatorname{Var}\left(\Pi_o^{*}\mid{\mathbf{\Pi}}^{*}\right).
$$
Because the noise  \({\mathbf{W}_I}\) and \(W_O\) is independent and Gaussian, it follows that
$$
\begin{aligned}
\operatorname{Var}\left(\Pi_o+W_O \mid \mathbf{\Pi}+\mathbf{W}_I\right) & =\operatorname{Var}\left(\Pi_o \mid \mathbf{\Pi}+\mathbf{W}_I\right)+\operatorname{Var}\left(W_O\right) \\
\operatorname{Var}\left(\Pi_o^*+W_O \mid \mathbf{\Pi}^*+\mathbf{W}_I\right) & =\operatorname{Var}\left(\Pi_o^* \mid \mathbf{\Pi}^*+\mathbf{W}_I\right)+\operatorname{Var}\left(W_O\right)
\end{aligned}
$$
Therefore, the ordering of the variances is preserved:
$$
\operatorname{Var}\left(\Pi_o+W_O\mid{\mathbf{\Pi}}+{\mathbf{W}_I}\right) \geq \operatorname{Var}\left(\Pi_o^{*}+W_O\mid{\mathbf{\Pi}}^{*}+{\mathbf{W}_I}\right),
$$
which implies
$$
h_\alpha\left(\Pi_o+W_O\mid{\mathbf{\Pi}}+{\mathbf{W}_I}\right) \geq h_\alpha\left(\Pi_o^{*}+W_O\mid{\mathbf{\Pi}}^{*}+{\mathbf{W}_I}\right).
$$
Therefore,
$$
\forall \Pi_o,{\mathbf{\Pi}} \quad  h_\alpha\left(\Pi_o \mid {\mathbf{\Pi}}\right) \geq h_\alpha\left(\Pi_o^* \mid {\mathbf{\Pi}}^*\right)
\Rightarrow
h_\alpha\left(\Pi_o+W_o \mid {\mathbf{\Pi}}+{\mathbf{W}_I}\right)\geq h_\alpha\left(\Pi_o^{*}+W_o \mid {\mathbf{\Pi}}^{*}+{\mathbf{W}_I}\right),
$$
which indicates 
$${\mathbf{\Pi}}^*, \Pi_o^*=\arg \min _{{\mathbf{\Pi}}, \Pi_o} \max_{\alpha} \left[\epsilon_{L B}\right]=\arg \min _{{\mathbf{\Pi}}+W_I, \Pi_o+W_O} \max_{\alpha} \left[\epsilon_{L B}\right] .$$
\end{proof}

\newpage
\subsection{{Effect of the order $\alpha$ in the Rényi entropy }}
\label{sec:sensitivity_alpha}
{Rényi entropy introduces the order $\alpha$ that
  generalizes the classic Shannon entropy, which corresponds to the
  special case $\alpha = 1$. The value of $\alpha$ controls the
  relative emphasis placed on different regions of the error
  distribution $\mathcal{E}$. Lower values of $\alpha$ assign greater
  weight to the tails---highlighting rare events---while higher values
  emphasize the mode of the distribution. This flexibility makes Rényi
  entropy particularly suitable when the conditional density
  $\rho(\Pi_o \mid \Pi)$ is heavy-tailed, asymmetric, or
  multi-modal. This is leveraged by IT-$\pi$ to estimate the
  irreducible error for different L$_p$-norms by finding the best
  value of $\alpha$ as part of the optimization problem.  When the
  error distribution is approximately Gaussian or concentrated, the
  influence of $\alpha$ is minimal, and standard Shannon entropy is
  typically sufficient.}

\begin{figure}[htbp]
    \centering
    \begin{subfigure}{0.3\textwidth}
        \includegraphics[width=\linewidth]{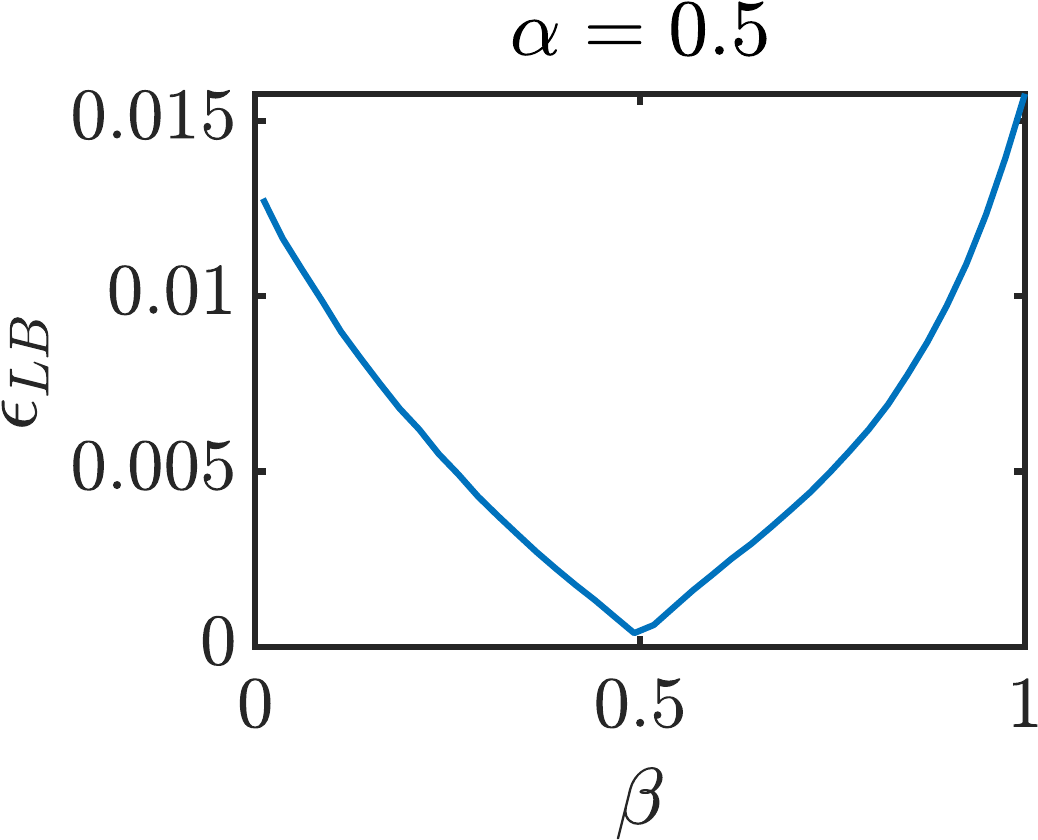}
    \end{subfigure}
    \hfill
    \begin{subfigure}{0.3\textwidth}
        \includegraphics[width=\linewidth]{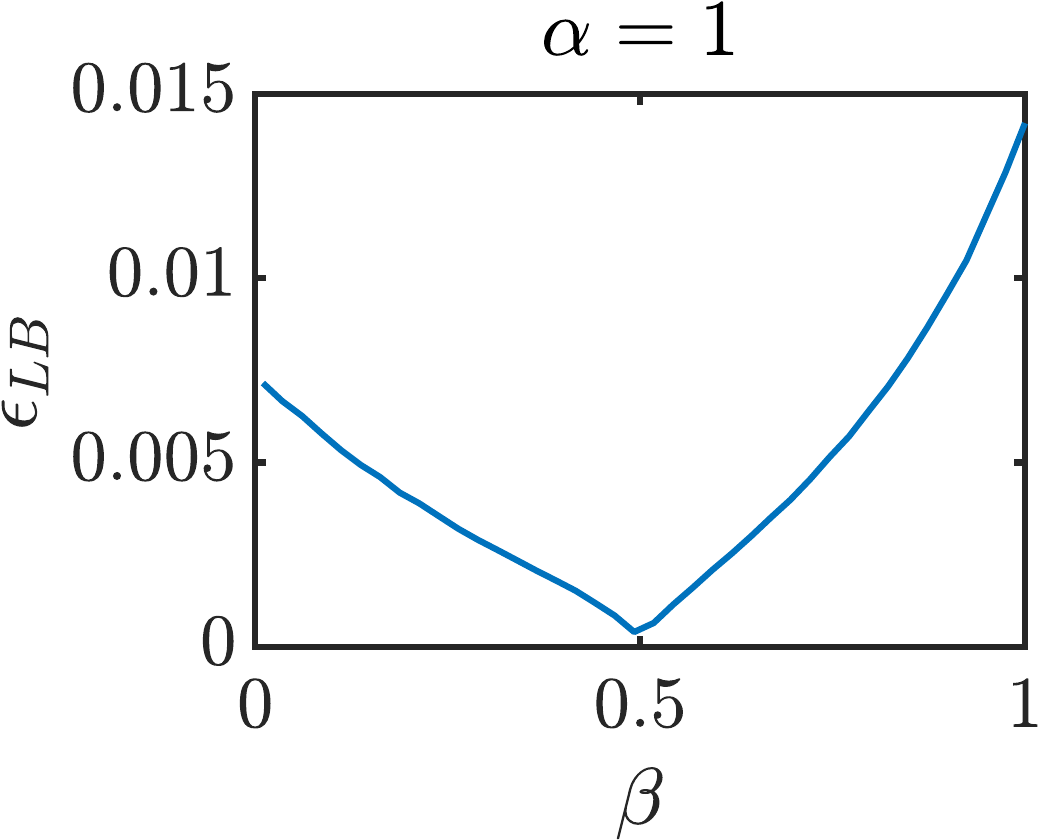}
    \end{subfigure}
    \hfill
    \begin{subfigure}{0.3\textwidth}
        \includegraphics[width=\linewidth]{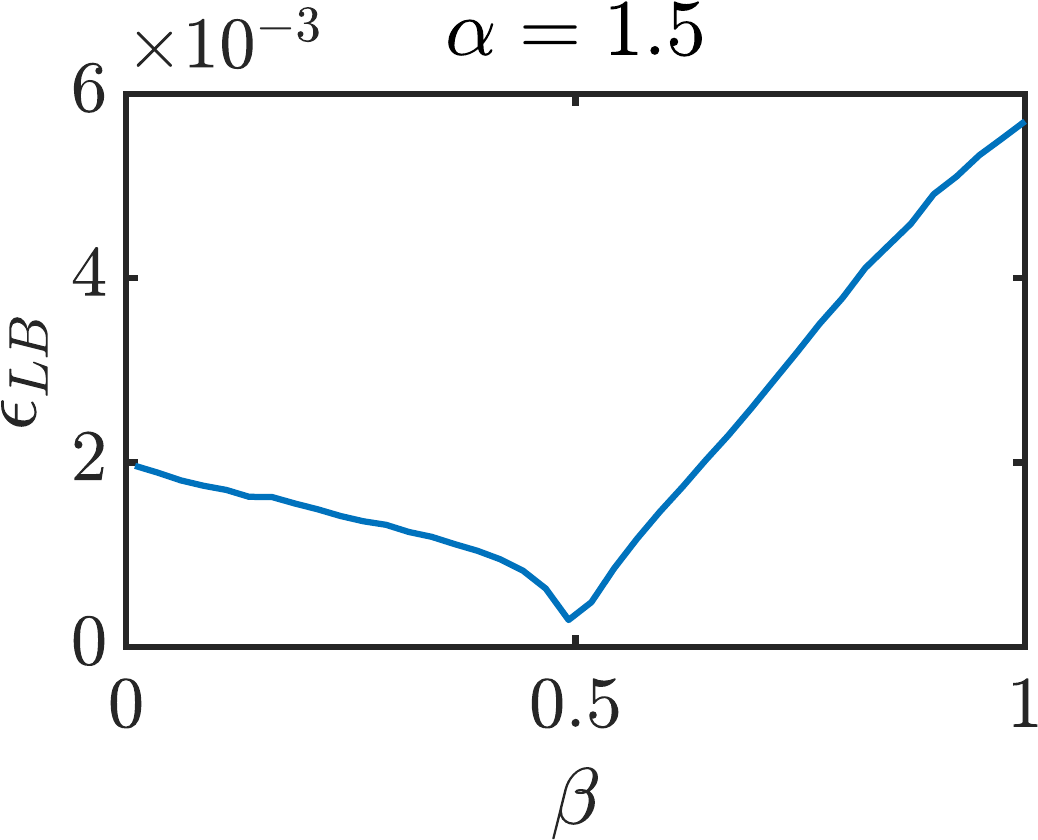}
    \end{subfigure}
    \caption{{Comparison of $\epsilon_{LB}$ as a
        function of $\beta$ for different values of $\alpha$ in the
        Rayleigh problem.}}
    \label{fig:elb_alpha_comparison}
\end{figure}

{In the cases discussed in the main text, where the
  irreducible error is close to zero and the error norm of interest is
  set to $p = 2$, varying the Rényi entropy order has small or no
  impact on the minimum irreducible error $\epsilon_{LB}^*$ and the
  resulting dimensionless variables $\Pi^*$. Hence, the standard
  Shannon entropy ($\alpha = 1$) is sufficient in such cases. To
  illustrate this, we consider the Rayleigh problem and plot
  $\epsilon_{LB}$ as a function of the coefficient $\beta$ for
  different values of $\alpha$=0.5, 1.0, and 1.5. The candidate
  dimensionless variable is expressed as
\[
\Pi = \sqrt{\frac{y}{Ut}} \left(\frac{Uy}{\mu/\rho}\right)^{\beta},
\]
where $\beta$ is the optimization parameter. As shown in
Figure~\ref{fig:elb_alpha_comparison}, across all tested values of
$\alpha$, the optimal coefficient remains $\beta^* = 0.5$, indicating
that the discovered variable is insensitive to the choice of $\alpha$
in this case.}

{In other situations, the choice of $\alpha$ plays a
  fundamental role in identifying the optimal dimensionless
  variables. This is demonstrated in \S\ref{sec:norm}, where we
  show that the optimal variables depend on whether the goal is to
  predict extreme events or more frequent occurrences, as determined
  by the choice of error norm. The reader is referred to that section
  for further details.  }

\newpage
\section{~Additional Validation}

\subsection{Optimal dimensionless inputs adapted to different error norms}
\label{sec:norm}

\begin{figure}[h]
    \centering
    \begin{subfigure}{0.7\textwidth}
        \centering
        \includegraphics[width=0.7  \linewidth,trim={0.0cm 0.0cm 0.0cm 0.0cm},clip]{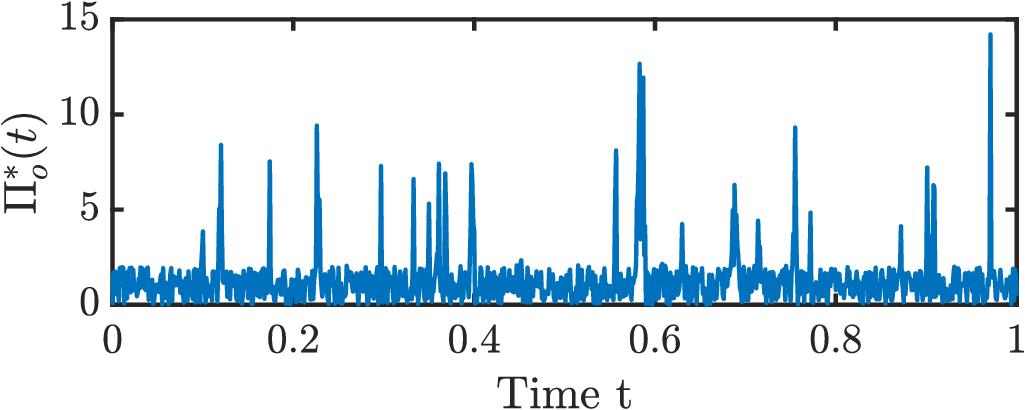} 
        \put(-260,80){$(a)$}
        \vspace{0.2cm}  
    \end{subfigure}
    \begin{subfigure}{0.35\textwidth}
        \centering
        \includegraphics[width=\linewidth,trim={0cm 0cm 0cm 0cm}]
        {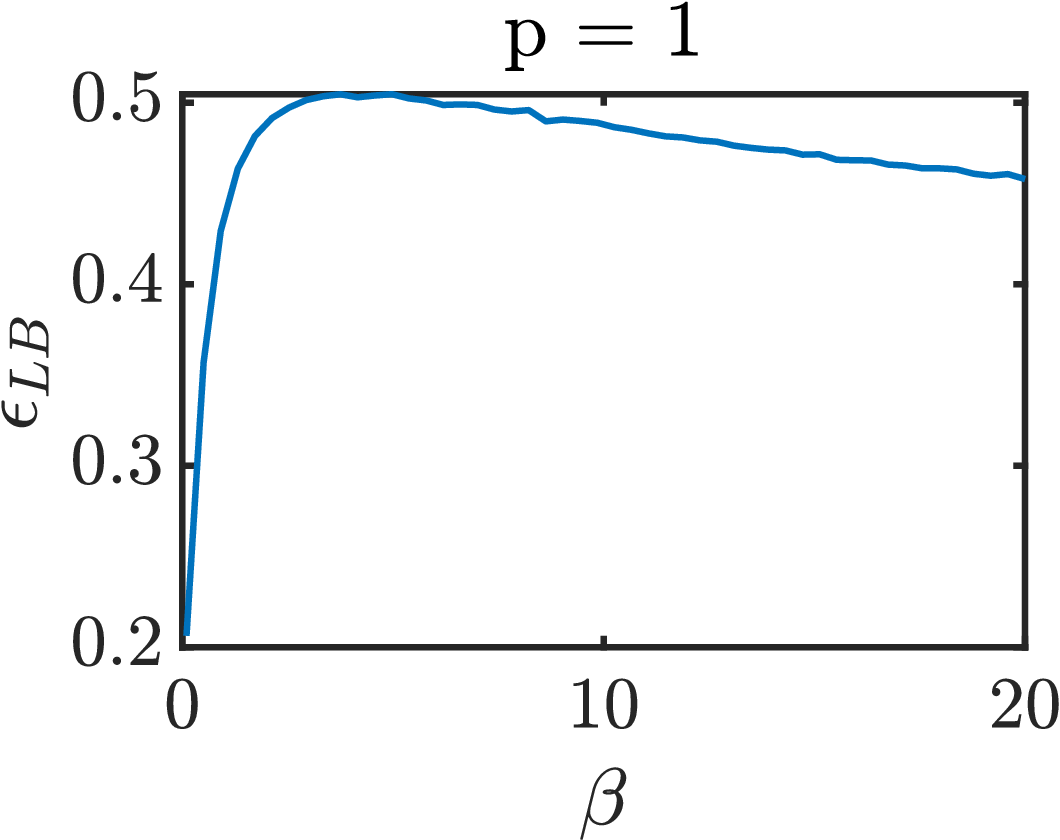}  
                \put(-180,120){$(b)$} 

    \end{subfigure}
    \hspace{0.05\textwidth}  
    \begin{subfigure}{0.35\textwidth}
        \centering
        \includegraphics[width=\linewidth,trim={0cm 0cm 0cm 0cm}]{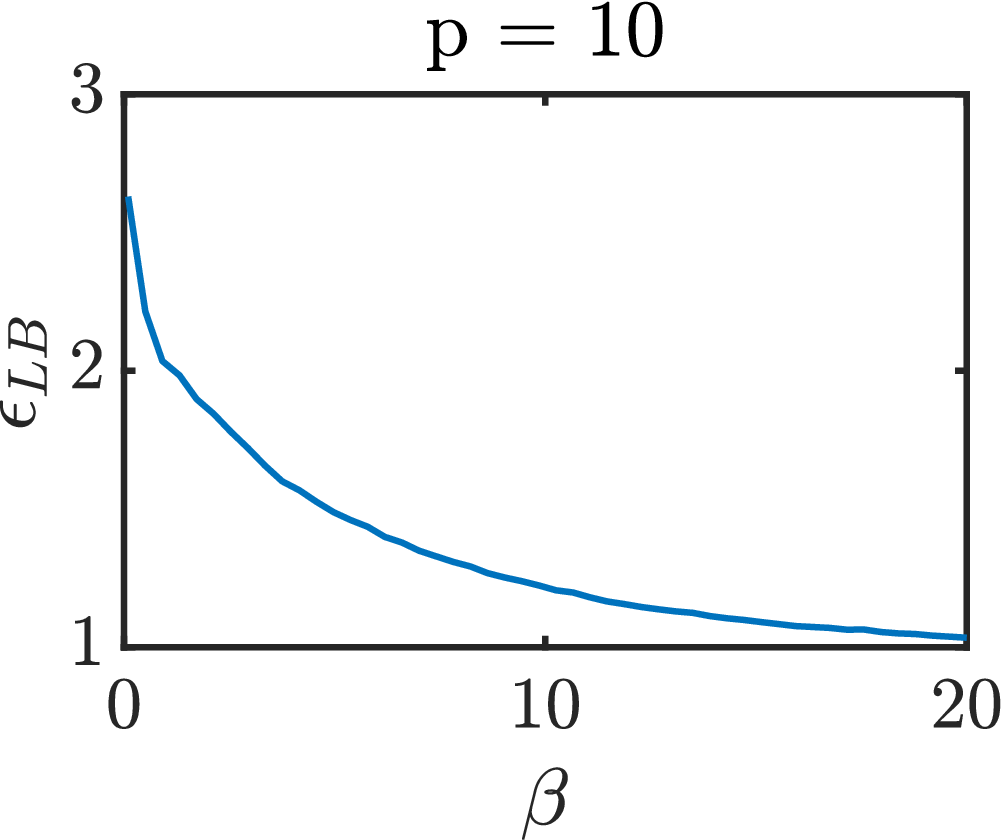}  
        \put(-180,120){$(c)$}
    \end{subfigure}

        \begin{subfigure}{0.35\textwidth}
        \centering
        \includegraphics[width=\linewidth,trim={0cm 0cm 0cm 0cm}]
        {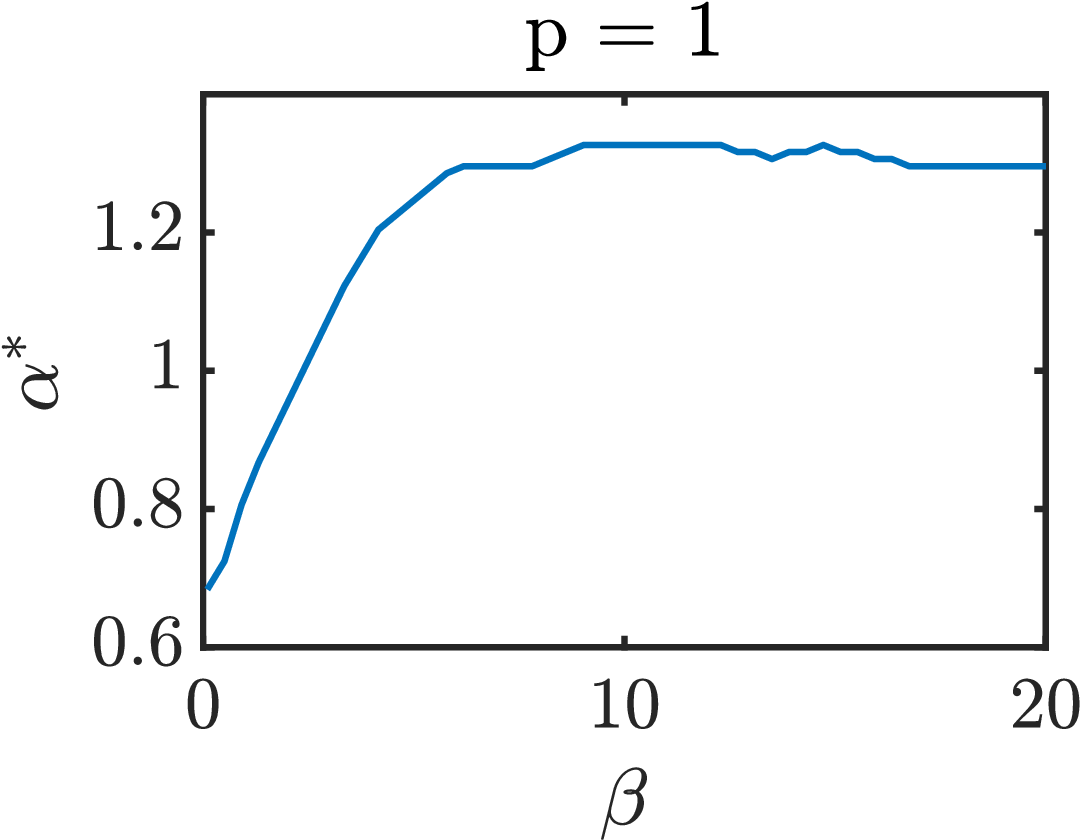}  
                \put(-180,120){$(d)$} 

    \end{subfigure}
    \hspace{0.05\textwidth}  
    \begin{subfigure}{0.35\textwidth}
        \centering
        \includegraphics[width=\linewidth,trim={0cm 0cm 0cm 0cm}]{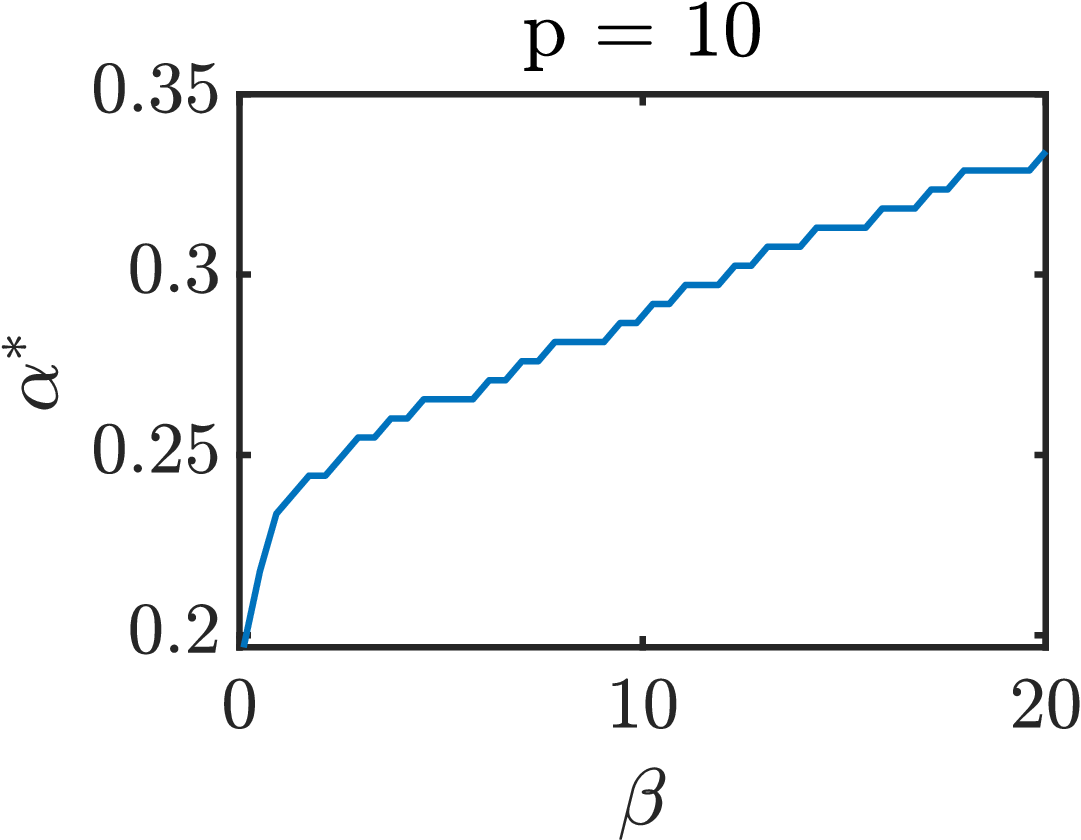}  
        \put(-180,120){$(e)$}
    \end{subfigure}
    \caption{(a) Time evolution of $\Pi_o^*(t)$. (b,c) Irreducible
      error $\epsilon_{LB}(\Pi)$ for $\Pi=\Pi_1\Pi_2^\beta$ as a
      function of $\beta$ for (b) $p=1$ and (c)
      $p=10$. {(d,e) $\alpha^*$ corresponds to the
        value that yields the tightest bound as a function of $\beta$
        for (d) $p=1$ and (e) $p=10$.}}
    \label{fig:p_norm_time_series}
\end{figure}

We present a validation case that examines how the optimal
dimensionless input $\Pi^*$, discovered by IT-$\pi$, varies depending
on the selected $L_p$-norm for the error. One scenario where this
behavior is particularly relevant is in the prediction of extreme
events, which is crucial for mitigating risks associated with rare but
high-impact occurrences.  As a representative example, we define the
dimensionless output $\Pi_o^*(t)$ as
$$
\Pi_o^*(t)=
\begin{cases}
\Pi_2(t) & \text{if } |\Pi_o(t-\delta t)| > 2, \\
\Pi_1(t) & \text{if } |\Pi_o(t-\delta t)| \le 2,
\end{cases}
$$ where $\delta t$ is the time step, and $\Pi_1$ and $\Pi_2$ denote
two dimensionless variables drawn from different
distributions. Specifically, $\Pi_1 \sim U(-2.1,2.1)$ is uniformly
distributed, while $\Pi_2 \sim \mathcal{N}(0,5)$ follows a normal
distribution, as shown in Figure \ref{fig:p_norm_time_series}(a). In
this example, the inputs are already dimensionless, allowing us to
focus on the ability of IT-$\pi$ to identify inputs optimized for a
given $L_p$ error norms. For large $L_p$-norms, the error is
predominantly driven by extreme values determined by the information
in $\Pi^* = \Pi_2$. Conversely, for small $L_p$-norms, the error is
mainly influenced by $\Pi^* = \Pi_1$, which governs the weak
variations in $\Pi_o^*$.

We apply IT-$\pi$ to identify optimal inputs for two norms, $p=1$ and
$p=10$. The candidate dimensionless variables can be expressed as $\Pi
= \Pi_1\,\Pi_2^\beta$, where $\beta$ is a free parameter. 
Figures
\ref{fig:p_norm_time_series}(b) and (c) illustrate how the irreducible
error depends on $\beta$ for the two error norms. 

 {To identify $\alpha^*$ that generates the tightest possible bound, we perform a golden-section search over $\alpha \in\left(\frac{1}{1+p}, 10\right]$.
As shown in Figure ~\ref{fig:p_norm_time_series}(d) and (e), the value of $\alpha^*$ that generates the tightest bound $\epsilon_{LB}$ varies with the norm used.}
For $p=1$ (panel
(b)), the optimal dimensionless variable is $\Pi^* = \Pi_1$,
corresponding to $\beta^* = 0$.  {The tightest bound is achieved using $\alpha^*=1.0$.} In contrast, for $p=10$ (panel (c)),
IT-$\pi$ identifies $\Pi^* = \Pi_2$, consistent with $\beta^* \to
\infty$.  {The tightest bound is achieved when $\alpha^*$ is close to 0.3.} This validation case demonstrates that IT-$\pi$ can
effectively tailor the optimal dimensionless input $\Pi^*$ to the
chosen $L_p$-norm.


\clearpage
\subsection{Details of the dimensionless input discovered by other methods}
 The dimensionless input discovered by other methods using the same output $\Pi_o^*$ across all validation and application cases in the main text, is summarized in
Table \ref{tab:comparative_details}.
\begin{table}[h!]
\centering
\setlength{\arrayrulewidth}{1.2 pt} 
\renewcommand{\arraystretch}{2}
\resizebox{\textwidth}{!}{%
{
\begin{tabular}{|l|c|c|c|c|}
\hline
\textbf{Method} & \textbf{Active Subspaces} & \textbf{PyDimension} & \textbf{BuckiNet} & \textbf{{BSM}} \\ 
\hline
 Rayleigh problem  & {${y}{t^{-0.5} \left(\mu/\rho\right)^{-0.5}}$} 
&{${y}{t^{-0.5} \left(\mu/\rho\right)^{-0.5}}$}
& {${y}{t^{-0.5} \left(\mu/\rho\right)^{-0.5}}$}
&{${y}{t^{-0.5} \left(\mu/\rho\right)^{-0.5}}$} \\
\hline
 Colebrook equation & $[\left(U \rho D/\mu\right)^{-0.3} (k/D)^{0.4},$
 & $ {\mu^{-0.8}}{U^{0.8} \rho^{0.8} D^{0.3} k^{0.5}}$  
 & $[U\rho k/ \mu,$
 & N/A  \\
 &$(U \rho D/\mu)^{-0.4} (k/D)^{-0.6} ]$
 &
 &$(U\rho D/ \mu) (D/k)]$
 &\\
\hline
 Malkus waterwheel & $[\frac{r^{1.6}  q_1^{1.6}  I^{1.2}  g^{1.9}  \omega^{0.3}  m_1^{1.0}}{v^{2.9}  K^{2.7}  m_2^{0.8}},$ 
 & $\frac{I^{0.8} g \omega^{0.2} m_1^{0.8} m_2^{0.4}}{r^{1.6} q_1^{1.5} \nu^{0.5} K^{0.2}}$ 
 & $[\frac{r I^{1.8} K^{2.5} \omega^{1.7} m_2^{1.25}}{q_1^{0.5} v^2 g^{0.8} m_1^{0.7}},$
 & N/A  \\
 & $\frac{r^{0.1}  v^{0.1}  K^{0.1}  g^{0.1}  m_1^{0.8}}{q_1^{0.3}  I^{0.2} \cdot \omega^{0.2}  m_2^{0.3}}]$
 & 
 & $\frac{r q_1 K^{0.5}}{v^{0.3} g^{0.7} m_1^{0.8}}]$
 &\\
\hline
 Rayleight-Bénard convection 
 & $\rho^{0.6} h^{2.1} \Delta T g^{0.7} \alpha /(\mu^{0.6} \kappa^{0.8})$
 & $\rho h^3 \Delta T g \alpha /(\mu \kappa)$
 & $h^{-3.6} \Delta T \alpha \mu^{9.3} g^{-1.2} \kappa^{-6.9} \rho^{-9.3}$ 
 & N/A \\
 \hline

Blasius boundary layer & $U^{0.5} y^{1.0} \rho^{0.5} /\left(\mu^{0.5} x^{0.5}\right)$
 & $U^{0.5} y^{1.0} \rho^{0.5} /\left(\mu^{0.5} x^{0.5}\right)$
 & $U^{0.5} y^{1.0} \rho^{0.5} /\left(\mu^{0.5} x^{0.5}\right)$ 
 & $U^{0.5} y^{1.0} \rho^{0.5} /\left(\mu^{0.5} x^{0.5}\right) $\\
 \hline
 
Velocity scaling 
& $(\rho/\rho_w)^{-0.25} (\mu_w/\mu)$ 
& ${\left(\rho / \rho_w\right)^{1}\left(\mu / \mu_w\right)^{0.9}}{\left(y \rho {\sqrt{\tau_w/\rho_w}} / \mu\right)^{-0.3}} $
& ${y \rho_w {\sqrt{\tau_w/\rho_w}}}{\mu^{-0.333} \mu_w^{-0.667}}$
& N/A  \\
\hline
 Wall flux 
 & $\mathbf{\Pi}_{\tau} =[\frac{{(T/T_w)}^{0.1} {(y/k_{rms})}^{0.7}  ES^{0.1}}{M^{0.1}  {(y/Ra)}^{0.7}},$
 & $\Pi_{\tau} = \frac{Pr^{0.1} \cdot {(y/Ra)}^{0.7}}{{(T/T_w)}^{0.2} \cdot Re^{0.8} \cdot {(y/k_{rms})}^{0.1} \cdot ES^{0.4}}$ 
 & $\mathbf{\Pi}_{\tau} = [\frac{{(T/T_w)}^{0.5} \cdot Re^{0.5} \cdot {(y/k_{rms})}^{0.3} \cdot {(y/Ra)}^{0.3} \cdot ES^{0.8}}{M^{0.1} \cdot Pr^{0.3}},$ 
 & N/A  \\
&$\frac{{(T/T_w)}^{0.6} \cdot Re^{0.1} \cdot {(y/Ra)}^{0.2} \cdot ES^{0.3}}{M^{0.2} \cdot {(y/k_{rms})}^{0.2}}]$
&
&$\frac{{(T/T_w)}^{0.6} \cdot Re^{0.1} \cdot M^{0.9} \cdot Pr^{0.4} \cdot {(y/Ra)}^{0.5}}{{(y/k_{rms})}^{0.8} \cdot ES^{0.8}}]$
&\\
&$\mathbf{\Pi}_{q} =[\frac{{(T/T_w)}^{0.7} \cdot {(y/k_{rms})}^{0.1}}{{(y/Ra)}^{0.1}},$
&$\Pi_{q} =\frac{Pr^{0.7} \cdot {(y/Ra)}^{0.8}}{Re^{0.6} \cdot {(y/k_{rms})}^{0.6}}$
&$\mathbf{\Pi}_{q} =[\frac{{(T/T_w)}^{1.0} \cdot Re^{0.5}}{M^{0.1} \cdot {(y/k_{rms})}^{0.5} \cdot ES^{0.2}},$
&\\
&$\frac{{(y/k_{rms})}^{0.7}}{{(T/T_w)}^{0.1} \cdot {(y/Ra)}^{0.7}}]$
&
&$\frac{{(T/T_w)}^{2.3} \cdot Re^{0.9} \cdot {(y/Ra)}^{0.1}}{M^{1.0} \cdot Pr^{0.1} \cdot ES^{2.5}}]$
&\\
\hline
Skin friction
& $U_{\infty}^{0.25} (\nu/\rho)^{-0.14} M^{-0.05} \delta^{0.12} {\delta^*} \theta^{-0.92} $
& $U_{\infty}^{0.8} (\nu/\rho)^{-0.2} M^{-0.3} \delta^{0.3} {\delta^*} \theta^{-0.8} $
&$U_{\infty}^{1.6} (\nu/\rho)^{1} M^{-1.3} \delta^{-1.9} {\delta^*} \theta^{1.2} $
& N/A
\\
\hline
MHD power generator
& ${l^3 \rho \frac{d p}{d x}}/{\mu^2}$
& ${l^3 \rho \frac{d p}{d x}}/{\mu^2}$
& ${l^3 {B_0}^{5.5} \eta^{-2.8} {\frac{d p}{d x}}^{-0.8}}{\mu^{-1} \rho^{-0.8}}$
& N/A
\\
\hline
Laser-metal interaction
& ${\eta P}^{-0.03}{\rho}^{0.03}{C_p}^{-0.41} ({T_1-T_0})^{-0.41} {r_0}^{0.81} {\alpha}^{0.08} $
& $\frac{\eta P}{\left(T_1-T_0\right)\rho C_{{p}} \sqrt{\alpha V_{{s}} r_0^3}}$
& ${\eta P}^{-0.1}{\rho}{C_p}^{0.1} ({T_1-T_0})^{0.5} {r_0}^{-1} {\alpha}^{-0.1} $
& N/A
\\
\hline
\end{tabular}%
}
}
\caption{Comparison of Active Subspaces, PyDimension, BuckiNet, and
  {BSM} across validation and application cases. The
  table presents the discovered inputs $\mathbf{\Pi}$ identified
  by each method. N/A refers to `not applicable'.}
\label{tab:comparative_details}
\end{table}

\newpage
\subsection{Comparison of optimal and suboptimal dimensionless variables}

We show the advantage of using IT-$\pi$ compared to a suboptimal
solution from Buckingham-$\pi$ theorem.  We illustrate this in two
different problems: the Rayleigh problem and dimensionless learning
for wall fluxes.\\

\begin{figure}
    \centering
    \begin{subfigure}[b]{0.48\linewidth}
        \centering
        \includegraphics[width=\linewidth]{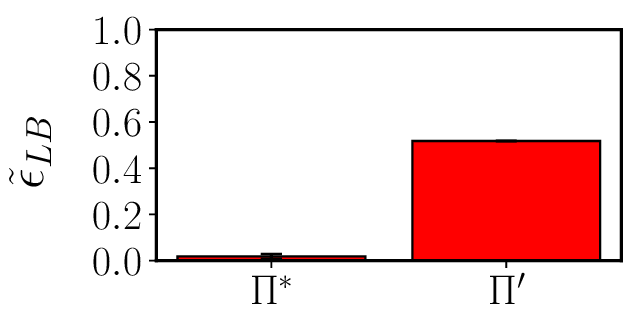}
        \put(-240,100){$(a)$}
    \end{subfigure}
    \hfill
    \begin{subfigure}[b]{0.48\linewidth}
        \centering
        \includegraphics[width=\linewidth]{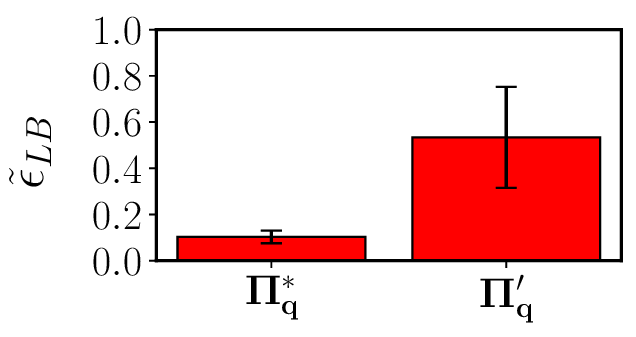}
                \put(-240,100){$(b)$}
    \end{subfigure}
    \caption{Comparison of optimal and suboptimal dimensionless
      variables in terms of the normalized irreducible error,
      $\tilde{\epsilon}_{LB}(\Pi)$, for (a) the Rayleigh problem using
      a single input $\Pi^*$ (optimal) and $\Pi'$ (suboptimal) and (b)
      dimensionless learning for wall fluxes using the two optimal
      inputs in $\mathbf{\Pi}^*$ and the four suboptimal inputs in
      $\mathbf{\Pi}'$.}
    \label{fig:rank_sub}
\end{figure}

{\bf The Rayleigh Problem}. Let us assume we want to explain the
dimensionless output $u/U$ using only one dimensionless variable.  A
simple suboptimal dimensionless input extracted from the
Buckingham-$\pi$ theorem is $\Pi' = y/(Ut)$. Figure
\ref{fig:rank_sub}(a) shows that $\Pi'$ yields a normalized
irreducible error of approximately 50\%, whereas the optimal
dimensionless variable from IT-$\pi$ achieves an irreducible error
close to 0\%. \\

{\bf Dimensionless learning for wall fluxes}. The suboptimal
dimensionless inputs, denoted as $\mathbf{\Pi_q}'$, are
constructed from the Buckingham-$\pi$ theorem as follows:
$$
\Pi_{q,1}' = \frac{y\,u\,\rho}{\mu}, \quad
\Pi_{q,2}' = \frac{u}{\sqrt{c_p\,T}}, \quad
\Pi_{q,3}' = \frac{c_p\,\mu}{\kappa}, \quad
\Pi_{q,4}' = \frac{y}{k_{\mathrm{rms}}}.
$$ These variables are chosen because they coincide with four
well-known dimensionless groups in the fluid dynamics community: the
local Reynolds number, Mach number, Prandtl number, and the relative
roughness height. Hence, they can be expected to be a reasonable
choice by experts in the field. Figure \ref{fig:rank_sub}(b) indicates
that, despite containing four variables, the suboptimal inputs
$\mathbf{\Pi_q}'$ yield an irreducible error of 0.6 compared to
0.1 when using the optimal inputs $\mathbf{\Pi_q}^*$, even though
the latter comprises only two dimensionless variables.
When training neural networks with the ANN$_3$ architecture, using
four suboptimal inputs results in a reduced efficiency of 82\% despite
using four inputs instead of two.

\subsection{{Comparison of gradient-based and IT-$\pi$ rankings}}
\label{sec:comparison_rank}

{We compare the sensitivity of the discovered
  dimensionless variables and their associated importance rankings
  between gradient-based methods and IT-$\pi$. Three challenging
  scenarios are considered: (i) when the list of dimensional input
  variables is incomplete, (ii) when the magnitudes of the input
  variables differ significantly, and (iii) when the available data
  are limited relative to the input dimensionality. We show that, in
  such cases, gradient-based methods may fail to accurately identify
  and rank the variables that are truly important for predicting the
  output. In contrast, IT-$\pi$ performs robustly, correctly
  identifying the most predictive input variables and their relative
  importance.  }

{As presented in the Results section of the main text,
  the ranking in IT-$\pi$ is constructed by ordering the normalized
  irreducible error associated with each individual dimensionless
  variable. For example, if
\[
\epsilon_{LB}(\Pi_2) \geq \epsilon_{LB}(\Pi_1) \geq \cdots \geq \epsilon_{LB}(\Pi_4),
\]
then $\Pi_4$ is identified as the most predictive variable and $\Pi_2$
as the least predictive. This ranking reflects the relative
informativeness of each variable in predicting the output.}

{In the gradient-based ranking frameworks, the
  importance of dimensionless variables is quantified using a
  derivative-based global sensitivity metric. This is achieved by
  expressing the physical relationship in a dimensionless form and
  applying a logarithmic transformation to the dimensionless
  inputs. Then, the importance of the $i$-th input variable is defined
  as
\begin{equation}
  \nu_i = \int \left( \frac{\partial}{\partial \gamma_i} g(\mathbf{\gamma}) \right)^2 \rho(\mathbf{\gamma}) \, d\mathbf{\gamma}
  = \mathbb{E} \left[ \left( \frac{\partial}{\partial \gamma_i} g(\mathbf{\gamma}) \right)^2 \right], \qquad i = 1, \dots, m,
\label{eq:importance}
\end{equation}
where $g(\mathbf{\gamma})$ is the response surface (i.e. model),
$\rho(\mathbf{\gamma})$ is a probability density over the input space,
and $\gamma_i = \log(\Pi_i)$ denotes the $i$-th log-transformed dimensionless
variable. The metric $\nu_i$ measures the average change in the output
$g$ as $\Pi_i$ is perturbed within the support of $\rho$. If $\nu_i >
\nu_j$, then perturbing $\Pi_i$ is expected to influence $g$ more than
perturbing $\Pi_j$, on average. This provides a natural definition of
variable importance and allows for a ranking based on the sensitivity
of the output to each input direction. 
For the response surface $g$, we employ a kernel regression model with a scaled radial basis function kernel, where the kernel width is treated as a learnable hyperparameter and adaptively optimized during training.
Next, we discuss several
challenging scenarios, each accompanied by an illustrative example.  }

\begin{figure}[h]
    \centering
    \includegraphics[width=0.9\linewidth]{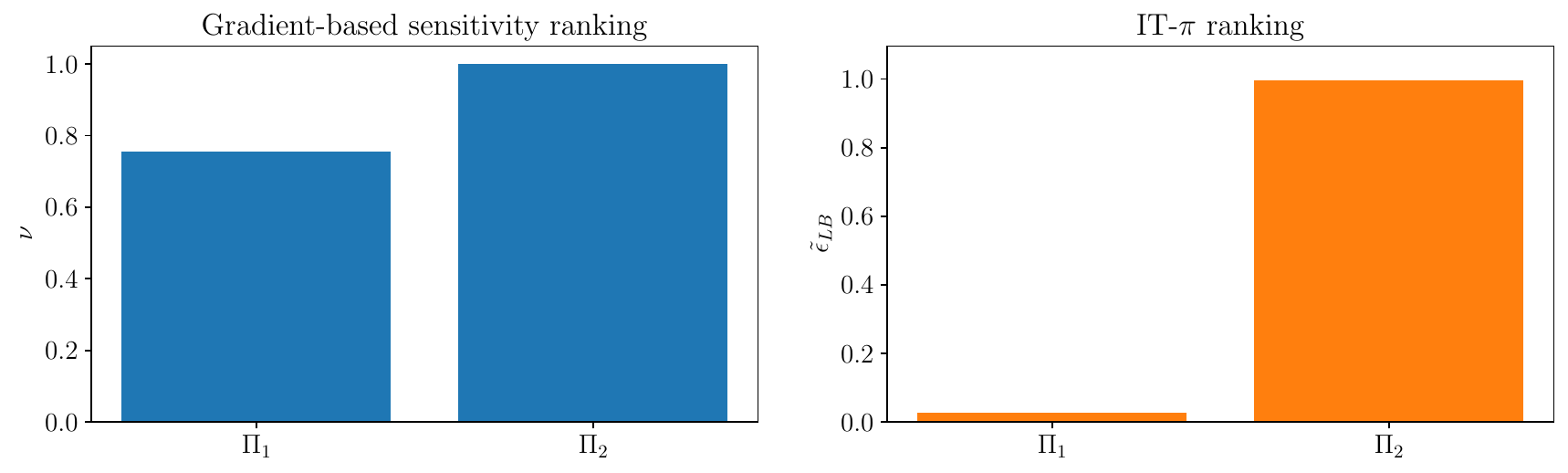}
    \caption{{Comparison between gradient-based and IT-$\pi$ rankings of input variables in the presence of large variations in input magnitudes. In the gradient-based ranking, higher values of $\nu$ indicate greater importance, whereas in the IT-$\pi$ ranking, lower values of $\epsilon_{LB}$ correspond to higher predictive relevance. }}
    \label{fig:ranking-comparison}
\end{figure}

 \begin{itemize}
 \item \textbf{{Incomplete list of dimensional input
     variables }}

{ We analyze the case in which one or more relevant
  dimensional variables are missing from the input set. This is a
  common situation that may arise when certain variables are not
  measurable, are unknown, or are discarded a priori under the
  assumption that they are irrelevant.  To illustrate this scenario,
  we construct an example in which one relevant dimensional variable
  is omitted from the input list. The output is defined as
\[
\Pi_o = \Pi_1^{1/3} \Pi_2^{2/3} \Pi_3,
\]
where
\[
\Pi_1 = \frac{u_1 y}{\mu / \rho}, \quad \Pi_2 = \frac{u_2 y}{\mu / \rho}, \quad \Pi_3 = \frac{u_3 y}{\mu / \rho},
\]
and $u_1$, $u_2$, $u_3$ are velocities, $y$ is a characteristic
length, $\mu$ is the dynamic viscosity, and $\rho$ is the density.
The available input variables are given by $\mathbf{q} =
\left[u_1, u_2, y, \mu, \rho\right]$, which excludes $u_3$---a
variable necessary for accurately predicting $\Pi_o$. The analytically
optimal dimensionless input based on the available set
$\mathbf{q}$ is:
\[
\Pi^{*, \text{Analytical}} = \Pi_1^{1/3} \Pi_2^{2/3} = \frac{u_1^{1/3} u_2^{2/3} y}{\mu / \rho}.
\]
We generate 1,000 samples of the dimensional inputs and the
corresponding output, and apply both gradient-based ranking and
IT-$\pi$ to identify the dominant dimensionless variable.  }

\noindent {\underline{Gradient-based ranking} results
  in the dominant input direction 
\[
\Pi^{*, \text{Gradient-based}} = \frac{u_1^{0.7} u_2^{0.26} y}{\mu / \rho},
\]
which does not recover the true analytical solution. The deficiency
arises from the fact that gradient-based methods rely on the gradient
of a model $g$, which approximates the true physical
relationship. However, when a key dimensional input is missing, the
model $g$ becomes a projection of the true relationship onto a
lower-dimensional subspace that excludes an essential direction. This
omission prevents the computed dimensionless variables from fully
spanning the relevant physical parameter space. As a result, the
gradient information becomes misleading: variables that are actually
important may appear unimportant, and vice versa. This distortion
compromises the reliability of gradient-based rankings, making them an
inaccurate reflection of the true influence of the underlying
dimensionless variables.}

\noindent {\underline{IT-$\pi$ ranking}  correctly
  identifies the optimal input,
\[
\Pi^* = \frac{u_1^{0.33} u_2^{0.67} y}{\mu / \rho} \approx \frac{u_1^{1/2} u_2^{2/3} y}{\mu / \rho},
\]
despite the missing variable. The reason for this success arises from
the fact that, unlike gradient-based methods---which rely on a
complete set of dimensional inputs to compute sensitivities—IT-$\pi$
does not rely on gradients or predefined model structures. Instead,
the discovery of predictive dimensionless variables is formulated as
an optimization problem, where the objective is to minimize the
irreducible error. This model-free, information-theoretic formulation
results in greater robustness of the method. As a result, when a
relevant input variable is missing, IT-$\pi$ is still able to identify
the best possible predictive dimensionless combination from the
available inputs---without generating distorted rankings or depending
on incomplete gradient information. This makes IT-$\pi$ particularly
well-suited for real-world scenarios in which some relevant physical
variables may be unmeasurable or unknown a priori. }


\item \textbf{{Magnitude of input variables differs
    significantly}}
 
{The second challenging situation arises when the
  range of values taken by a variable differs significantly from that
  of others, even though this disparity is not reflected in the
  governing equation itself. To illustrate this case, we construct an
  analytical example in which the output $\Pi_o$ is related to the inputs $\mathbf{\Pi}= [\Pi_1, \Pi_2]$ as
  follows:
$$
\Pi_o = \sin\left(\log \left(\Pi_1 \right) \right) + \log(\Pi_2) = \sin(\gamma_1) + \gamma_2,
$$
where
$$
\Pi_1 = e^{\gamma_1}, \quad \Pi_2 =  e^{\gamma_2},
$$
and
$$
\gamma_1 = \log(\Pi_1) \sim \operatorname{Uniform}[0,1], \quad \gamma_2 = \log(\Pi_2) \sim \operatorname{Uniform}[0,0.02].
$$
Note that the values taken by $\Pi_2$ are much smaller in magnitude
than those of $\Pi_1$. In this example, $\Pi_1$ is significantly more
informative, as it explains nearly all of the variance in the
output. In contrast, $\Pi_2$ contributes only a small perturbation and
effectively behaves as noise. }
%

\noindent {\underline{Gradient-based ranking.}
The gradient-based sensitivities are given by
\[
\begin{gathered}
\nu_1 =\mathbb{E} \left[ \left( \frac{\partial \Pi_o}{\partial \gamma_1} ) \right)^2 \right] =
\mathbb{E}\left[\cos^2(\gamma_1)\right] = \frac{1}{2} + \frac{\sin(2)}{4} \approx 0.727, \\
\nu_2 =\mathbb{E} \left[ \left( \frac{\partial \Pi_o}{\partial \gamma_2} ) \right)^2 \right] = \mathbb{E}\left[1^2\right] = 1.
\end{gathered}
\]
As shown in the left panel of Figure~\ref{fig:ranking-comparison}, the
gradient-based criterion incorrectly ranks $\Pi_2$ as more important
than $\Pi_1$. This outcome arises because the gradient-based metric
does not account for the significantly narrower range of values of
$\Pi_2$, which contributes negligibly to the overall variability of
$\Pi_o$. Despite its high gradient magnitude, $\Pi_2$ behaves
effectively as noise, while $\Pi_1$ governs most of the output
variance.  }\\

\noindent { \underline{IT-$\pi$ ranking}.  The ranking
  fromIT-$\pi$ is based on the irreducible error of each individual
  variable
\[
{\tilde \epsilon_{LB}(\Pi_1) = 0.03, } \quad 
{\tilde \epsilon_{LB}(\Pi_2) = 0.99.}        
\]
As shown in the right panel of Figure~\ref{fig:ranking-comparison},
IT-$\pi$ correctly identifies $\Pi_1$ as the dominant
contributor. This example highlights that gradient-based rankings must
be applied with caution when input variables differ significantly in
scale.  This limitation can be easily mitigated by standardizing the
dimensional input variables to lie within similar ranges before model
fitting and gradient computation. Nonetheless, IT-$\pi$ does not
require such preprocessing as its formulation is inherently invariant
to input scaling.}


\item \textbf{{Available data are limited
  relative to the input dimensionality }}

  {The third challenging scenario arises when only a
    limited number of samples are available for analysis. To
    illustrate this case, we revisit the Rayleigh problem under
    data-scarce conditions, specifically considering a setting with
    only 27 samples. As shown in Table~\ref{tab:limited_data},
    IT-$\pi$ successfully recovers the known optimal dimensionless
    variable, whereas the gradient-based method (Active Subspaces)
    fails to identify the correct ranking and instead produces a
    spurious leading variable. This failure stems from the fact that,
    in high-dimensional settings with limited data, the surrogate
    response surface fitted by Active Subspaces is poorly constrained
    and does not faithfully represent the underlying physical
    relationship. Although IT-$\pi$ is also susceptible to errors when
    data are extremely sparse, it tends to perform more reliably by
    avoiding the use of gradient estimates from surrogate models,
    resulting in greater robustness under limited data conditions.  }
\end{itemize}


\subsection{{Data partitioning and generalization in heat flux ANN }}
\label{sec:data_partitioning}

\begin{figure}[H]
    \centering
    \includegraphics[width=0.5\linewidth]{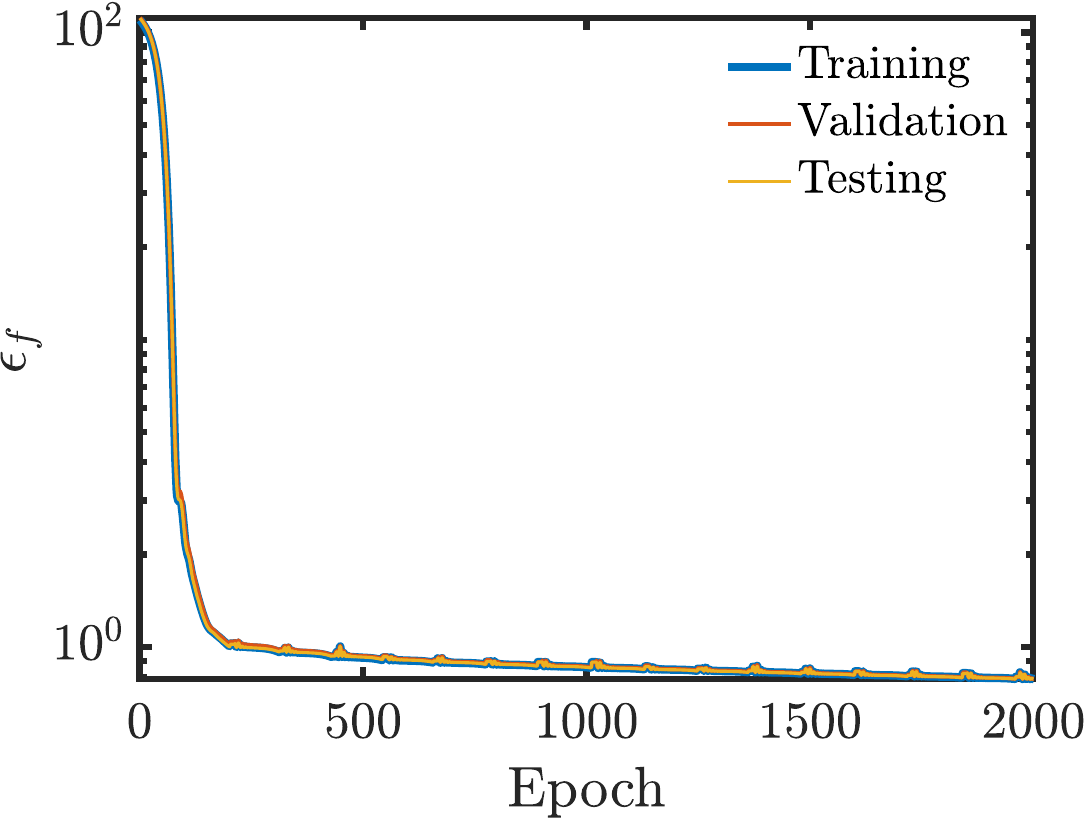}
    \caption{{Model error of $\mathrm{ANN}_3$ across
        training, validation, and test datasets as a function of
        training epochs.}}
    \label{fig:training_validation_error}
\end{figure}

{ As discussed in \S\ref{sec:challenges}, a central
  challenge in data-driven dimensionless learning is the
  transferability and generalizability of the discovered dimensionless
  variables. We argued that this limitation often stems not from the
  learning method itself, but from the selection of the training
  data. Here, we validate the generalizability of the dimensionless
  variables identified for the heat flux model using the ANN described
  in the Applications section of the manuscript.

The model was constructed by partitioning the dataset into training
(70\%), validation (15\%), and testing (15\%) subsets. The
dimensionless variables identified by IT-$\pi$ were derived using only
the training data and validated on the held-out validation set; the
test set remained completely unused during the discovery phase. Final
performance was assessed using the testing set, which contains
previously unseen cases not involved in model training.

Table~\ref{tab:dataset_splits} summarizes the wall roughness topology
(denoted by ``Roughness ID''), Mach number ($M_c$), Reynolds number
($Re_c$), and set assignment.  The ``Roughness ID'' labels each of the
ten irregular, multiscale rough surfaces, which are generated using
Gaussian probability density function.  Additional details about the
dataset are provided in
Ref.~\cite{ma2024building}. Figure~\ref{fig:training_validation_error}
shows the epoch-by-epoch errors for training, validation, and test
sets. The similarity of these curves indicates the generalization
within the sampled $\Pi$-space. 
  At convergence, the training, validation, and test mean squared
  errors correspond to approximately $0.63 \%, 0.64 \%$, and $0.63 \%$
  of the variance of the target variable, respectively.
  These results confirm the ability of the
model to generalize and accurately predict unseen cases with
performance comparable to that on the training data.}
\begin{table}[htbp]
{
  \centering
  \caption{{Dataset splits of ANN model for wall heat flux.}}
  \label{tab:dataset_splits}
  \begin{subtable}[t]{0.48\linewidth}
    \centering
    \begin{filecontents*}{Table/IT_PI_review_part1.csv}
ID,Mach,Re,Set
1,1,4000,validation
1,1,8000,training
1,1,16000,training
1,2,8000,training
1,2,16000,training
1,4,8000,test
1,4,16000,test
2,1,4000,training
2,1,8000,validation
2,1,16000,training
2,2,4000,training
2,2,8000,test
2,2,16000,validation
2,4,8000,training
2,4,16000,training
3,1,4000,training
3,1,8000,training
3,1,16000,training
3,2,8000,training
3,2,16000,training
3,4,8000,training
3,4,16000,training
4,1,4000,validation
4,1,8000,training
4,1,16000,training
4,2,8000,test
4,2,16000,training
\end{filecontents*}
\csvreader[
      tabular=cccc,
      table head=\toprule Roughness ID & $M_c$ & $Re_c$ & Set \\ \midrule,
      late after last line=\\\bottomrule
    ]{Table/IT_PI_review_part1.csv}%
    {ID=\ID,Mach=\Mach,Re=\Re,Set=\Set}%
    {\ID & \Mach & \Re & \Set}
  \end{subtable}%
  \hfill
  \begin{subtable}[t]{0.48\linewidth}
    \centering
    \begin{filecontents*}{Table/IT_PI_review_part2.csv}
ID,Mach,Re,Set
4,4,8000,training
4,4,16000,training
5,1,8000,training
5,1,16000,training
5,2,8000,training
5,2,16000,training
5,4,8000,training
5,4,16000,training
6,2,8000,training
6,2,16000,training
6,4,8000,training
6,4,16000,training
7,2,8000,test
7,2,16000,training
7,4,8000,training
7,4,16000,training
8,2,8000,training
8,2,16000,training
8,4,8000,validation
8,4,16000,test
9,2,8000,training
9,2,16000,training
9,4,8000,training
9,4,16000,validation
10,2,8000,training
10,2,16000,validation
10,4,8000,training
10,4,16000,test
\end{filecontents*}
\csvreader[
      tabular=cccc,
      table head=\toprule Roughness ID & $M_c$ & $Re_c$ & Set \\ \midrule,
      late after last line=\\\bottomrule
    ]{Table/IT_PI_review_part2.csv}%
    {ID=\ID,Mach=\Mach,Re=\Re,Set=\Set}%
    {\ID & \Mach & \Re & \Set}
  \end{subtable}
  }
\end{table}

\newpage

\end{document}